\documentclass[manuscript,screen]{acmart}

\AtBeginDocument{%
  }

 \setcopyright{rightsretained}
\acmJournal{PACMHCI}
\acmYear{2025} \acmVolume{10} \acmNumber{CSCW1} \acmArticle{} \acmMonth{6} \acmPrice{}\acmDOI{TBD}
\makeatletter
\def\@copyrightpermission{
  © {Owner/Author | ACM} {2025}. This is the author's version of the work. 
  It is posted here for your personal use. Not for redistribution.
  The definitive Version of Record will be published soon.}
\makeatother

\usepackage{verbatim}
\usepackage{xcolor}
\usepackage{wrapfig}

\usepackage{graphicx}

\usepackage{array}
\usepackage{longtable}
\usepackage{adjustbox}
\usepackage{float}

\begin{document}

\title[Toward the Expansion of 
Chaplaincy Care into Online Spiritual Care Communities]{Meeting Patients Where They're At: Toward the Expansion of 
Chaplaincy Care into Online Spiritual Care Communities}

\author{Alemitu Bezabih}
\orcid{0000-0001-9603-8537}
\affiliation{%
  \institution{Colorado School of Mines, Department of Computer Science}
  \city{Golden}
  \state{CO}
  \postcode{80401}
  \country{USA}}
\email{alemitubezabih@mines.edu}

\author{Shadi Nourriz}
\orcid{0009-0004-1873-6750}
\affiliation{%
  \institution{Colorado School of Mines, Department of Computer Science}
  \city{Golden}
  \state{CO}
  \country{USA}}
\email{shadinourriz@mines.edu} 

\author{Anne-Marie Snider}
\orcid{0000-0002-3362-929X}
\affiliation{%
  \institution{California State University at Chico, Department of Sociology}
  \city{Chico}
  \state{CA}
  \country{USA}}
  \postcode{95929}
\email{asnider@csuchico.edu} 

\author{Rosalie Rauenzahn}
\orcid{0009-0004-6073-8709}
\affiliation{%
  \institution{Colorado School of Mines, Department of Engineering, Design, \& Society}
  \city{Golden}
  \state{Colorado}
  \country{USA}
  \postcode{80401}}
\email{rauenzahn@mines.edu} 

\author{George Handzo}
\orcid{0000-0001-6683-7303}
\affiliation{%
  \institution{HealthCare Chaplaincy Network}
  \city{New York}
  \state{New York}
  \country{USA}
  \postcode{10018}}
\email{ghandzo@healthcarechaplaincy.org} 

\author{C. Estelle Smith}
\orcid{0000-0002-4981-7105}
\affiliation{%
  \institution{Colorado School of Mines, Department of Computer Science}
  \city{Golden}
  \state{Colorado}
  \country{USA}
  \postcode{80401}}
\email{estellesmith@mines.edu}

\renewcommand{\shortauthors}{Bezabih et al.}

\begin{abstract}
Despite a growing need for spiritual care in the US, it is often under-served,  inaccessible, or misunderstood, while almost no prior work in CSCW/HCI research has engaged with professional chaplains and spiritual care providers. This interdisciplinary study aims to develop a foundational understanding of how spiritual care may (or may not) be expanded into online spaces---especially focusing on anonymous, asynchronous, and text-based online communities. We conducted an exploratory mixed-methods study with chaplains ($N=22$) involving interviews and user testing sessions centered around Reddit support communities to understand participants' perspectives on technology and their ideations about the role of chaplaincy in prospective Online Spiritual Care Communities (OSCCs). Our Grounded Theory Method analysis highlighted benefits of OSCCs including: meeting patients where they are at; accessibility and scalability; and facilitating patient-initiated care. Chaplains highlighted how their presence in OSCCs could help with shaping peer interactions, moderation, synchronous chats for group care, and redirecting to external resources, while also raising important feasibility concerns, risks, and needs for future design and research. We used an existing taxonomy of chaplaincy techniques to show that some spiritual care strategies may be amenable to online spaces, yet we also exposed the limitations of technology to fully mediate spiritual care and the need to develop new online chaplaincy interventions. Based on these findings, we contribute the model of a ``Care Loop'' between institutionally-based formal care and platform-based community care to expand access and drive greater awareness and utilization of spiritual care. We also contribute design implications to guide future work in online spiritual care. 
\end{abstract}




\maketitle

\newcolumntype{A}{>{\raggedright\arraybackslash}m{.3cm}}
\newcolumntype{Z}{>{\raggedright\arraybackslash}m{2.5cm}}
\newcolumntype{Y}{>{\raggedright\arraybackslash}m{2.8cm}}
\newcolumntype{D}{>{\raggedright\arraybackslash}m{0.7cm}}
\newcolumntype{B}{>{\raggedright\arraybackslash}m{1.8cm}}
\newcolumntype{C}{>{\raggedright\arraybackslash}m{1.1cm}}

\newcommand{\participants}{
\begin{table}
\footnotesize
    
    \centering
    
    \begin{tabular}{A  Z   Y   D   B   B   D}
    \hline
    \textbf{ID} & \textbf{Professional Title} & \textbf{Credentials} 
    & \textbf{Years} & \textbf{R/S Affiliation} & \textbf{Location} & \textbf{Reddit} \\
    \hline
    \hline
    P01 & Professor & Ordained Minister, Licensed Psychologist & \textgreater 10 & Christianity & Western USA & N \\
    \hline
    P02 & Chaplain & M.Div. & 1-5 & Islam & Southern USA & Y \\
    \hline
    P03 & Chaplain, Spiritual Leader & M.Div. & \textgreater 10 & Christianity & Western USA & N \\
    \hline
    P04 & Chaplain, Spiritual Care Counselor, Spiritual Director & M.Div. & 5-10 & Christianity & Western USA & Y \\
    \hline
    P05 & Spiritual Director, Spiritual Leader & M.S. in Pastoral \& Spiritual Care & \textgreater 10 & Christianity & Western USA & N \\
    \hline
    P06 & Director of Case Management & Licensed Social Worker 
    & 1-5 & Christianity & Western USA & Y \\
    \hline
    P07 & Chaplain & M.Div. & \textless 1 & Christianity & Western USA & N \\
    \hline
   P08 & Chaplain & Pending ACPE  
   & 1-5 & Humanist & Western USA & Y \\
    \hline
    P09 & Physician & M.D. & \textgreater 10 & Christianity & East Coast & N \\
    \hline
    P10 & Chaplain, Director, Spiritual Care \& Education & BCC, Advanced Practice BCC & 5-10 & Not Disclosed & East Coast & Y \\
    \hline
    P11 & Manager, Chaplaincy Department & M.Div., BCC, ACPE  
    & \textgreater 10 & Not Disclosed & East Coast & Y \\
    \hline
    P12 & Chaplain & BCC & \textgreater 10 & Christianity & East Coast & Y \\
    \hline
    P13 & Chaplain & BCC & \textgreater 10 & Islam & East Coast & N \\
    \hline
    P14 & Chaplain, Spiritual Director & M.S. in Buddhist Chaplaincy & \textgreater 10 & Buddhism & East Coast & N \\
    \hline
    P15 & Seminarian & Certified Death Doula & 1-5 & Buddhism & East Coast & N \\
    \hline
    P16 & Chaplain, Physician, Professor, Spiritual Care Specialist & M.Div., BCC & \textgreater 10 & Judaism & West Coast & Y \\
    \hline
    P17 & Chaplain, Professor & M.Div., BCC & \textgreater 10 & Judaism & West Coast & Y \\
    \hline
    P18 & Vice President of Spiritual Care & M.Div., ACPE 
    & \textgreater 10 & Christianity & West Coast & N \\
    \hline
    P19 & Chaplain, Mindfulness Educator & BCC & \textgreater 10 & Not Disclosed & West Coast & N \\
    \hline
    P20 & Chaplain & Doctorate in Ministry, BCC & \textgreater 10 & Christianity & Southern USA & N \\
    \hline
    P21 & Executive Director, Spiritual Care Counselor & M.Div., ACPE  
    & \textgreater 10 & Christianity & Southern USA & N \\
    \hline
    P22 & Chaplain, ACPE Educator & ACPE 
    & \textgreater 10 & Christianity & Western USA & N \\
    \hline
    \end{tabular}
    \caption{Participant Table. The ``Reddit'' column indicates participants' self-reported prior familiarity (Yes/Y) or lack of familiarity (No/N) with the Reddit platform. M.Div. denotes a Master's in Divinity; BCC, Board Certified Chaplain; M.S., Master's of Science; ACPE, Association for Clinical Pastoral Education Certification.}
    \label{tab:participants}
\end{table}
}

\newcommand{\subreddits}{
\begin{table}
    
    \centering
    \vspace{-15pt}
    \begin{tabular}{l c l}
    \hline
    \textbf{Subreddit} & \textbf{Frequency} & \textbf{Participants Visited} \\
    \hline
     \hline
    r/cancer & 9 & P04, P05, P06, P11, P12, P13, P17, P18, P22 \\
    \hline
    r/lonely & 9 & P01, P02, P03, P05, P06, P07, P10, P12, P22 \\
    \hline
    r/hospice & 5 & P16, P17, P19, P20, P21 \\
    \hline
    r/suicidewatch & 4 & P01, P03, P04, P11 \\
    \hline
    r/spirituality & 3 & P17, P18, P20 \\
    \hline
    r/ptsd & 2 & P08, P10 \\
    \hline
    r/bpdlovedones & 1 & P15 \\
    \hline
    r/chaplaincy & 1 & P20 \\
    \hline
    r/griefsupport & 1 & P18 \\
    \hline
    r/swresources & 1 & P04 \\
    \hline
    r/tfmr\_support & 1 & P08 \\
    \hline
    r/psychedelicspiritualy & 1 & P21 \\
    \hline
    
    r/trulyoffmychest & 1 & P17 \\
    \hline
    
    \end{tabular}
    \caption{Summary of Subreddits Visited by Participants.}
    \label{tab:subreddits}
    \vspace{-20pt}
\end{table}
}

\newcolumntype{J}{>{\raggedright\arraybackslash}m{2.9cm}}
\newcolumntype{K}{>{\raggedright\arraybackslash}m{3.8cm}}
\newcolumntype{L}{>{\raggedright\arraybackslash}m{6cm}}

\newcommand{\taxonomies}{
\begin{longtable}[t]{J K L}
\hline
\textbf{Taxonomy Item (Freq., Cat.)} & \textbf{Description} & \textbf{Example Quote} \\
\hline
\hline
\endfirsthead

\multicolumn{3}{c}{{\bfseries \tablename\ \thetable{} -- continued from previous page}} \\
\hline
\textbf{Taxonomy Item (Freq., Cat.)} & \textbf{Description} & \textbf{Example Quote} \\
\hline
\hline
\endhead

\hline 
\multicolumn{3}{|r|}{{Continued on next page}} \\ \hline
\caption[]{AHCTC taxonomy items identified from chaplains' verbal or typed replies to OPs on visited support subreddits; IE = Intended Effect, M = Method, I = Intervention.} \\
\endfoot

\caption{AHCTC taxonomy items identified from chaplains' verbal or typed replies to OPs on visited support subreddits; IE = Intended Effect, M = Method, I = Intervention.} \\
\endlastfoot
\\

\multicolumn{3}{l}{\textbf{Applicable Online}} 
\\
\hline \hline
Offer emotional support (11, M)  & Attending to emotional needs & \textit{``Maybe knowing you are not alone is the place to start, and by this thread you can see how many are with you.''} (P14, Typed)\\
\hline
Active Listening (8, I) & Purposeful listening to what someone is saying & \textit{``Listen to and observe someone's messaging and someone's statements, [understand] what's beneath it and kind of trying to read a little bit deeper.''} (P02, Verbal)\\
\hline
Demonstrate caring and concern (8, IE) & E.g., via offering emotional support \& encouraging self-care. & \textit{``This sounds really tough. Staying at the bedside all the time is totaly exhausting physically and emotionally.''} (P14, Typed)\\
\hline
Acknowledging response to difficult experience (7, I) & Normalizing emotions \& personal responses to difficult experience. & \textit{``I'm so sorry and I see the complex mix of emotions you're feeling amid this grief.''} (P20, Verbal)\\
\hline
Encourage sharing of feelings (6, M) & E.g., inviting to reminisce \& asking guided questions. & \textit{``[I'd ask OP] what feelings come up for you around that? Are you feeling sad, scared, angry, or peaceful about this?''} (P18, Verbal)\\
\hline
Encourage story-telling (6, M) & Surfacing meaning \& feelings via purposeful telling of stories \& life experiences. & \textit{``If you're interested in sharing a bit more about your experience, I'd be happy to just listen.''} (P15, Typed)\\
\hline
Ask guided questions (5, I) & Ask leading questions to understand the situation and bring particular responses. & \textit{``What have you done to get over something like this, or have you felt something like this before? Is there something specific weighing on you?''} 
 (P12, Typed)\\
\hline    
Prayer for healing (4, I) & A prayer with the intent to bring about any form of healing. & \textit{``I will light a candle for you today in your honor. Know you are cared for here.''}(P14, Typed)\\
\hline
Establish Rapport and connectedness (4, IE) & Establish an initial relationship. & \textit{``I would say hi, I'm a chaplain, a professional healthcare chaplain. I read your post, and I often support people like yourself in experiences, as you described.''} (P17, Verbal)\\
\hline
Encourage end-of-life review (4, M) & 
Facilitate preparation for the process of dying 
& \textit{``[I'd say] there're many resources available for you, but you know, we could go down the hospice road, we could go down the palliative care road.''} (P04, Verbal)\\
\hline
Provide sacred reading(s) (2, I) & Providing a Bible, Quran, Torah, etc. & \textit{``[Providing] this poem or painting, instead of you [users] scroll down to all the problems.''} (P22, Verbal)\\
\hline
Encourage self-care (1, M) & Encouraging emotional, physical, or spiritual self-care & \textit{``You definitely have to take care of yourself and if that means stepping out occasionally that is okay.''} (P12, Typed)\\
\hline
Collaborate with care team member (1, M) & Partner with interdisciplinary care team. & \textit{``It may be helpful if moderators could point lonely people who come to this subreddit to resources, URLs, -- support groups, etc.''} (P03, Typed)\\
\hline
\\
\multicolumn{3}{l}{\textbf{Challenging to Apply Online}} 
\\
\hline \hline
Convey a calming presence (8, IE) & Being present in a non-anxious manner. & \textit{``Really hard to replicate [in-person presence] in an online environment. ... It's hard to connect meaningfully via screen.''} (P15, Verbal)\\
\hline    
Provide compassionate touch (3, I) & Bodily touch to demonstrate compassion. & \textit{``I think the danger is only connecting by technology, not connecting in person because people still need human touch.''} (P22, Verbal)\\
\hline
\multicolumn{3}{l}{\textbf{New Techniques, not Included in AHCTC}} \\
\hline \hline
Encourage to seek support (2,--) & Encourage telling concerns to close people. & \textit{``I think that talking to a friend is rarely a bad idea. Our friends often want to be able to help us more than we allow, and being vulnerable with someone is a gift if they are trustworthy.''} (P08, Typed)\\
\hline
Provide practical advice (1, --) & Provide advice, but \textit{only} when it is specifically requested. & \textit{``Take the day off--no plans of any kind. Order food that feeds your soul. Sit outside. Feel alive as much as you can in this body to remind you that you are not in that moment anymore.''} (P08, Typed) \\    
\hline
Share own experiences \& stories (1,--) & Share relatable personal experiences. & \textit{``When I was close to your age, I also wrestled with thoughts of suicide. Today, I am in my mid-50s.''} (P03, Typed)\\
\hline
Educate on reply behaviors (1,--) & Teach users how to reply. & \textit{``Speaking [writing] to everyone else, ... I would say, it's a lesson that I've learned that I have to be as aware of the impact of what I do [though] I [have good] intentions behind it.''} (P18, Verbal)\\
\hline
Apologize (1, --) & Use apologies on behalf of others' unhealthy behaviors. & \textit{``I would apologize for humanity in some, you know, generic way. I'm really sorry that the comments though well intended, had that impact.''} (P18, Verbal)\\
\hline
\label{tab:taxonomies}
\end{longtable}
}

\newcolumntype{M}{>{\raggedright\arraybackslash}m{2.5cm}}
\newcolumntype{N}{>{\raggedright\arraybackslash}m{4.5cm}}
\newcolumntype{O}{>{\raggedright\arraybackslash}m{7cm}}

\newcommand{\screenshot}{
\begin{figure}[t]
  \centering
        \includegraphics[width=\textwidth]{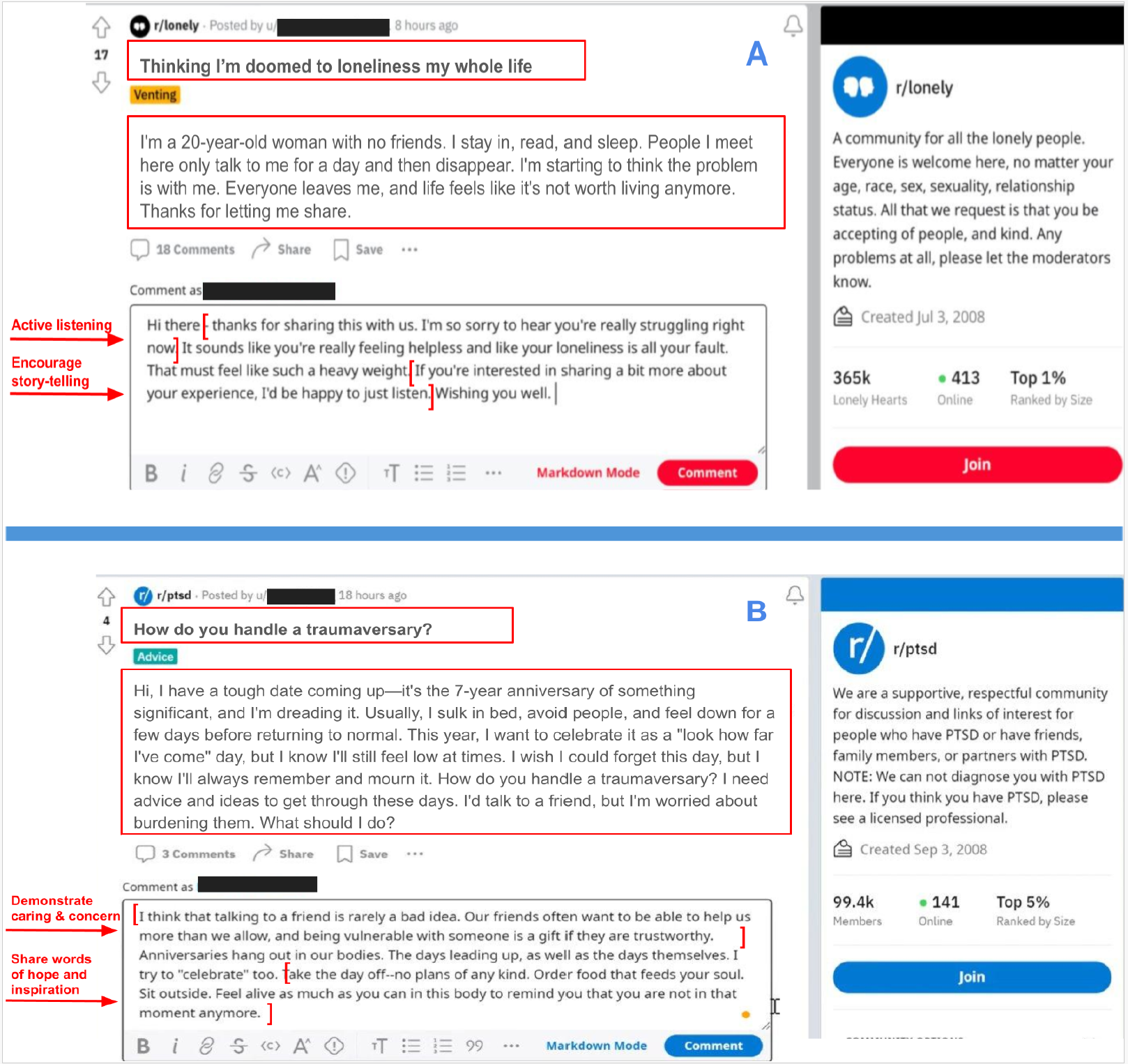}
        \captionof{figure}{Example screenshots depicting participants' prospective responses to live original posts. All text in red boxes is paraphrased from original public posts for user privacy. \textbf{A} (top) shows a reply from P15 to an OP on r/lonely. \textbf{B} (bottom) shows a reply from P08 to an OP on r/ptsd. Red labeled arrows (left) indicate which chaplaincy taxonomy items were applied to specific parts of their reply.}
        \label{fig:screenshot}
\end{figure}

 }

\newcommand{\taxonomy}{
\begin{figure}[b]
    \centering
    \includegraphics[width=\textwidth]{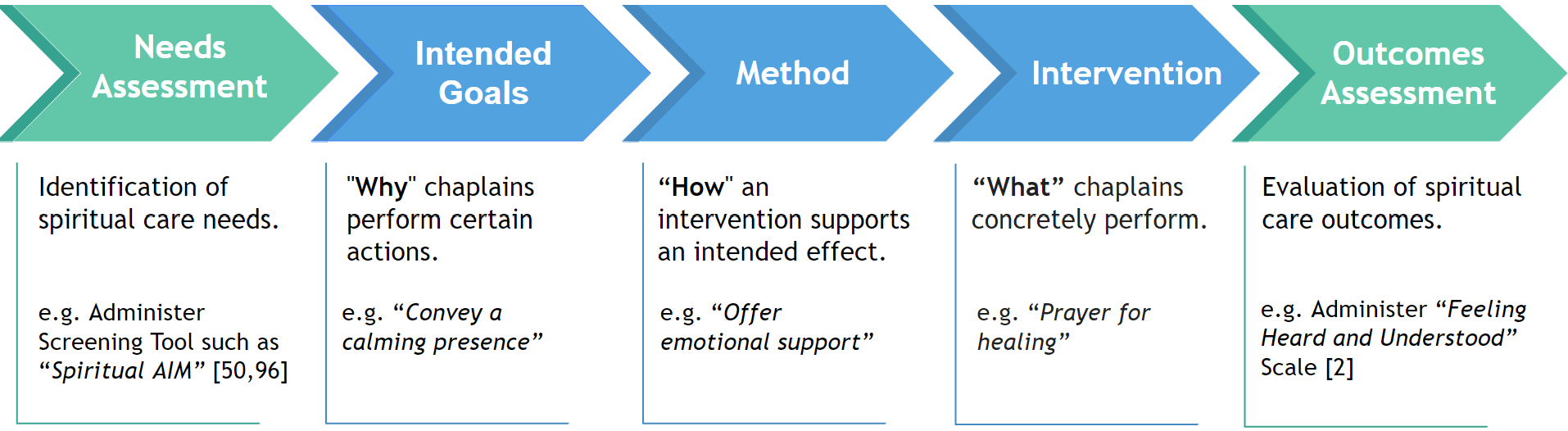}
    \caption{The Advocate Health Care Taxonomy of Chaplaincy (AHCTC) Framework, with Proposed Extensions. Blue categories were identified in the original AHCTC~\cite{massey_what_2015}. Green categories represent needed extensions to this framework for future clinical utility.}
    \label{fig:taxonomy}
\end{figure}
}

\newcommand{\careloop}{
\begin{figure}
    \centering
    \includegraphics[width=\textwidth]{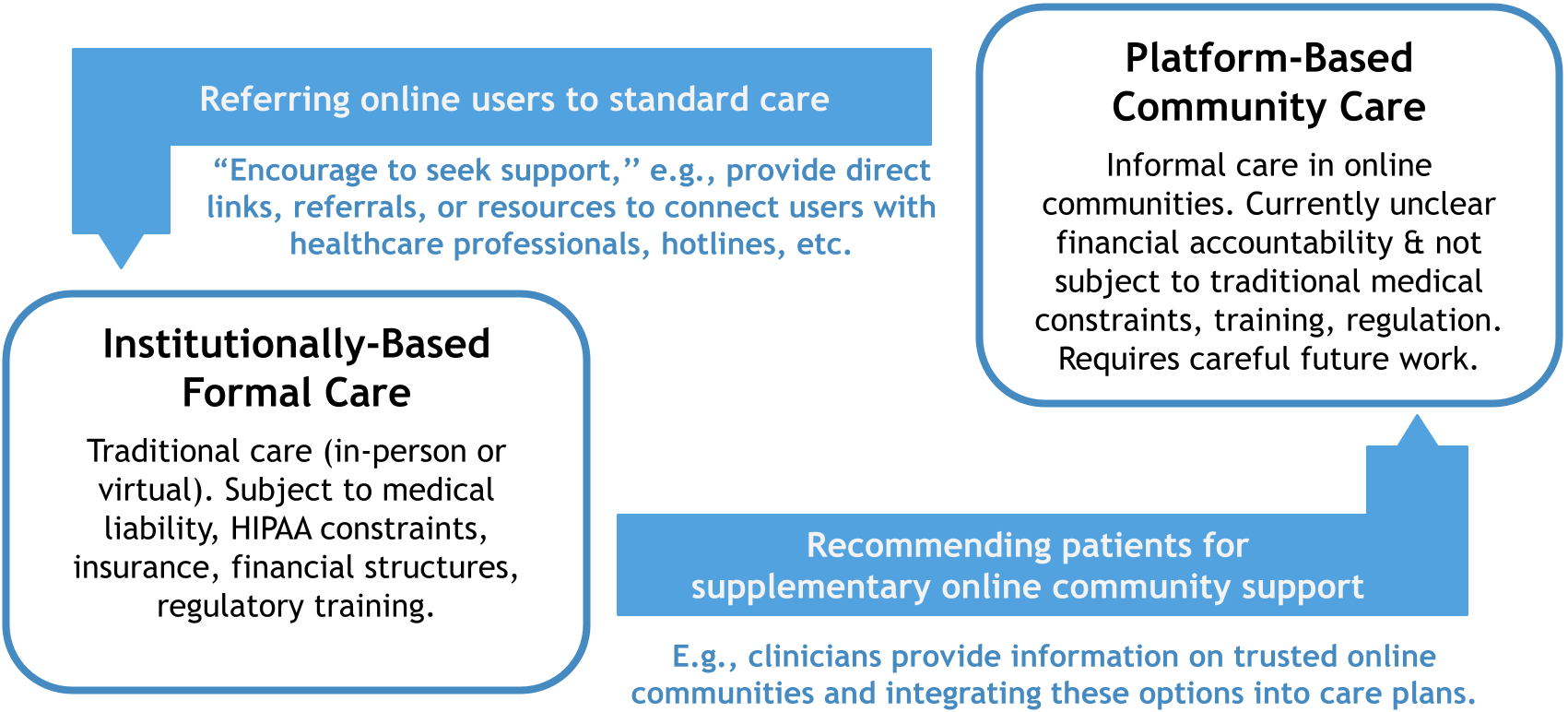}
    \caption{The Care Loop Model. Integrating institutionally-based formal care (left)  with platform-based community care (right) requires two-way flow, allowing patients to transition between formal and online community care settings. The model addresses limitations of both settings by enhancing accessibility of spiritual care, raising awareness of chaplaincy services, and providing continuous support. HIPAA: Health Insurance Portability and Accountability Act.}
    \label{fig:careloop}
\end{figure}
}

\newcommand{\baylor}{
\begin{figure}[b]
    \centering
    \includegraphics[width=0.7\textwidth]{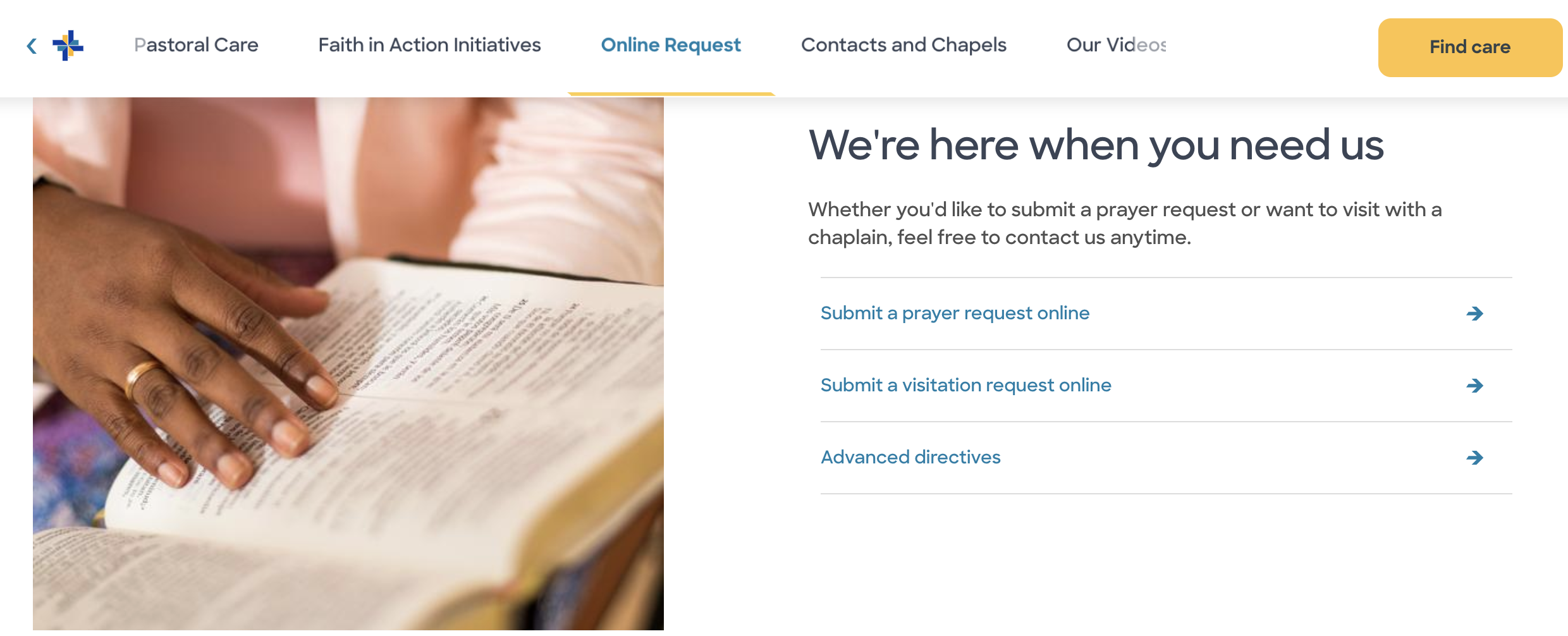}
    \caption{Baylor Scott \& White Interface for Requesting Prayer or Chaplain Visit. Screenshot captured June 2024 from \href{https://www.bswhealth.com/support/spiritual-care}{www.bswhealth.com/support/spiritual-care}.}
    \label{fig:baylor}
\end{figure}
}

\section{Introduction}

Spiritual care is a discipline within healthcare that specifically treats the \textit{human spirit}, often during periods of distress induced by disruptive life challenges, traumas, or health conditions~\cite{puchalski_improving_2009}. Spiritual care addresses fundamental human needs for: (1) meaning and purpose; (2) belonging and connection; and (3) loving and being loved~\cite{shields_spiritual_2015}. Addressing these needs may or may \textit{not} be related to religious beliefs, rituals, and techniques~\cite{massey_what_2015}. In the US, most of the healthcare system is predicated upon diagnosis, treatment, prognosis, and recovery from physical or mental health conditions (or less frequently, preventive medicine~\cite{mckee_holistic_1988}). Unlike the rest of healthcare, spiritual care takes a wholly \textbf{non-prescriptive} approach---it is \textit{not} about fixing, curing, or preventing disease, but rather about providing loving companionship and attuned, accepting presence to the painful reality of a patient's spiritual or existential distress~\cite{puchalski_improving_2009}. Spiritual care professionals often focus on creating connection with patients (sometimes through silent presence, gentle physical touch, or guided questions) so that clients can feel seen and cared for while finding their own answers from within~\cite{massey_what_2015}. It can enable patients to find \textit{``solace, comfort, connection, meaning, and purpose amid suffering, disarray, and pain,''} even with acceptance of the loss of social productivity and eventual death~\cite{puchalski_improving_2009}. Although palliative care is the most common medical context associated with spiritual care, chaplains or other spiritual care providers work across many disease contexts (e.g., episodic, chronic, or terminal care) and settings (e.g., hospitals, clinics, educational institutions, military, prison). Medical research has firmly established the importance of spiritual care for individuals’ health and wellbeing~\cite{hodge_template_2006, balboni_spirituality_2022, long_spirituality_2024,yaden_meta-analysis_2022}. In practice, however, spiritual care is often underserved to the detriment of patient wellbeing and quality of life due to challenges such as insufficient staffing, geographical disparities, and lack of assessment of diverse spiritual needs~\cite{rushton_what_2014, de_diego-cordero_more_2023, long_spirituality_2024}.

Technology has great potential to address these challenges, but spiritual care lags behind other healthcare domains in adopting technology~\cite{winiger_changing_2022}. This could be due to conventional pre-pandemic perspectives within chaplaincy that spiritual care should \textit{only} be provided in person~\cite{ferrell_urgency_2020} and low public consciousness about the importance of spiritual care, including in CSCW/HCI communities~\cite{smith_sacred_2022}. However, the nature of professional chaplaincy care delivery is now evolving, e.g., through increasing \textit{telechaplaincy} practices~\cite{winiger_changing_2022, winiger_navigating_2023}. The COVID-19 outbreak majorly contributed to this move by highlighting a need to explore new models of delivery that do not depend upon in-person bedside care~\cite{ferrell_urgency_2020, winiger_changing_2022, sprik_chaplains_2022}.
Likewise, recent CSCW research has simultaneously demonstrated that in online health community contexts such as~\href{https://www.caringbridge.org/}{CaringBridge}, ``spiritual support'' is the underlying motivation of supportive behaviors~\cite{smith_what_2021,smith_sacred_2022}, and ``prayer'' is the most frequent and important form of support to users~\cite{smith_i_2020}.
Beyond these emerging trends and papers, there is limited research in the fields of spiritual care and CSCW on how professional spiritual care work can be effectively mediated via technology. Thus, this study contributes initial work toward expanding models of delivery for professional spiritual care into the context of text-based, asynchronous, anonymous, and public online health communities. \textbf{We aim to learn how chaplains believe spiritual care may or may \textit{not} be expanded into online spaces.} To serve this goal, we have formed an interdisciplinary research team (\S~\ref{sec:team}), and we center the perspectives of professional chaplains as participants in an exploratory study of prospective Online Spiritual Care Communities (OSCCs). 
Given the lack of prior HCI work in chaplaincy care, we engaged healthcare chaplains across diverse geographies and care settings (e.g., surgical, chronic, palliative, community-based settings) to provide broad, foundational contributions, and we ask the following research questions:

\begin{itemize}
    \item \textit{\textbf{RQ1}: What are the perspectives of 
    healthcare 
    chaplains toward technology as a medium of chaplaincy work?}
    \item \textit{\textbf{RQ2}: 
    What types of chaplaincy work can and cannot be facilitated via online communities, considering the nature of the spiritual care profession and the limitations of technology?
    }
    \item \textit{\textbf{RQ3}: 
    How do chaplains imagine their prospective role and presence in online support communities---especially anonymous, text-based, and asynchronous spaces?
    }
\end{itemize}

We completed exploratory interviews to ask about the role of technology in chaplaincy care. Next, we conducted user testing sessions in which we showed chaplains large, active support communities and asked how they felt professional chaplaincy may play a role. We selected \href{www.Reddit.com}{Reddit} as our study site because Reddit hosts numerous large, active communities dedicated to supporting physical and mental illnesses~\cite{prinster_community_2024}, and Reddit profiles are generally (pseudo)anonymous~\cite{van_meerveld_role_2017}, enabling deeper levels of sensitive disclosure~\cite{choudhury_mental_2014,ammari_self-declared_2019, zou_self-disclosure_2024}. There are, however, serious risks of harm for users on Reddit, including abuse, trolling, doxxing, misinformation, encouraging symptomatic behaviors~\cite{chancellor_suicide_2021,nourriz_design_2024}, etc. If we are to design OSCCs, we must develop strategies to ensure safety and efficacy for users, while also instilling best practices of spiritual care. Our findings provide formative insights from chaplaincy care toward these goals.

Overall, our work makes three main contributions: (1) We introduce the field of spiritual care to the CSCW and HCI community as an important research area, exploring why it matters and clarifying how it differs from other forms of healthcare; (2) We present empirical insights on the perspectives of professional spiritual care workers toward technology and prospective OSCCs; and (3) We contribute the ``Care Loop'' model of integrating online spiritual care with existing clinical care, along with design implications that can lay the foundations for future work to build and maintain new or existing online care communities.

\section{Background on Professional Chaplaincy Care}

\subsection{Origins and Contexts of Chaplaincy Care}

Hospital chaplaincy has a long history dating back to its origination within the ancient and medieval Christian church, which also fostered the creation of hospitals themselves~\cite{buklijas_medicine_2008}. In US history, chaplaincy was first embraced by the military to provide spiritual care to its service members, who often lacked community support during traumatizing and life-altering wartime experiences~\cite{plummer_chaplaincy_1996, gordon_military_2023, chaplain_corps_us_2015, besterman-dahan_role_2012, 
otis_overview_2009}. Tracing back to 1775~\cite{gordon_military_2023}, military chaplaincy continued through various wars and conflicts until today~\cite{morgan_help-seeking_2016, besterman-dahan_role_2012, howard_collaborative_2008}. 
In the late 1800s, Anton Boisen became the first US hospital chaplain, where he situated chaplaincy within psychiatric care units as a complementary form of care. As a patient with schizophrenia himself, Boisen advocated for mentally ill  patients to define their own experiences of ``madness'' \textit{fully} on their own terms, and to have support from professionals with religious/spiritual (R/S) training to bear witness and help make sense of their experiences in ways that their doctors and therapists could and would not provide~\cite{boisen_exploration_1936}. In the 20th century, chaplaincy has since been instated within hospitals~\cite{pesut_health_2016, jacobs_what_2008, ford_development_2006} and settings like workplaces~\cite{wolf_workplace_2018, seales_corporate_2012}, schools~\cite{norman_personal_2002}, sports~\cite{waller_collegiate_2023}, and prisons~\cite{denney_prison_2018}. Regardless of context, chaplains' primary role involves providing a compassionate presence and support through crises, facilitating coping strategies, and promoting well-being~\cite{handzo_outcomes_2014, white_provision_2023}. This study focuses on healthcare chaplaincy across a variety of settings to examine prospective roles of chaplains in online health spaces.

\subsection{
Healthcare Chaplaincy
}\label{sec:differentiation}

Healthcare chaplains serve patients of all faiths (including no faith, agnosticism, or atheism), offering support in transcendent issues of identity, meaning, and purpose~\cite{jacobs_what_2008}. Their role often includes translating between patients and medical teams, participating in ethical decision-making, and contributing to patient satisfaction~\cite{timmins_role_2018, pesut_health_2016, raymond_de_vries_lost_2008}. Chaplaincy focuses on providing what has been described as \textit{``characterizing all human contact''} in a medical system~\cite{nolan_hope_2011}. For example, chaplains explain their work as humanizing activities such as a \textit{``being-with''} and/or accompanying presence~\cite{nolan_hope_2011,olsman_witnesses_2022}, or a \textit{``gentle presence''}~\cite{sinclair_spiritual_2012}. Rather than trying to fix problems, find solutions, or in some cases, extend life, the work of chaplains centers around immediate human connection through deep listening~\cite{nolan_hope_2011,olsman_witnesses_2022}, non-verbal communication, including silence~\cite{bassett_contribution_2018}, a sitting with the patient and witnessing all the changes they encounter through illness, disaster response, distress, and/or  dying,~\cite{bassett_contribution_2018,nolan_hope_2011,olsman_witnesses_2022,sinclair_spiritual_2012}, and ritual support if requested~\cite{jeuland_chaplains_2017}.

As one chaplain participant in a palliative care context explained, \textit{``Chaplains are the least as possible focused on finding solutions. ... They rather ask, `What does this mean to you?' and `How do you deal with that?' and `What’s your path in this?' So, bearing witness to the difficult situation''}~\cite{olsman_witnesses_2022}. This quote demonstrates the verbal and embodied strategies of a chaplain's practice, in seeing and hearing patients through the process of being treated in a medical context, when both physical and psychological pains such as despair, fear, and isolation are present. Some research suggests that chaplains adjust verbal and embodied practices according to differing religious practices~\cite{klitzman_muslim_2023}, however there is a lack of studies that explore the diversity of verbal and embodied practices of spiritual care, and almost no work explores \textit{non}-verbal tools or text-based applications that could enhance, detract or expand upon communication practices of chaplains. We aim to address this gap.

\taxonomy

\subsection{The Advocate Health Care Taxonomy of Chaplaincy}\label{sec:AHCTC}

There are no universally adopted standards of what chaplaincy care exactly is or what its specific clinical duties entail. However, to demonstrate the value of chaplaincy care, a common language is needed~\cite{handzo_outcomes_2014}. Thus, the Advocate Health Care Taxonomy of Chaplaincy (AHCTC) was developed to describe what healthcare chaplains do~\cite{massey_growing_2012,massey_what_2015}. It was created by using previously published lists to generate a preliminary set of 348 items~\cite{massey_growing_2012,massey_what_2015}. Using a robust mixed-method approach including literature reviews, retrospective medical record reviews, chaplain focus groups, self-observations, and experience sampling of chaplains in the course of clinical work, the team arrived at 100 final items. Consistent with the use of clinical pathways for nursing and other interprofessional healthcare communication in the US, these items were separated into three groups of terms:

\begin{enumerate}
    \item \textit{\textbf{``Intended Effects''}} are the desired contributing outcome~\cite{vandecreek_discipline_2014} a chaplain is striving to help address or meet--i.e., ``Why'' chaplains did what they did.
    \item \textit{\textbf{``Methods''}} describe how specific interventions support intended effects--i.e., the ``How'' of encounters.
    \item \textit{\textbf{``Interventions''}} are concrete gestures, actions, or activities in a chaplaincy visit--i.e., the ``What'' of encounters.
\end{enumerate}

The AHCTC was made freely available with adaptation allowed (or even encouraged) to suit contextually-situated clinical needs. Despite some initial success (e.g., an adaptation at the UK Birmingham Children’s Hospital~\cite{nash_adapting_2019}), it did not gain traction for reasons that remain unclear. Although the final list in AHCTC is a major concentration of precursor lists, its 100 items may be intimidating for users. Moreover, to completely integrate AHCTC within clinical use, assessment tools such as FICA~\cite{borneman_evaluation_2010}, HOPE~\cite{anandarajah_spirituality_2001}, Spiritual AIM~\cite{shields_spiritual_2015,kestenbaum_spiritual_2022}, PC-7~\cite{fitchett_development_2019}, or ONC-5~\cite{labuschagne_development_2024} would need to be connected to the front end of the taxonomy, with a set of value-based outcome measures developed and attached at the end (e.g., the Feeling Heard and Understood scale~\cite{ahluwalia_macra_2022}). Figure~\ref{fig:taxonomy} depicts an overview of the existing AHCTC framework (in blue), along with these two prospective integrations needed for future clinical use (in green). Despite its limitations (including the fact that it has not been updated since 2015), the AHCTC remains a foundational document for healthcare chaplaincy. To the best of our knowledge, no other taxonomies of this nature are available. Since professional opinions diverge on the exact duties of chaplaincy (including among our participants), our methods adopt the AHCTC to ground our findings in the most recent and carefully-derived taxonomy available in the chaplaincy literature. We use AHCTC terms to denote which existing chaplaincy practices may or may not be applicable online. Our results also suggest new chaplaincy techniques (i.e., \textit{not} noted in the AHCTC) that are needed for online contexts. We acknowledge the limitations of using the AHCTC in \S\ref{sec:revisiting}, while our results suggest a need for either significantly updated or new spiritual care taxonomies moving forward.

\subsection{
Differentiating Mental Health and Spiritual Care
}

Within healthcare, people most often conflate mental health and spiritual care, however these are distinct (yet complementary~\cite{cooper_military_2023}) fields with different background training and approaches. Spiritual care targets people's \textit{spiritual} needs. Spiritual care \textbf{\textit{specialists}} often have titles like chaplains, spiritual counselors, or spiritual directors. In the US, full-time chaplains receive specialized training in accredited training programs and can next pursue a Board Certified Chaplain (BCC) credential. 
Chaplains often have \textit{separate} roles as religious or faith leaders. For example, those serving in spiritual leadership roles may have a Masters of Divinity (M.Div.) seminary degree, thus BCC and M.Div. can co-occur.
However, the nature of their work in faith leadership is distinct in nature and performance from their chaplaincy roles. Moreover, nurses (or other clinicians) may pursue continuing education credits for spiritual care \textbf{\textit{generalist}} training as a \textit{supplementary} aspect of their care work (e.g., through organizations like the \href{https://www.spiritualcareassociation.org/}{Spiritual Care Association}).

On the other hand, mental health professionals target people's \textit{cognitive} or \textit{mental} health needs through fields like psychiatry, neurology, psychology, and psychotherapy that have different types of trainings and credentials---e.g., Medical Doctor (MD), Licensed Mental Health Counselor (LMHC), Licensed Social Worker (LSW), Master of Social Work (MSW), etc. Some care techniques \textit{do} overlap between mental health and spiritual care---e.g., active listening, empathetic presence, or guided question asking---but the overall \textbf{approach} and \textbf{goals} of these two disciplines are distinct. For example, psychiatry has occasionally evaluated how aspects of R/S (e.g., prayer, church attendance, spiritual experiences, etc.) may improve psychological symptoms or outcomes~\cite{bonelli_religious_2012,braam_religion_2019,rosmarin_religious_2013}, but critically, psychiatry does \textit{not} directly address R/S needs. Mental health providers diagnose a mental health condition or issue, and then use pharmacological, behavioral, cognitive, or social interventions to attempt to prevent or improve symptoms~\cite{pinho_editorial_2024, coventry_psychological_2020}, which are measured using psychometrics (i.e. \textit{quantitative} survey-based measures) such as the PHQ-9 for depression~\cite{costantini_screening_2021} or the GAD-7 for anxiety~\cite{johnson_psychometric_2019}. A therapist might occasionally discuss R/S-related issues with patients and provide their own interpretations, analyses, or recommendations for a client's situation, but they do \textit{not} provide R/S rituals, blessings, or sacraments during a session, nor is the purpose of the therapist simply ``to be with'' their client to provide a sense of loving companionship. Their purpose is to assess, address, improve, or prevent symptoms of mental illness.

Spiritual care, on the other hand, often uses spiritual needs assessment tools (such as the FICA, HOPE, and Spiritual AIM~\cite{borneman_evaluation_2010,anandarajah_spirituality_2001,shields_spiritual_2015}) that provide clinicians with questions that \textit{qualitatively} assess a patient's R/S beliefs, evaluate whether these beliefs are working well or causing distress, and determine how the care plan should be sensitized accordingly.\footnote{The ONC-5 and PC-7~\cite{fitchett_development_2019,labuschagne_development_2024} are relatively newer quantitative psychometric instruments for rating spiritual distress, however they have not yet been validated or implemented broadly within clinical settings yet.} Without any prescriptive nature or any interest in a singular disease, disorder, or set of symptoms~\cite{feldman_spirit_2011}, spiritual care \textit{directly} addresses patients' R/S needs and values. Moreover, spiritual care providers are trained to \textit{never} impose their own beliefs on patients, or even to offer their own interpretations or analyses of a patient's situation. Instead, they simply listen compassionately, ask questions that help patients to uncover and articulate their own meaning-making, or if requested, provide R/S rituals. Moreover, chaplains are attuned to recognize spiritual \textit{v.s.} mental health needs, and to refer patients to mental health professionals as appropriate~\cite{nieuwsma_collaborating_2014, cooper_military_2023}.

Mental health is already a major research area in CSCW. For example, researchers have explored novel sociotechnical mechanisms for expanding models of delivery through mental telehealth~\cite{langarizadeh_telemental_2017} or evidence-based psychotherapy techniques like cognitive reappraisal~\cite{smith_effective_2021,morris_crowd-powered_2014,morris_efficacy_2015,kitson_supporting_2024}, motivational interviewing~\cite{shah_modeling_2022,nurmi_engaging_2020}, and other Cognitive or Dialectical Behavioral Therapy skills ~\cite{thieme_designing_2023,schroeder_pocket_2018}. Frameworks are now available to support the design~\cite{slovak_hci_2024}, uptake~\cite{gemert-pijnen_holistic_2011} and evaluation~\cite{agarwal_evaluation_2022} of mental health technologies, some of which have undergone clinical trials~\cite{kruzan_social_2022,balcombe_human-computer_2022}. Inspired by HCI work across mental health and other forms of healthcare, our paper provides initial foundations for CSCW/HCI to embrace technology for expanding models of delivery for spiritual care.

\section{Related Literature}

\subsection{Technology \& Telechaplaincy in Professional Spiritual Care Work}
Chaplaincy has historically considered itself an in-person care profession, given the essential importance of presence, energetic attunement to patients, and `being with' as described above. However, through telehealth offerings, healthcare is now increasingly being delivered beyond the traditional brick \& mortar confines of clinic walls~\cite{winiger_changing_2022}--especially driven by the COVID-19 pandemic~\cite{ferrell_urgency_2020}. Care teams typically engage with spiritual care providers via referrals through Electronic Health Record (EHR) systems~\cite{campbell_standardized_2023}, clinical notes~\cite{woggon_development_2022}, and assisted decision-making systems, especially in palliative care~\cite{bush_systematic_2018}. Most often, care team members initiate chaplain referrals after they observe that a patient is in serious distress or has become non-compliant. Chaplains may also drop-in to patient rooms while making rounds of a clinical unit to see if services are desired. Less frequently, patients may initiate contact by requesting chaplaincy services. For example, Figure~\ref{fig:baylor} shows a public clinical interface where patients can request prayer or chaplaincy services.

\baylor

However, spiritual care has lagged behind other medical disciplines in its adoption of telehealth and remote care options~\cite{winiger_changing_2022,sprik_chaplains_2022}. Preliminary data suggest that patient experiences with emerging telechaplaincy services are overwhelmingly positive, and that patients significantly prefer anonymous emails (>70\% of contacts) and prayer requests (>70\%) over phone calls (~20\%) or video calls (~1\%)~\cite{fleischman_telechaplaincy_2016,fleischman_telechaplaincy_2017}. These data point toward anonymity as a potential benefit in the design of digital tools for spiritual care and contributed to our selection of Reddit as a study site.
 
Other emerging studies of telechaplaincy describe the need for hybrid care given the changing nature of the world and generational differences~\cite{rice-jalloh_zooming_2022}, as well as systemic challenges of maintaining professional standards and equitably serving patient populations in digital care settings~\cite{winiger_navigating_2023}. A few studies (published in medical journals) also examine new digital tools for chaplaincy care. For example, \citeauthor{kelly_single-arm_2024} piloted a web app called ``MyInspiration'' for providing oncology patients with spiritual care resources during cancer treatment~\cite{kelly_single-arm_2024}, and \citeauthor{calder_improving_2023} created a tool to digitize tracking of pastoral care services in a healthcare setting~\cite{calder_improving_2023}. Prior work also discusses  prospective roles and ethics of AI ``carebots'' or virtual pastors in spiritual care, arguing that even though generative AI-based technologies might serve some needed functions, they are not effective replacements for human care and connection~\cite{young_virtual_2022,brown_will_2023,trothen_replika_2022}. 
 
 Prior studies examine palliative care~\cite{yang_role_2019, wang-qin_nurse_2018, maccaull_using_2010, kawsar_e-esas_2012,schulte_hospital_2020}, but do not focus on spiritual care, except for a few HCI studies that mentioned spiritual needs~\cite{oleary_something_2024, ahmadpour_how_2023, haque_findings_2014}. Beyond these examples, we are unaware of any other design work at the intersection of clinical spiritual care and HCI. As spiritual care expands into digital spaces, HCI and social computing expertise will be of great benefit to spiritual care. Therefore, in the future, one branch of HCI research should investigate within-clinic applications for spiritual care (such as the examples above), while another should explore in-the-wild peer-provisioned opportunities for support; our study contributes to the small but growing body of work on the latter.

\subsection{Toward the Design of Online Spiritual Care Communities}\label{sec:OSCCs}
Online health communities provide needed social support for patients and caregivers, and take a variety of forms and formats across different platforms. In anonymous, asynchronous, and primarily text-based communities on platforms like PatientsLikeMe, Cancer Survivors Network, or Reddit, prior work suggests that patients tend to seek social and \textit{especially} informational and emotional support from others with similar conditions~\cite{nambisan_information_2011, record_i_2018, zhao_online_2022, duimel_profiling_2023, colineau_talking_2010, lee_why_2010}. In non-anonymous online communities, the emphasis or type of support may shift. For example, on the platform CaringBridge, a health blogging platform used by patients, caregivers, and known members of their support networks during life-threatening illness~\cite{ma_write_2017}, \citeauthor{smith_i_2020} found that informational support was rare, while prayer was both the most common \textit{and} most valued form of support for users~\cite{smith_i_2020}. Additionally, peoples' underlying religious or spiritual beliefs and communities influenced how many users chose to exchange support on the platform~\cite{smith_what_2021}. This research on CaringBridge derived the following definition of spiritual support for HCI:

\begin{description}
    \item[Spiritual Support] 
    is an integral dimension that underlies and can be expressed through every category of social support, including informational, emotional, instrumental, network, esteem~\cite{cutrona_controllability_1992}, and prayer~\cite{smith_i_2020} support. This dimension creates a triadic relationship between a recipient, a provider, and the sacred or significant, with the purpose of helping recipients and providers experience a mutually positive presence with each other, and with the sacred or significant.
\end{description}

We adopt this definition in our methods, since it offers a real world proof-of-concept that spiritual support can emerge organically in online communities~\cite{smith_i_2020,smith_what_2021,smith_thoughts_2023,smith_sacred_2022}. Elsewhere, digital prayer~\cite{kaur_sway_2021,kwon_spiritual_2024}, online worship~\cite{claisse_keeping_2023,wolf_spirituality_2022,wolf_god-i-box_2023}, and tangible R/S digital artifacts~\cite{markum_mediating_2024} are also now gaining more traction in HCI. 

Peer-provisioned online support could potentially fill gaps in the availability of professional care, while also retaining human connection (rather than, for example, replacing humans with chatbots~\cite{trothen_replika_2022}). Thus our work asks, how might spiritual support be conscientiously expanded into anonymous communities, where users are often isolated offline, but can access thousands or millions of peers online? And what can HCI learn from  chaplaincy care to help improve the safety and efficacy of these communities? To address these research concerns, we used Reddit to examine spiritual care provider interactions with online communities.

\subsection{Reddit as a Research Site}
Reddit is a frequent research site in HCI~\cite{proferes_studying_2021}, where researchers often explore questions related to moderation and governance~\cite{seering_metaphors_2022,zhang_policykit_2020}, rules and norms~\cite{fiesler_reddit_2018,chandrasekharan_internets_2018}, and computational social science~\cite{bao_conversations_2021}. Medical research has also previously explored subreddits, given the degree to which users often self-diagnose~\cite{giles_self-_2011}, explore possible treatment or ``DIY'' treatment options~\cite{buntinx-krieg_dermatology_2017,edenfield_always_2019}, and discuss anxiety and depression~\cite{low_natural_2020}. We selected Reddit as our study site because of its existing support subreddits, including those listed in Table~\ref{tab:subreddits}. Some of these subreddits (e.g.,  r/SuicideWatch~\cite{zhu_user_2024,liu_time--event_2022, kavuluru_classification_2016}, r/lonely~\cite{jiang_many_2022,andy_studying_2021}, etc.) have been frequently examined in HCI, albeit without any focus on spiritual care. Reddit also offers (pseudo)anonymous profiles that can offer more comfort with sensitive disclosures~\cite{choudhury_mental_2014,ammari_self-declared_2019, zou_self-disclosure_2024}. Subreddits are astonishingly diverse, despite the fact that they share a common set of UI/UX features and affordances~\cite{prinster_community_2024}. For our study, we use Reddit primarily to spur \textit{reflection} on the current realities of online anonymous support communities, as well as \textit{ideation} on what chaplains feel would be necessary in the future to realize the potential for expanding evidenced-spiritual care and support within these spaces. 
\section{Methods}\label{sec:methods}
We conducted empirical formative work with professional spiritual care stakeholders to understand their perspectives and ideations about Online Spiritual Care Communities (OSCCs), and the prospective roles of professional chaplains within them. Our research comprised semi-structured interviews followed by user testing sessions on Reddit with participants located across the USA.   
This study was reviewed by the Human Subjects Research Board at the Colorado School of Mines and deemed exempt from IRB review. Our study pre-registration at Open Science Framework is available at \href{https://osf.io/524by}{https://osf.io/524by}. All de-identified data is publicly available for re-use or re-analysis through the Qualitative Data Repository at Syracuse University at \href{https://doi.org/10.5064/F6R7J9HL}{https://doi.org/10.5064/F6R7J9HL}.

\subsection{Recruitment and Participant Sample}
Recruitment began by contacting people within our existing professional networks. We then employed snowball sampling and cold-emailing to identify additional participants. Eligibility criteria included adults 18+ years of age with professional training or credentialing in chaplaincy. We recruited $N=22$ participants meeting these criteria; complete sample characteristics are in Table~\ref{tab:participants}. To ensure that we captured generalized principles of spiritual care, we intentionally sought participants from diverse R/S backgrounds: 14 participants self-identified with Christianity (55\%), 2 each (9\%) with Islam, Judaism, and Buddhism, and 1 with Humanism; 3 additional participants did not disclose any R/S background (14\%). All participants are in the USA, including 4 on the West Coast (18\%), 8 in the Western USA (36\%), 3 in the Southern USA (14\%), and 7 on the East Coast (32\%). 11 participants self-identified as female (50\%), 10 as male (45\%), and 1 as non-binary (5\%). Most  participants ($N=15$, 68\%) are highly experienced with 10+ years working in clinical or community-based care settings. Most also had no familiarity with Reddit ($N=13$, 59\%) prior to our study. Through travel for data collection and geographical dispersion of the research team, 16 sessions were conducted in-person; the remaining 6 were conducted over Zoom. Data were collected over approximately 4 months in early 2024.

\participants

\subsection{Protocol}\label{sec:protocol}

\subreddits

\paragraph{Part I: Semi-structured Interviews}
Data collection sessions included two parts with a 5-minute break between them. To understand participants' perspectives on technology in general and social media in particular, interviews focused on their day-to-day professional activities, how/whether there is an integration of technology in their practice, and their experiences and perspectives on using social media for spiritual care. We started by showing participants the definition of spiritual support from~\cite{smith_what_2021} (\S~\ref{sec:OSCCs}) to ground this discussion in prior HCI literature. Next, we asked questions such as: \textit{What are your day-to-day responsibilities? What types of technology have been integrated into your practice? How have you observed clients interacting with social media during their health journeys?} A complete list of interview questions is available in Appendix~\ref{sec:interviewquestions}.

\paragraph{Part II: User Testing Sessions}
During subsequent user testing sessions, participants visited, used, and reflected upon live support-related subreddits; the complete user testing protocol is available in Appendix~\ref{sec:usertestingsessions}. All sessions were audio-recorded along with screen recordings of user testing sessions. Sessions ranged from 48 minutes to 3 hours and 38 minutes, with an average duration of 1.4 hours. Participants were compensated with a \$50 Amazon e-gift card.

Participants who were unfamiliar with Reddit were shown a brief introductory slideshow, including descriptions of Reddit's major affordances (e.g., original posts (OPs), replies, up/down votes, etc.) and examples of both \textit{non-}support and support-related subreddits to provide context on the platform in general. Participants were then invited to choose and explore one or two support-related subreddits, either from among provided examples or from a search related to the participants' patient group of interest, depending on the time available (see Table~\ref{tab:subreddits}). On the researcher's computer, participants were guided to click on links to subreddits and to explore the interface, including community description, rules, moderators, and recent posts/replies. To understand their impression about subreddit pages and overall user interactions, we posed questions such as, \textit{What is your first impression about this page? What are your perspectives on the idea that professional spiritual care providers participate in online spaces like these? In what ways might professional spiritual care providers intervene in these spaces to maximize benefits to users? Would you recommend your clients visit this subreddit, why or why not?} To understand how spiritual care providers might behave when providing online care, we also asked participants to click on any given OP and to compose a response in the commenting box. To avoid direct impacts of our research on the communities, these comments were not submitted but were screen recorded for later analysis. See Figure~\ref{fig:screenshot} for examples of typed responses to OPs. Some participants preferred to speak out responses verbally rather than typing; Table~\ref{tab:taxonomies} summarizes both typed and verbal reply comments.

\screenshot 

\subsection{Data Analysis}

\subsubsection{Qualitative Analysis using Grounded Theory Method}
We used Grounded Theory Method (GTM)~\cite{muller_curiosity_2014} to guide our data collection and analysis. Recordings were auto-transcribed with Microsoft Word and double-checked and corrected against audio by the research team. Throughout the data collection period, all data were inductively open-coded by the first and second authors in an ongoing manner. Next, affinity mapping with the online tool \href{https://miro.com/}{Miro} was used to organize codes into axial categories. Through iterative refining of codes and categories, continuous comparison of new data with existing codes, memoing of emergent themes, modification of interview questions, and discussions with the full research team, we continued recruiting participants until data saturation was achieved. Results presented in \S~\ref{sec:RQ1} 
(RQ1)  
and \S~\ref{sec:RQ3} 
(RQ3) 
describe purely inductive themes. To address RQ2 (\S~\ref{sec:RQ2}), we realized a need to strengthen our analysis by using the Advocacy Healthcare Taxonomy of Chaplaincy (AHCTC)~\cite{massey_what_2015,advocate_health_care_chaplaincy_2014} (see \S~\ref{sec:AHCTC}) to align the terms of our analysis with those from professional chaplaincy care. More specifically, clusters of spiritual care behaviors initially emerged inductively in our analysis of participants' verbal or typed replies to OPs. Rather than relying upon our own verbiage to label clusters, we used AHCTC item labels (either ``intended effects'', ``methods'', or ``interventions'') to deductively re-cluster only this portion of the data. This re-clustering yielded a set of spiritual care activities that emerged organically during data collection---i.e., we did not systematically ask about or observe all AHCTC items. For some clusters, there did not exist suitable items in the AHCTC. We created new labels for those cases only.

\subsubsection{Characterizing Chaplain Behaviors in Replies to OPs}\label{sec:counting}
We collected a discrete set of participants' replies to OPs, while our GTM process generated a set of labels that described these behaviors. Therefore, we decided to return to the data and count how often these behaviors occurred in order to provide a preliminary descriptive glimpse of emergent chaplain activities online. The first two authors met to review all verbal or typed responses to OPs, discussing each example until reaching consensus on which label(s) to apply. Table~\ref{tab:taxonomies} reports the frequency counts of all online spiritual care activities observed. (Although fully offline activities were occasionally mentioned, we did not include these since they do not relate to our RQs.)

\section{Results}\label{sec:results}
Participants highlighted the unique nature of spiritual care relative to the rest of healthcare and emphasized how chaplains complement other types of care through addressing the \textit{whole} human. For instance, P07 provided a definitional account of spiritual care, describing how, \textit{``Everything else that we do [i.e., the rest of healthcare, including] social work, medical care aims for repairing the body, repairing mental health, but not really for dealing with the psyche---the internal portions of the person which need to be cared for.''} Confirming the notion that spiritual care targets the \textit{spirit} rather than the physical body or mind, P11 described how spiritual care is always a complement rather than a substitute for other medical or therapeutic care: \textit{``Being a chaplain, as soon as we run into suicidality, we certainly provide spiritual care, but we also immediately refer to mental health resources.''} Embracing the unique nature and capacity for chaplaincy care to improve whole-human centered care, we have organized our findings according to our three research questions.

\subsection{Chaplains' Perspectives Toward Technology as a Medium of Chaplaincy Work (RQ1)}\label{sec:RQ1}
This section provides examples of chaplains' current tech use (\S~\ref{sec:tool}) and highlights their perspectives on the usefulness of online spaces for spiritual care (\S~\ref{sec:perspectives}). It concludes with an overview of concerns about online care (\S~\ref{sec:considerations}). 

\subsubsection{Existing Technology Use in Spiritual Care}\label{sec:tool}
Participants often highlighted how their one-on-one care provision shifted to phone calls or video conferencing during COVID-19. Several participants also described how AI/ML analysis of EHR data became particularly useful for finding patients. For example, chat histories became helpful for identifying spiritual care needs. As P09 noted, \textit{``AI systems can analyze patient EPIC chats, filter those in need of spiritual care, and refer them to chaplains. We use such systems and do a similar kind of analysis in [Name of Hospital].''} P11 similarly informed us that marginalized populations are currently under-served and under-referred to chaplaincy care due to systematic biases in current referral practices; locating and serving un-referred patients by applying AI/ML/data science techniques over EHR data could identify candidates in need of spiritual care. Participants also noted using technologies such as apps, websites, and social media as \textit{recommended resources} or for asynchronous communication with patients and their family members. For example, P05 expressed how social media was a helpful tool for sharing spiritual care content, noting how they posted daily scriptures on a church's Instagram page to better serve followers, \textit{``We are on Instagram, I take the words from the Daily Scripture ... to kind of engage people.''} P04 also shared about recommending visual video aids, e.g., \textit{``There's YouTube for guided imagery.''} 

Overall, participants reflected receptivity and positive attitudes toward technology. Many shared concerns that relative to other medical disciplines, chaplaincy is significantly behind in technology adoption (e.g., \textit{``We are probably behind other fields. I feel like we are behind.''} (P11)), and they expressed a need to catch up. Nonetheless, participants currently use technological tools for telechaplaincy appointments, EHR referrals, connecting with underserved patients, connecting patients with their families, and connecting with a wider audience with spiritual care needs. 

\subsubsection{Chaplains' Perspectives on Online Spaces}\label{sec:perspectives}
\paragraph{Meeting People Where They Are At.} Participants viewed social media as an opportunity that could give chaplaincy an extra-dimensional space to extend professional spiritual care. First, participants acknowledged that a lot of people are turning to the Internet to seek online support for various physical, mental, and spiritual health problems. As P02 said, \textit{``People already turn to Google/Internet/Reddit when feeling confused about their faith tradition, or when having whatever issues or concerns to find answers. ... Obviously there's a lot of people on Reddit and a core element of spiritual support, at least as I understand it, is meeting people where they are.''} P08 commented on personal observations that more people are seeking help online than in churches: \textit{``Thousands of people are here [online], saying hey, I need support. But my pastor friends are sitting in empty desks.''} P16 underscored how online spaces could be the modern space to reach out to people and fulfill professional responsibility: 

\begin{quote}
    Reddit would be the place where one is called to move into and offer a little. [Similar to] what's been known as street ministry, this could be a future opportunity in spiritual care---Reddit ministry, 
    a chaplain assigned to Reddit, a Reddit chaplain.
\end{quote}

P13 took it as a space to fulfill community service, like serving in-person communities: \textit{``You can work on both levels: institution-wise and community-wise, but this [subreddit] is a community.''} 

\paragraph{Scalability \& Accessibility.}
Participants highlighted that social media could be a great extension to outpatient treatment. For instance, P17 said, \textit{``We have all these outpatient people now that only do telehealth, and so they don't ever see people in person. For them, it's a great extension to what they're already doing because they don't even ever get a chance to see someone one-on-one.''} Online spaces could also be a place to reach people who are otherwise inaccessibile. As P02 shared, \textit{``Being out there [on Reddit] makes chaplaincy services available and accessible. ... I feel the accessibility of it for people who have different blinders, different barriers, different reasons that might hinder them from seeking the support in person or other avenues.''} Moreover, participants anticipated that Reddit communities might be the only way of getting support for some people. For example, \textit{``Many people on this online community may not have support systems ... probably haven't been to the hospital and how do we communicate with them to let them know they're not alone.''} (P13); \textit{``My gut instinct is that part of the reason this Reddit page is so helpful to these people is that so many of them don't actually feel like they have a choice''} (P08). Interestingly, P16 carefully tracked an OP's number and timing of posts to see how they heavily rely on online help, especially during times of crises: 

\begin{quote}
    And this was a day ago [looking into OP's post] and so all this conversation took place over the same day, so it looks like he went to Reddit for immediate support. Let's see, he arrives on the 13th, so he was there for several weeks. And then things went bad and he turned to Reddit for support. It sounds like this is a role that Reddit plays to be there for people in an immediate way when they don't know who else to turn to.
\end{quote} 

\paragraph{Increased Patient Agency in Initiating Care.}
Participants appreciated that social media-based care could change the direction of care initiation. In the standard care context (e.g., hospital), spiritual care is often provider-initiated. P15 described how a shift to providing care via social media spaces could shift care dynamics away from ``exploratory'' drop-ins from chaplains toward more targeted user-initiated inquiries:

\begin{quote}
    Often [in the hospital], you're just going room-to-room sort of offering your services to folks who didn't ask for them.
    Here, they [users] have something in their mind that they'd like to speak to you about. ... like, `hey, I have these three questions, I'm struggling with X Y, and Z.    
\end{quote}

This flexibility allows patients to choose when to seek spiritual care on topics they desire, makes the client-chaplain conversation engaging, and provides a space to open up more. As P12 said:

\begin{quote}
    It's really led by the patient, or whoever seeking something. So they are choosing to engage. People would open up more in online spaces...[For example] this person is saying a bunch of stuff here in this post that they might not say to me at the bedside, ... 
    Here in the hospital the patient really doesn't have say over when I come into the room. A lot of my visits are just cold calls. 
\end{quote}

Participants also think that online contexts afford new innovative types of engagement, e.g., \textit{You can do one-on-one sort of care, obviously you can do group stuff online. Yeah, it can just be a little bit more dynamic and you can design spaces that are a little bit more creative, targeted, and focused. I've seen it to be like an effective medium.''} (P15)

\subsubsection{Concerns with Online Spiritual Care.}\label{sec:considerations}
In addition to outlining the opportunities online spaces provide, participants highlighted several concerns that require careful consideration. 

\paragraph{A Need For Trust.}
Participants would be very cautious when referring patients to online spaces and would do so only if they perceived sites as trustworthy. P01 would like to check the credibility of the people behind those communities, even if they were professionals: \textit{``I would suggest subreddits to my clients if I know who the moderators were and who developed it, because even in the arena of spiritually oriented care, there are certain people I trust and people I am less trusting of.''} Similarly, P14 said, \textit{``I would only make referrals to something that I trust. And that I know is coming from a beneficial, compassionate place.''} To determine trustworthiness, some participants would require evidence-based evaluation metrics that demonstrated that the subreddit had been \textit{``constructed with full awareness of current research''} (P01). Others would seek less formal methods such as prior experiences or witnesses from other people, e.g., positive reports from patients. As P16 said:
\begin{quote}
    I'll recommend people to CaringBridge, we've used it a lot and I've joined. ... My patients use it, it works for them. I used it and tried it, works for me and other patients. So that's enough evidence for me. ... Yes, I'll do research and draw from research and be scholarly, but I'm a reflective practitioner.
\end{quote}

Generally, most participants are hesitant to recommend visited subreddits. However, they have developed trust for certain online spaces that are more focused and contextually situated toward healthcare and wellbeing, including CaringBridge or other online communities by third parties---e.g., P15 said, \textit{``More care-centered spaces like [name of a website that hosts small spiritual groups] or whatever firms might be more comfortable than just the wild [social media].''}

\paragraph{Platform-Based Community Care is Insufficient.} 
For the sake of the opportunities in \S~\ref{sec:perspectives}, participants felt that standard clinical care needs to be extended into online spaces; likewise spiritual care in online spaces should cycle back toward non-anonymous standard care, whether in-person and/or virtually (videoconferencing). For example, P16 suggested a \textit{``continuing type of contact from online and anonymous to in-person, intimate, ongoing therapeutic kind of relationship.''} Participants especially mentioned this need when exposed to posts in which people shared genuine, sensitive, and hard spiritual care needs (e.g., suicidal thoughts) that might not be handled effectively through online communities. Hence, they suggested redirecting users to standard care by providing hotline numbers (e.g., \textit{``if you're feeling this way, call here.''} (P04)) or links to facilities (e.g., \textit{``here's how you can make an appointment and all the logistical things.''} (P02)). Participants proposed that online spaces/communities could be spots for finding/locating such people and referring them to standard care. As P15 said:
\begin{quote}
    It's like someone's basically raising their hand in a room [online] saying, I'm really suffering! And so, you can meet them there and be like, I hear that, and happy to talk privately if you'd like, which just feels like a natural need for most people who are suffering.
\end{quote}

They suggested that identifying and/or referring those people could be one of the roles of chaplains' online presence, or even that an AI system could support this process. \textit{``There could be a creative way of working with AI to pick up keywords and then maybe get them to professional help. ... If someone's using negative words and suicidal, then AI could be used to alert a professional.'' }

\paragraph{Risks of Scams, Predators, Misinformation, Hate Speech, or Imposing Religious Beliefs.} Participants were very concerned with the risks of social media, as patients coming to online communities are uniquely vulnerable. P02 said, \textit{``You're dealing with people that are very vulnerable, somebody may say something that may cause someone to harm themselves.''} P06 was particularly concerned for seniors, \textit{``I would be worried about them running into scams or being taken advantage of.''} They also mention concerns such as hate speech or incivility (P16), misinformation (P09), and the risk of imposing religio-political beliefs, e.g., \textit{``White Christian nationalism is a real risk in the U.S.''} (P01). In addition to possible harms, participants also suggested  solutions such as constant disclaimers, such as \textit{``safety reminders pinned at the top''} (P06), teaching skills to\textit{``help people come to realize the harm''} (P02), or using AI to filter toxic posts (P22).

\paragraph{Impacts on Marginalized Groups.} Despite acknowledging how technology can improve access, participants also noted that it can marginalize certain patient groups such as seniors, people without access to technology, or people who are less tech-savvy or seriously ill. P17 said, \textit{``By instituting AI and also apps and other things in healthcare, we are totally leaving people who are not able [behind], either ageist or who are not technologically savvy.''} P06 was particularly concerned about homeless people being left out, \textit{``[Homeless people] lose their phones or they get stolen or don't have them. Really, lack of access is a big one.''} Hence, they advised careful judgment in when to adopt technologies or not. \textit{Careful discernment is needed when to use technology and when not to go virtual/online. ... I think it's person by person, time by time, not all or nothing''} (P22).

\subsection{Types of Chaplaincy Care Work Online (RQ2)}\label{sec:RQ2} 
This section focuses on our analysis of chaplains' prospective behaviors online, as captured through asking them to reply to OPs in visited subreddits (see \S~\ref{sec:counting}) and provides a preliminary understanding of the types of spiritual care techniques that emerged according to AHCTC terms (\S~\ref{sec:counting}). Table~\ref{tab:taxonomies} summarizes 20 items, divided into three categories: applicable online (\S~\ref{sec:applicable}); challenging to apply online (\S~\ref{sec:challenging}); and new techniques (\S~\ref{sec:newtechniques}). Out of 100 possible AHCTC items, 15 items were detected; we also added 5 new items. Some responses were coded for multiple items, as in Figure~\ref{fig:screenshot}. For each item, a brief description is provided by consulting the user manual from Advocate Health Care~\cite{advocate_health_care_chaplaincy_2014}, along with an example quote from our data.

\taxonomies

\subsubsection{Applicable Online}\label{sec:applicable}
We observed that participants exercised great caution while composing their replies and took take time to prepare ahead (e.g., P07 first drafted their response on scratch paper); some became visibly emotional in response to difficult content on the subreddits (e.g., 1 participant cried briefly). These considerations of time and emotional tolls must be considered in future design efforts~\cite{nourriz_design_2024}.

Overall, thirteen chaplaincy care activities were labeled as applicable for Reddit, though the \textit{efficacy} of these online activities compared to in-person services is unknown. For instance, related to `Active Listening', participants thought online venues could serve as \textit{``a space for being seen and heard''} (P16), but they highlighted skills on \textit{``what to listen for, what to look (read) for when someone posts something''} (P02). Regarding `Ask guided questions', most participants preferred seeking further clarification to better understand the situation rather than providing direct responses to OPs. For the complete list of the remaining items see Table~\ref{tab:taxonomies}.

\subsubsection{Challenging to Apply Online}\label{sec:challenging}
Two spiritual care activities, 'Convey a calming presence' and 'Provide compassionate touch', were perceived as challenging to be delivered in virtual online contexts. This makes sense because both activities demand body-to-body in-person human connection which is peculiar to the unique nature of spiritual care provision. Participants doubted the possibility of fully `being with' others through online spaces because they thought \textit{`` there's a loss of human connection''} (P17) and \textit{``really hard to replicate [in-person presence] in an online environment''}(P15). For others, however, human connection is \textit{``dependent [only] on the heart''} (P14), and being anonymous or remote was perceived as \textit{``no different than being at the bedside''} (P16).

\subsubsection{New Techniques}\label{sec:newtechniques}
Interestingly, five new chaplaincy techniques were identified which didn't fit into the taxonomy. These include: ``Encourage to seek support,'' ``Provide practical advice,'' ``Share own experiences,'' ``Educate on reply behaviors,'' and ``Apology for other users behaviors.'' The emergence of new techniques might be associated with the unique nature of the new online context which may require new chaplaincy approaches. For example, encouraging online users to seek in-person support relates with the inadequate nature of virtual spiritual care and the limitations of technology to fully provide human connection---which was one of the concerns discussed by participants( see section\S~\ref{sec:considerations}), indicating referral needs back to the physical care (see also the care-loop in Figure~\ref{fig:careloop} we proposed which complements one-on-one care with online mass care). Other taxonomy items ($N = 10$) were observed on the intermediate clusters of data on Miro, but were not included in our list because they were discussed in the in-person hospital context, rather than online.\footnote{Other chaplaincy taxonomy items not included in the list: Align care plan with patients' values, Explore R/S beliefs, Explore faith \& values, Providing religious item(s), Encourage Self-reflection, Explore faith \& values, Encourage someone to recognize their strength, Encourage R/S practices, Meaning-Making, De-escalate emotionally charged situations.}

\subsection{Prospective Roles for Chaplaincy in Online Spiritual Care Communities (RQ3)}\label{sec:RQ3}
In this section, we first summarize how chaplains felt that healthcare systems should engage in online care communities (\S~\ref{sec:shouldthey}), and next describe what roles chaplains could hold within them (\S~\ref{sec:roles}), along with perceived barriers to fulfilling these roles online (\S~\ref{sec:barriers}).

\subsubsection{Should Professional Chaplains and Healthcare Systems Attend to Online Communities?}\label{sec:shouldthey}
Most participants found the idea of professional chaplains participating in online care communities a \textit{``worthwhile investment''} (P02). Especially following the COVID-19 pandemic, P02 emphasized its relevance \textit{`` because a lot more folks who are already boxed out from communities even further boxed out''}. After exposure to the subreddits during a live demo, P05 expressed a change of mind toward online care communities: \textit{``I hate social media, but this [interview] has opened my mind to the possibility of doing good work [there].''} However, a few prioritize their current organizations' missions: as P12 said, \textit{``Most of us have full-time [responsibilities]. Why would we be engaging [online] in something that's not related to the mission of our organization?''} In line with this, P09 questioned our research focus in anonymous online communities, and advised that research should \textit{``focus on improving spiritual care provision in hospitals, through already existing and controlled hospital systems like EPIC and patient portals, instead of working on uncontrolled social media forums.''} 

\subsubsection{Roles of Online Chaplains}\label{sec:roles}
Participants discussed several contributions they could make if they were to join online care communities. 

\paragraph{Shaping Peer Interactions.} Participants felt that professional replies to OPs could inspire, model, and eventually teach other users how to offer appropriate support: \textit{``We facilitate and lead people towards a specific type of behavior. ....And this is one of the ways in which we might shape social media experiences that might have a great impact on improving human relations''} (P07). Participants also believed that through their participation professional care work skills could be diffused into the online space, i.e., \textit{``Educate others how to be a spiritual companion''} (P18) and also \textit{``...teaching people how to ask better (deeper) questions and teaching others how to respond''} (P05). Participants (e.g., P12, P15) also proposed design mechanisms for achieving these goals such as using flairs indicating credentials and pinning professionals' responses to appear at the top of the discussion thread.

\paragraph{Moderation \& Training.} Chaplains believe they could support online communities through moderating them, training user-moderators, and crafting community rules. When asked who should moderate, P15 responded, \textit{``Ideally, professionals.''} However, acknowledging the limited time and availability of professional chaplains, participants also saw the value of training other volunteers without this background in order to cultivate an effective and supportive safe space. As P01 put it, \textit{``People with expertise around spiritual care could help to train and support the moderators and they could help construct the rules.''} 

\paragraph{Hosting Synchronous Chats.} Some participants suggested that professionals could improve engagement and provoke discussion through offering synchronous chat sessions, similar to ``ask me anything'' (ama) threads that are commonplace in some subreddits. For instance, P10 raised the considerable problem of failure for an OP to receive replies due to the asynchronous nature of communication online: \textit{``I don't know how you get over that problem of people just typing something and then no one responding. That almost needs like a hey, we're going to have a group chat at this time, join the group chat''}. Offering synchronous sessions with professionals guarantees that someone is there to reply. 

\paragraph{Redirecting to External Resources.} Finally, participants mentioned that they could re-direct people to external resources. For example, P02 shared, \textit{``All these different [social media] tools can be very helpful and you as an institution or as a care provider or whatnot can really do a lot of good by providing resources and yourself being a resource.''}

\subsubsection{Barriers}\label{sec:barriers}

\paragraph{Time \& Payment.} Chaplains acknowledged that their presence in online communities could be beneficial, yet it would require a significant time investment that cannot be managed alongside their current responsibilities: \textit{``The hardest thing is this is a time sucker, there is no ending to this. I know that when I'm sitting with someone I have an hour''} (P05). Hence, they proposed the importance of establishing a compensation model: \textit{``It's something I think we have to get into in the future. But you have to have a person who's committed to that,''} (P13) and \textit{``That person would have to be paid.''} (P05). 

\paragraph{Disrupting Community Dynamics.} Moreover, participants were concerned about how chaplains' presence could influence or disrupt natural dynamics in online communities, e.g., (un)intentional self-advertising. For example, P17 described how professional presence might induce discomfort: \textit{``So you're wondering, are they sharing this because they want me to know who they are so I buy their services? So there is an ethically weird feeling''}.

P02 also expressed concerns that \textit{``providers' content may naturally get more attention, and that leads easily to branding.''} Due to their concerns about perceived self-advertising, P02 calls for appropriate balancing of the strong value of chaplains' presence in online spaces versus \textit{``appropriate limits that are set to minimize harm and maximize care.''}

\section{Discussion}\label{sec:discussion}
Our findings underscore how professional spiritual care providers think of and engage with the possibility of Online Spiritual Care Communities (OSCCs). Given the complexity, expense, and daunting nature of the US healthcare system, many people turn to the Internet to seek health support~\cite{wang_online_2021, naslund_social_2020}. Users of Reddit and other online health spaces may or may not have already visited formal healthcare providers, but may lack complete and holistic end-to-end care---especially regarding care beyond physical needs~\cite{mount_whole_1993}. Individuals may encounter problems like internalized stigma~\cite{arnaez_association_2020}, linguistic discrepancies~\cite{al_shamsi_implications_2020}, and healthcare accessibility, affordability, and medical debt~\cite{coombs_barriers_2021, cyr_access_2019, kluender_medical_2021} that prevent them from seeking formal institution-based care or continuing care. Hence, extending professional spiritual care to online health communities can help address unmet care needs and locate new people needing care. 

This section discusses three design and social implications of our work that can serve as a foundation for establishing and maintaining new OSCCs or existing online care communities: positioning online spiritual care within healthcare (\S~\ref{sec:positioning}); acknowledging the unique nature of spiritual care that cannot be easily translated into technology (\S~\ref{sec:uniquenature}); bringing spiritual care into the public consciousness (\S~\ref{sec:consciousness}). We conclude by discussing limitations of this work and future research directions of value for HCI and spiritual care.

\subsection{Positioning of Online Professional Spiritual Care Within Healthcare}\label{sec:positioning}
Based on the opportunities and concerns of online spiritual care that were discussed, we propose a new \textbf{``Care Loop''} model that could improve awareness, knowledge, utilization, and provision of spiritual care by leveraging social media (Figure~\ref{fig:careloop}). We aim to improve the attention and resources given to the spiritual well-being of patients by incorporating voluntary, online community care hosted on social media platforms as a supplementary ``adjuvant'' to existing institutional care. Below, we unpack the systemic requirements of this model, its underlying sociotechnical implications, and future questions it opens. 

\careloop

\subsubsection{A ``Care Loop'' Between Institutionally-Based Formal Care and Platform-Based Community Care}\label{sec:careloop}
Our results suggest the need for a strategically implemented ``Care Loop'' (Figure~\ref{fig:careloop}) between institutionally-based formal care and platform-based community care. In this model, patients could encounter spiritual care either in clinical settings or in-the-wild online, and harmoniously transition between them to address deficits of either setting independently. We found evidence that anonymous, online communal spaces may provide agency for patients to start conversations, which could in turn increase public awareness of chaplaincy. In addition to increasing accessibility, the presence of chaplains in online spaces could provide a paradigm shift for spiritual care initiation, which has previously almost always come from a care team~\cite{ho_spiritual_2018, gijsberts_spiritual_2019, rykkje_educational_2022, paal_spiritual_2015}. Our findings align with prior work showing that  housing chaplaincy work within the medical context alone can be problematic due to issues with referral biases (e.g., too late or never in palliative care)~\cite{laranjeira_perceived_2023}, lack of awareness from care providers and patients of the nature and value of chaplaincy~\cite{poncin_what_2020}, and professional burnout and inaccessibility of services~\cite{hall_valuing_2021}. Furthermore, discrete appointment-based sessions allocate \textit{limited} time for patients due to insufficient integration of spiritual care in healthcare and a lack of adequate professional human capital. Hence, we propose backing institutionally-based formal care with continuously available community-based spiritual care by leveraging the affordances of social media platforms.

Extending professional spiritual care online requires building and maintaining \textbf{trustworthiness} of OSCCs equivalent to standard care settings. Almost all participants hesitated about referring their patients to the visited subreddits because they did not know the moderators and the background of these communities. Here, prior work on technical requirements for designing trust (e.g.,  graphic design\cite{kuzheleva-sagan_designing_2016, lowry_picture_2014}; structure, content, and social-cue design~\cite{wang_overview_2005}; and personalization~\cite{urban_online_2009}) is relevant and needed. 
Our results point toward an additional need for a systemic level verification of OSCC credibility, especially through affiliations with trusted institutions such as hospitals, R/S communities, or trustworthy third parties such as nonprofits led by established experts. Yet \textit{how} to successfully form and support such affiliations remains an open question.

Another concern is the \textbf{insufficiency} of platform-based community care on its own, hence the need to refer individuals back to institutionally-based formal care. Our findings call for a system of double and complementary referral points. Patients in regular care settings could be sent to \textit{trusted} OSCCs for ongoing support, while users could be referred to standard care when online care is inadequate. To resolve these concerns, appropriately integrating online professional care into the healthcare system is required. These insights informed us to propose the Care Loop model, emphasizing the complementary nature of online and offline care, and guiding the design of hybrid systems that leverage both modalities' strengths to better support individuals' spiritual needs. 

\subsubsection{Design Implications for OSCCs as Platform-Based Community Care}
Once the relationship of OSCCs within the healthcare system is defined, the following community design implications related to moderation, verification of credentials, and UI/UX would be essential. 

\paragraph{Moderation.}
First, people should be trained to maintain OSCCs (e.g., \textit{``a Reddit Chaplain''} (P16)), either by creating and moderating new communities or engaging with the moderators of existing communities. Professional chaplains could directly moderate, or they could assist in training lay volunteers; in either case, training in \textit{how} to moderate effectively~\cite{seering_metaphors_2022}, choose effective rules and governance strategies~\cite{smith_governance_2024}, and configure the right technical solutions and bots~\cite{jhaver_human-machine_2019, smith_impact_2022} will be needed. Most subreddits have few moderators who wield a great deal of power~\cite{schneider_admins_2022}, resulting in a huge moderator-user ratio where moderators generally cannot address concerns synchronously, especially time-critical issues such as suicidal thoughts (e.g., r/SuicideWatch had nearly half a million members and 10 moderators at time of writing). Dedicated professional moderators could serve as synchronous resources, either in an ongoing basis or during scheduled ``group chats'' like ``ask-me-anything'' (ama) events, for real-time replies. They could do this by identifying users needing immediate care (possibly using AI) and referring them to one-on-one care after an online ``first-aid.'' Reddit moderation is conventionally a voluntary, unpaid responsibility, however it generates substantial financial value~\cite{li_measuring_2022}; unlike the existing volunteer model, our participants called for developing financial models to compensate the professional activities of chaplains in these spaces---whether on Reddit or other, future platforms. 

\paragraph{Verified Credentials.} To establish trust and transparently describe institutional affiliations, chaplains' accounts could be verified with their credentials. Recent work discusses the intricacies of user verification practices on pseudo-anonymous platforms like Reddit~\cite{smith_governance_2024}, which are currently manual, time-intensive, and bias-prone processes. However, as in~\cite{smith_governance_2024}, our participants suggested design solutions such as verified user flairs (supplied by trusted moderators) that indicate which users are trained professionals (without necessarily disclosing names/affiliations) \textit{or} profiles that showed moderators' real names. However, questions remain as to how to technically achieve such verification, and whether chaplains should \textit{always} be identified, given the possibility of disturbing the dynamics of natural user interaction.  

\paragraph{UI/UX for Shaping Norms.}
Professionals could also participate in user interactions by providing professional responses to concerns online, as demonstrated in Table~\ref{tab:taxonomies} and Figure~\ref{fig:screenshot}. Such engagement could positively shape user interactions and improve the quality of support, either through visibility into who is and is not a professional. Or, rather than relying on popularity and upvotes of replies, the UI/UX could elevate the visibility of professionals' comments by ranking or placing them at the top of threads. Moreover,  UI/UX should improve the visibility of rules--such as by applying safety labels or warnings or presenting the rules alongside the commenting interface to ensure that they are not missed. Finally, delivering spiritual care resources and educational content in the sidebar (or possibly via AI recommendations to pre-selected and verified articles, videos, interactive online workshops, etc.) can help demystify spiritual care, explaining what it is, how it works, and its benefits. Beyond this high-level guidance, numerous design questions must be addressed through future work, as outlined in \S~\ref{sec:futureOSCCroles}.

\subsection{The Unique Nature of Spiritual Care Cannot be Easily Translated into Technology}\label{sec:uniquenature}
The unique nature of spiritual care lies in its deeply personal, relational, energetic, and empathetic qualities, which cannot be easily replicated or translated into technology. Unlike many services that can be automated or digitized, spiritual care requires profound \textit{human} connection, intuitive understanding, and the ability to offer nuanced, compassionate responses for an individual's specific emotional and spiritual needs. Rather than the prescriptive nature of other types of healthcare which seek to ``fix'' problems, spiritual care is centered around active listening, loving presence, and a sense of shared humanity that fosters trust and vulnerability~\cite{nolan_hope_2011,olsman_witnesses_2022}. While technology can facilitate access to spiritual resources and support communities, it lacks the inherent ``human touch'' that is essential to genuine spiritual care. 

Given the significance of spiritual care for health and wellbeing, the scarcity of professional spiritual care providers~\cite{hall_valuing_2021}, and current trends toward automation (e.g., AI ``companions''), there is a real risk of delegating spiritual care services to machines and agents. Therefore, extreme caution must be exercised to carefully discern which spiritual care activities need to remain in-person and one-on-one, and when, what, and how others can be mediated through digital technologies like OSCCs. In this work, we explored this question by using the AHCTC~\cite{massey_what_2015} to categorize spiritual care activities that our participants provided or suggested during user testing sessions. We found that not all spiritual care activities can be easily or fully delivered through technology. Considering the unique importance of human connection, touch, and presence, future research should continue to investigate to what extent spiritual care can be mediated through technology, and the limitations of technology to fully transmit these experiences.  

\subsubsection{The Unique Nature of Online Settings Need New Chaplaincy Approaches}
The identification of five techniques that are new to chaplaincy in our results---``Encourage to seek support'', ``Provide practical advice'' (\textit{only} when requested), ``Share own experiences'', ``Educate on reply behaviors'', and ``Apologize for other users''---highlights the need to revisit existing taxonomies and to tailor new interventions for online spiritual care that can dynamically adapt to the unique challenges and opportunities of virtual environments. For instance, the recommendation to seek in-person support underscores the importance of recognizing the limitations of online interactions and hearkens back to the Care Loop model  (\S~\ref{sec:careloop}). Additionally, techniques such as chaplains' sharing of their own personal experiences and educating users on appropriate reply behaviors points to the necessity for these platforms to foster a sense of community and reciprocal engagement, which can enhance the support network available to users~\cite{kairam_social-ecological_2022}.

\subsubsection{A Need for an Interdisciplinary Approach In Connecting Professional Spiritual Care with HCI Efforts}
Our findings suggested the need to revisit the AHCTC taxonomy to better serve the evolving landscape of professional spiritual care (see Figure~\ref{fig:taxonomy}), not only by adding new items, but by extending its scope to upfront spiritual needs assessment and later outcomes assessments. Regarding needs assessment, some participants had used data science with AI/ML techniques to identify patients in greatest need of spiritual care through analysis of EHR data or chat histories in patient portals. In the context of OSCCs, chaplains similarly described AI-based opportunities to identify and serve users in the greatest need. Moreover, in online communities, participants also described a need for following-up with users to assess whether their needs had been met and how effectively. Such practices are also needed in clinical contexts. Although it is currently unclear how to achieve this effectively in anonymous, asynchronous online spaces, it is an important future goal. Most importantly, the taxonomy should be simplified for clinical application and adapted to encompass new forms of care, such as those emerging in online communities. This revitalization will require a concerted interdisciplinary effort to harmonize HCI methodologies with spiritual care practice, and to maximize benefits for users and society. 

\subsection{Bringing Professional Spiritual Care into Public Consciousness}\label{sec:consciousness}

Stigma associated with discussions around seeking and receiving care for stigmatized health conditions (e.g., physical disabilities, HIV/AIDS, mental illness)
has a long history. The idea of becoming what Erving Goffman describes as a ``discredited'' person includes hiding a knowledge of what stigmatized diseases' care entails, how it works, and how to find providers and avoid negative labels~\cite{weinstein_stigma_1965}. However, in recent years, there has been a significant shift. For example, public conversations about mental health have increased, leading to a de-stigmatization around the discussion of seeking mental health services~\cite{mclellan_talking_2022}. Despite these positive changes, many people still do not know how to pursue the services available, or how to navigate the complex mental healthcare systems in the United States. An informed and empathetic public dialogue is crucial for raising awareness; Reddit is one such location where users \textit{do} have these discussions~\cite{low_natural_2020}. 

Spiritual care faces similar challenges of under-utilization and a lack of awareness of how to navigate and locate spiritual care within a complex medical system. Many people do not understand what spiritual care is, how it works, why it matters, or where to find it. Just as mental healthcare has benefited from increased visibility and open discussions, spiritual care requires this same attention. By bringing the work of spiritual care into the public eye, explaining its importance, clarifying its functions, and changing how and where the public learns about the activities of spiritual care, society can more easily recognize and utilize these services. This increased visibility can help demystify spiritual care, encouraging individuals to seek the support they need and fostering a more holistic approach to wellbeing.

Our work has explored the possibility of implementing OSCCs in public online spaces on Reddit, 
which could substantially increase the visibility of spiritual care. Online spiritual care would be a departure from traditional practices and represents a transformative shift in the role and perception of chaplains. 
Our findings indicate that through online care, chaplains could engage with public education about their roles and services, which contrasts with the typical U.S. medical system where knowledge of chaplaincy often only arises when it is directly needed. The ability to explore chaplaincy online could allow individuals to understand and appreciate this form of care before requiring it, offering a pivotal moment of exposure for the profession, 
enhancing its accessibility and relevance, and encouraging proactive engagement with spiritual care. 

\subsection{Limitations \& Future Work}\label{sec:future}
\subsubsection{Sample Limitations} 
Although we successfully recruited diverse participants with Christian, Muslim, Jewish, Buddhist, Humanist, and undisclosed backgrounds, our sample did not include all possible R/S backgrounds. In particular, we intentionally sought Indigenous participants but were unsuccessful. Likewise, participants were from diverse locations across the US, but our sample did not include all geographic regions. We acknowledge that there are numerous R/S beliefs, geographies, and support subreddits beyond those mentioned here. Future work can complement our efforts by exploring additional subreddits and perspectives from additional R/S groups within the US, as well as across diverse international contexts with vastly different healthcare structures and cultures. Moreover, our sample of 22 participants provided an in-depth qualitative exploration of chaplains' attitudes and perspectives. However, this small sample may or may not be representative of the broader US or of the broader professional spiritual care community. Our findings suggest a need for verification with quantitative methods that can suggest how or whether these insights generalize across larger samples. Therefore, the next stages of our research will involve national surveys of chaplains to address the limitations of small sample sizes in qualitative work.

\subsubsection{Limitations of the AHCTC Taxonomy}\label{sec:revisiting}
Although to our knowledge, the AHCTC is the most rigorous and recent taxonomy currently available for professional spiritual care, we acknowledge that it is nonetheless an imperfect and outdated tool, and that some of its terms seem to be overlapping or incomplete. In this paper, we used the existing taxonomy \textit{only} to analyze our data related to chaplains' responses to OPs; we did not systematically query whether the existing AHCTC fully encompassed the in-person work of chaplains or whether every AHCTC item could be applicable (or not) online. Our results suggest a need for future work to derive either an updated or entirely new taxonomy for spiritual care that more thoroughly reflects contemporary practices within both in-person care settings \textit{and} digitized care settings (e.g., tele-chaplaincy, texting, phone calls, asynchronous online communities). We encourage future research to systematically identify which elements of the original framework are still useful, which elements require modification or removal, which elements might be safely combined to eliminate redundancy, and which new techniques should be added. Moreover, future research should systematically interrogate the efficacy of such a taxonomy, ensuring it remains relevant and effective across these diverse settings.
Future work should also continue connecting work in digital spiritual care to chaplaincy practices, and to understand what types of spiritual care behaviors and techniques are effective and needed online.

\subsubsection{Ethical Reflections}
 Exposing participants to unfiltered, sensitive, asynchronous posts posed unique and unanticipated methodological challenges in our study. Although our participants are professionally trained to manage emotional distress from exposure to sensitive patient situations, our methods nonetheless induced discomfort for some participants, raising awareness of an ethical issue our team had not previously encountered. For example, one participant encountered a post about a suicide plan with a timeline that had already passed; this evoked significant distress due to their inability to intervene. In face-to-face care, immediate intervention is possible, and chaplains' existing training prepares them well to provide an appropriate response. However, chaplains have never received training for how to appropriately manage \textit{asynchronous} interactions. Thus, exposure to such posts may adversely affect participants' well-being by inducing a sense of professional helplessness or emotional overwhelm. We offer three reflections on this ethical issue. First, future research methods should involve measures beyond standard debriefing during informed consent, such as explicit content warnings about the nature of \textit{asynchronous} posts that could be encountered during the study, and care resources available for participants if needed. Second, this nuanced interaction highlights precisely why new spiritual care \textit{trainings} must be developed for any new online chaplaincy care techniques that may be introduced into practice, as well as new \textit{technical approaches} for intervening if or when asynchronous content is viewed after a significant amount of time has passed since a post was made. Finally,  future work should likewise identify ways to protect \textit{users} from traumatic content that could impede healing, as has been suggested in recent work about online suicide bereavement~\cite{doyle_i_2024} and empathy fatigue~\cite{nourriz_design_2024}. 

\subsubsection{A Need for HCI Expertise in Clinical Spiritual Care}
Although beyond the scope of this present study, a quote from P09 in \S~\ref{sec:shouldthey} emphasizes that researchers need to \textit{``focus on improving spiritual care provision in hospitals, through already existing and controlled hospital systems like EPIC and patient portals.''} HCI expertise will also be invaluable in this regard, thus future research should also carry forth interdisciplinary work to improve sociotechnical systems for spiritual care within clinical and hospital settings.

\subsubsection{Open Questions on Designing Future OSCCs}\label{sec:futureOSCCroles}
While defining the role of OSCCs within healthcare could address concerns raised by participants, the practical establishment and maintenance of this relationship needs further research. How would institutional affiliations (such as with hospitals or religious institutions) work, or can third parties endorsed by trusted entities manage OSCCs? What are the implications of these affiliation arrangements on professional time compensation models? Is Reddit the right platform for OSCCs, and should research focus on supporting existing communities, or building new ones? Moreover, this study focused on \textit{chaplains'} perspectives on how spiritual care can be expanded online, but participants also shared observations on lay user behaviors and their prospective roles in OSCCs that were beyond scope for the present RQs. We plan to report on this set of findings in future work. However, our findings and design directions were nonetheless distilled from a single-sided perspective. Therefore, we also plan to collect additional data from members of online support communities (e.g., lay users, moderators) and \textit{prospective} users who could benefit from participation in OSCCs---e.g., patients, caregivers, supporters, etc. What are these users' top needs and priorities for spiritual care? Are they interested in affiliated OSCCs, or would they prefer that OSCCs remain unaffiliated? Should community membership remain free and public, or should it be private and subscription-based? We intend to address all of these questions in our short- and long-term research plans.

\subsubsection{The Need to Investigate Spiritual and/or Religious Experiences}
Although our participants perceived that people's concerns in online support communities are genuine spiritual care needs, discussions of explicitly R/S experiences were surprisingly uncommon in our participants' interactions. However, prior research indicates that people dealing with terminal illnesses (e.g., cancer) often turn to spiritual coping mechanisms~\cite{balboni_religiousness_2007}, while prior work has reported the use of other online spaces to facilitate R/S activities or experiences, such as prayer~\cite{smith_i_2020,smith_thoughts_2023} or Q\&A about religious beliefs~\cite{scardigno_authentic_2020}. Many AHCTC items also relate to specific R/S rituals, needs, and concerns~\cite{massey_what_2015}. One reason why R/S-related discussions may be lacking in our data might be due to the illness-centered, secular nature of the selected support subreddits. In other corners of Reddit, spiritual experiences and beliefs are more overtly discussed (e.g., r/Spirituality, r/Religion, or subreddits for specific religions, psychedelic experiences, etc.). What types of R/S ideologies do users of these support communities hold? What is the value of these ideologies in relation to their physical, mental, and spiritual health concerns? What types of R/S experiences or activities are most essential to their needs for spiritual care, given the changing dynamics of R/S in the modern world? Future research can explore explicit R/S-based communities and investigate the nature of users' concerns in those communities and the nature of spiritual care provision happening there. 

\subsection{Positionality Statement}\label{sec:team} 
Our research team is comprised of four researchers with an HCI or design background, one medical sociologist, and one researcher in professional chaplaincy care who is a director at a national chaplaincy-related organization. Our personal R/S beliefs vary from agnostic or spiritual and non-religious (or \textit{ex-}religious) to devout, practicing Christian. We acknowledge that personal beliefs may have influenced this work. We took care to mitigate biases by conducting reflexive group discussions before and throughout the project, recruiting participants from as many different R/S backgrounds as possible, honoring all participant perspectives, and using the non-denominational AHCTC taxonomy to shape our language choices in our analysis and presentation of this work.

\subsection{Conclusion}
In this study, we conducted interviews and user testing sessions with professional spiritual care providers to explore the potential of online spiritual care in anonymous, text-based, and asynchronous spaces such as Reddit. We found that most participants supported the use of technology as a tool of chaplaincy practice. They also advocated for the value of expanding chaplaincy care into online communities, while their perspectives shifted during our study to increase support for healthcare systems to implement and provide training and policies for Online Spiritual Care Communities. However, they were concerned that current online communities are not yet ready for clinical integration or participation. Our work provides foundational guidelines and design recommendations for future work to study and build these communities, in service of bringing spiritual care into great public consciousness and expanding access to care.

\begin{acks}
We extend our sincerest appreciation to all of the wonderful chaplain and spiritual care provider participants who contributed their time and insights to this project. This work was generously funded by John Templeton Foundation grant \href{https://www.templeton.org/grant/expanding-models-of-delivery-for-online-spiritual-care}{\#62930}.
\end{acks}


\bibliographystyle{ACM-Reference-Format}
\bibliography{JTF}


\begin{thebibliography}{165}


\ifx \showCODEN    \undefined \def \showCODEN     #1{\unskip}     \fi
\ifx \showDOI      \undefined \def \showDOI       #1{#1}\fi
\ifx \showISBNx    \undefined \def \showISBNx     #1{\unskip}     \fi
\ifx \showISBNxiii \undefined \def \showISBNxiii  #1{\unskip}     \fi
\ifx \showISSN     \undefined \def \showISSN      #1{\unskip}     \fi
\ifx \showLCCN     \undefined \def \showLCCN      #1{\unskip}     \fi
\ifx \shownote     \undefined \def \shownote      #1{#1}          \fi
\ifx \showarticletitle \undefined \def \showarticletitle #1{#1}   \fi
\ifx \showURL      \undefined \def \showURL       {\relax}        \fi
\providecommand\bibfield[2]{#2}
\providecommand\bibinfo[2]{#2}
\providecommand\natexlab[1]{#1}
\providecommand\showeprint[2][]{arXiv:#2}

\bibitem[Agarwal et~al\mbox{.}(2022)]%
        {agarwal_evaluation_2022}
\bibfield{author}{\bibinfo{person}{Smisha Agarwal}, \bibinfo{person}{Madhu Jalan}, \bibinfo{person}{Holly~C. Wilcox}, \bibinfo{person}{Ritu Sharma}, \bibinfo{person}{Rachel Hill}, \bibinfo{person}{Emily Pantalone}, \bibinfo{person}{Johannes Thrul}, \bibinfo{person}{Jacob~C. Rainey}, {and} \bibinfo{person}{Karen~A. Robinson}.} \bibinfo{year}{2022}\natexlab{}.
\newblock \bibinfo{booktitle}{\emph{Evaluation of {Mental} {Health} {Mobile} {Applications}}}.
\newblock \bibinfo{type}{{T}echnical {R}eport}. \bibinfo{institution}{Agency for Healthcare Research and Quality (AHRQ)}.
\newblock
\urldef\tempurl%
\url{https://doi.org/10.23970/AHRQEPCTB41}
\showDOI{\tempurl}


\bibitem[Ahluwalia et~al\mbox{.}(2022)]%
        {ahluwalia_macra_2022}
\bibfield{author}{\bibinfo{person}{Sangeeta~C. Ahluwalia}, \bibinfo{person}{Brian~G. Vegetabile}, \bibinfo{person}{Maria~Orlando Edelen}, \bibinfo{person}{Claude~Messan Setodji}, \bibinfo{person}{Anthony Rodriguez}, \bibinfo{person}{Adam Scherling}, \bibinfo{person}{Jessica Phillips}, \bibinfo{person}{Carrie~M. Farmer}, \bibinfo{person}{Jordan~M. Harrison}, \bibinfo{person}{Julia Bandini}, \bibinfo{person}{Wenjing Huang}, \bibinfo{person}{Lucy Schulson}, \bibinfo{person}{Anne Walling}, \bibinfo{person}{Sarah Dalton}, \bibinfo{person}{Monique Martineau}, \bibinfo{person}{Owen Hall}, \bibinfo{person}{Danielle Schlang}, \bibinfo{person}{Melissa~A. Bradley}, {and} \bibinfo{person}{Katherine Ast}.} \bibinfo{year}{2022}\natexlab{}.
\newblock \showarticletitle{{MACRA} {Palliative} {Care} {Quality} {Measure} {Development}—{Testing} {Summary} {Report}}.
\newblock \bibinfo{journal}{\emph{Rand Health Quarterly}} \bibinfo{volume}{9}, \bibinfo{number}{3} (\bibinfo{date}{June} \bibinfo{year}{2022}), \bibinfo{pages}{3}.
\newblock
\showISSN{2162-8254}
\urldef\tempurl%
\url{https://www.ncbi.nlm.nih.gov/pmc/articles/PMC9242575/}
\showURL{%
\tempurl}


\bibitem[Ahmadpour et~al\mbox{.}(2023)]%
        {ahmadpour_how_2023}
\bibfield{author}{\bibinfo{person}{Naseem Ahmadpour}, \bibinfo{person}{Phillip Gough}, \bibinfo{person}{Melanie Lovell}, \bibinfo{person}{Philip Austin}, \bibinfo{person}{Philip Poronnik}, \bibinfo{person}{Wendy~Qi Zhang}, \bibinfo{person}{Judy Kay}, \bibinfo{person}{Bob Kummerfeld}, \bibinfo{person}{Tim Luckett}, \bibinfo{person}{Martin Brown}, \bibinfo{person}{Jane~L. Phillips}, {and} \bibinfo{person}{Meera Agar}.} \bibinfo{year}{2023}\natexlab{}.
\newblock \showarticletitle{How can {HCI} support end-of-life care? {Critical} perspectives on sociotechnical imaginaries for palliative care}. In \bibinfo{booktitle}{\emph{Extended {Abstracts} of the 2023 {CHI} {Conference} on {Human} {Factors} in {Computing} {Systems}}} \emph{(\bibinfo{series}{{CHI} {EA} '23})}. \bibinfo{publisher}{Association for Computing Machinery}, \bibinfo{address}{New York, NY, USA}, \bibinfo{pages}{1--7}.
\newblock
\showISBNx{978-1-4503-9422-2}
\urldef\tempurl%
\url{https://doi.org/10.1145/3544549.3585768}
\showDOI{\tempurl}


\bibitem[Al~Shamsi et~al\mbox{.}(2020)]%
        {al_shamsi_implications_2020}
\bibfield{author}{\bibinfo{person}{Hilal Al~Shamsi}, \bibinfo{person}{Abdullah~G. Almutairi}, \bibinfo{person}{Sulaiman Al~Mashrafi}, {and} \bibinfo{person}{Talib Al~Kalbani}.} \bibinfo{year}{2020}\natexlab{}.
\newblock \showarticletitle{Implications of {Language} {Barriers} for {Healthcare}: {A} {Systematic} {Review}}.
\newblock \bibinfo{journal}{\emph{Oman Medical Journal}} \bibinfo{volume}{35}, \bibinfo{number}{2} (\bibinfo{date}{April} \bibinfo{year}{2020}), \bibinfo{pages}{e122}.
\newblock
\showISSN{1999-768X}
\urldef\tempurl%
\url{https://doi.org/10.5001/omj.2020.40}
\showDOI{\tempurl}


\bibitem[Ammari et~al\mbox{.}(2019)]%
        {ammari_self-declared_2019}
\bibfield{author}{\bibinfo{person}{Tawfiq Ammari}, \bibinfo{person}{Sarita Schoenebeck}, {and} \bibinfo{person}{Daniel Romero}.} \bibinfo{year}{2019}\natexlab{}.
\newblock \showarticletitle{Self-declared {Throwaway} {Accounts} on {Reddit}: {How} {Platform} {Affordances} and {Shared} {Norms} enable {Parenting} {Disclosure} and {Support}}.
\newblock \bibinfo{journal}{\emph{Proc. ACM Hum.-Comput. Interact.}} \bibinfo{volume}{3}, \bibinfo{number}{CSCW} (\bibinfo{date}{Nov.} \bibinfo{year}{2019}), \bibinfo{pages}{135:1--135:30}.
\newblock
\urldef\tempurl%
\url{https://doi.org/10.1145/3359237}
\showDOI{\tempurl}


\bibitem[Anandarajah and Hight(2001)]%
        {anandarajah_spirituality_2001}
\bibfield{author}{\bibinfo{person}{Gowri Anandarajah} {and} \bibinfo{person}{Ellen Hight}.} \bibinfo{year}{2001}\natexlab{}.
\newblock \showarticletitle{Spirituality and {Medical} {Practice}: {Using} the {HOPE} {Questions} as a {Practical} {Tool} for {Spiritual} {Assessment}}.
\newblock \bibinfo{journal}{\emph{American Family Physician}} \bibinfo{volume}{63}, \bibinfo{number}{1} (\bibinfo{date}{Jan.} \bibinfo{year}{2001}), \bibinfo{pages}{81--89}.
\newblock
\urldef\tempurl%
\url{https://www.aafp.org/pubs/afp/issues/2001/0101/p81.html}
\showURL{%
\tempurl}


\bibitem[Andy(2021)]%
        {andy_studying_2021}
\bibfield{author}{\bibinfo{person}{Anietie Andy}.} \bibinfo{year}{2021}\natexlab{}.
\newblock \showarticletitle{Studying {How} {Individuals} {Who} {Express} the {Feeling} of {Loneliness} in an {Online} {Loneliness} {Forum} {Communicate} in a {Nonloneliness} {Forum}: {Observational} {Study}}.
\newblock \bibinfo{journal}{\emph{JMIR Formative Research}} \bibinfo{volume}{5}, \bibinfo{number}{7} (\bibinfo{date}{July} \bibinfo{year}{2021}), \bibinfo{pages}{e28738}.
\newblock
\showISSN{2561-326X}
\urldef\tempurl%
\url{https://doi.org/10.2196/28738}
\showDOI{\tempurl}


\bibitem[Arnaez et~al\mbox{.}(2020)]%
        {arnaez_association_2020}
\bibfield{author}{\bibinfo{person}{James~Marcus Arnaez}, \bibinfo{person}{Anne~C. Krendl}, \bibinfo{person}{Bryan~P. McCormick}, \bibinfo{person}{Zhongxue Chen}, {and} \bibinfo{person}{Andrea~K. Chomistek}.} \bibinfo{year}{2020}\natexlab{}.
\newblock \showarticletitle{The association of depression stigma with barriers to seeking mental health care: a cross-sectional analysis}.
\newblock \bibinfo{journal}{\emph{Journal of Mental Health}} \bibinfo{volume}{29}, \bibinfo{number}{2} (\bibinfo{date}{March} \bibinfo{year}{2020}), \bibinfo{pages}{182--190}.
\newblock
\showISSN{0963-8237}
\urldef\tempurl%
\url{https://doi.org/10.1080/09638237.2019.1644494}
\showDOI{\tempurl}
\newblock
\shownote{Publisher: Routledge \_eprint: https://doi.org/10.1080/09638237.2019.1644494}.


\bibitem[Balboni et~al\mbox{.}(2022)]%
        {balboni_spirituality_2022}
\bibfield{author}{\bibinfo{person}{Tracy~A. Balboni}, \bibinfo{person}{Tyler~J. VanderWeele}, \bibinfo{person}{Stephanie~D. Doan-Soares}, \bibinfo{person}{Katelyn N.~G. Long}, \bibinfo{person}{Betty~R. Ferrell}, \bibinfo{person}{George Fitchett}, \bibinfo{person}{Harold~G. Koenig}, \bibinfo{person}{Paul~A. Bain}, \bibinfo{person}{Christina Puchalski}, \bibinfo{person}{Karen~E. Steinhauser}, \bibinfo{person}{Daniel~P. Sulmasy}, {and} \bibinfo{person}{Howard~K. Koh}.} \bibinfo{year}{2022}\natexlab{}.
\newblock \showarticletitle{Spirituality in {Serious} {Illness} and {Health}}.
\newblock \bibinfo{journal}{\emph{JAMA}} \bibinfo{volume}{328}, \bibinfo{number}{2} (\bibinfo{date}{July} \bibinfo{year}{2022}), \bibinfo{pages}{184}.
\newblock
\showISSN{0098-7484}
\urldef\tempurl%
\url{https://doi.org/10.1001/jama.2022.11086}
\showDOI{\tempurl}


\bibitem[Balboni et~al\mbox{.}(2007)]%
        {balboni_religiousness_2007}
\bibfield{author}{\bibinfo{person}{Tracy~A. Balboni}, \bibinfo{person}{Lauren~C. Vanderwerker}, \bibinfo{person}{Susan~D. Block}, \bibinfo{person}{M.~Elizabeth Paulk}, \bibinfo{person}{Christopher~S. Lathan}, \bibinfo{person}{John~R. Peteet}, {and} \bibinfo{person}{Holly~G. Prigerson}.} \bibinfo{year}{2007}\natexlab{}.
\newblock \showarticletitle{Religiousness and {Spiritual} {Support} {Among} {Advanced} {Cancer} {Patients} and {Associations} {With} {End}-of-{Life} {Treatment} {Preferences} and {Quality} of {Life}}.
\newblock \bibinfo{journal}{\emph{Journal of Clinical Oncology}} \bibinfo{volume}{25}, \bibinfo{number}{5} (\bibinfo{date}{Feb.} \bibinfo{year}{2007}), \bibinfo{pages}{555--560}.
\newblock
\showISSN{0732-183X}
\urldef\tempurl%
\url{https://doi.org/10.1200/JCO.2006.07.9046}
\showDOI{\tempurl}
\newblock
\shownote{Publisher: Wolters Kluwer}.


\bibitem[Balcombe and De~Leo(2022)]%
        {balcombe_human-computer_2022}
\bibfield{author}{\bibinfo{person}{Luke Balcombe} {and} \bibinfo{person}{Diego De~Leo}.} \bibinfo{year}{2022}\natexlab{}.
\newblock \showarticletitle{Human-{Computer} {Interaction} in {Digital} {Mental} {Health}}.
\newblock \bibinfo{journal}{\emph{Informatics}} \bibinfo{volume}{9}, \bibinfo{number}{1} (\bibinfo{date}{March} \bibinfo{year}{2022}), \bibinfo{pages}{14}.
\newblock
\showISSN{2227-9709}
\urldef\tempurl%
\url{https://doi.org/10.3390/informatics9010014}
\showDOI{\tempurl}
\newblock
\shownote{Number: 1 Publisher: Multidisciplinary Digital Publishing Institute}.


\bibitem[Bao et~al\mbox{.}(2021)]%
        {bao_conversations_2021}
\bibfield{author}{\bibinfo{person}{Jiajun Bao}, \bibinfo{person}{Junjie Wu}, \bibinfo{person}{Yiming Zhang}, \bibinfo{person}{Eshwar Chandrasekharan}, {and} \bibinfo{person}{David Jurgens}.} \bibinfo{year}{2021}\natexlab{}.
\newblock \showarticletitle{Conversations {Gone} {Alright}: {Quantifying} and {Predicting} {Prosocial} {Outcomes} in {Online} {Conversations}}.
\newblock In \bibinfo{booktitle}{\emph{Proceedings of the {Web} {Conference} 2021}}. \bibinfo{publisher}{Association for Computing Machinery}, \bibinfo{address}{New York, NY, USA}, \bibinfo{pages}{1134--1145}.
\newblock
\showISBNx{978-1-4503-8312-7}
\urldef\tempurl%
\url{https://doi.org/10.1145/3442381.3450122}
\showURL{%
\tempurl}


\bibitem[Bassett et~al\mbox{.}(2018)]%
        {bassett_contribution_2018}
\bibfield{author}{\bibinfo{person}{Lynn Bassett}, \bibinfo{person}{Amanda~F. Bingley}, {and} \bibinfo{person}{Sarah~G. Brearley}.} \bibinfo{year}{2018}\natexlab{}.
\newblock \showarticletitle{The contribution of silence to spiritual care at the end of life: a phenomenological exploration from the experience of palliative care chaplains}.
\newblock \bibinfo{journal}{\emph{Journal for the Study of Spirituality}} \bibinfo{volume}{8}, \bibinfo{number}{1} (\bibinfo{date}{Jan.} \bibinfo{year}{2018}), \bibinfo{pages}{34--48}.
\newblock
\showISSN{2044-0243, 2044-0251}
\urldef\tempurl%
\url{https://doi.org/10.1080/20440243.2018.1431034}
\showDOI{\tempurl}


\bibitem[Besterman-Dahan et~al\mbox{.}(2012)]%
        {besterman-dahan_role_2012}
\bibfield{author}{\bibinfo{person}{Karen Besterman-Dahan}, \bibinfo{person}{Susanne~W. Gibbons}, \bibinfo{person}{Scott~D. Barnett}, {and} \bibinfo{person}{Edward~J. Hickling}.} \bibinfo{year}{2012}\natexlab{}.
\newblock \showarticletitle{The {Role} of {Military} {Chaplains} in {Mental} {Health} {Care} of the {Deployed} {Service} {Member}}.
\newblock \bibinfo{journal}{\emph{Military Medicine}} \bibinfo{volume}{177}, \bibinfo{number}{9} (\bibinfo{date}{Sept.} \bibinfo{year}{2012}), \bibinfo{pages}{1028--1033}.
\newblock
\showISSN{0026-4075}
\urldef\tempurl%
\url{https://doi.org/10.7205/MILMED-D-12-00071}
\showDOI{\tempurl}


\bibitem[Boisen(1936)]%
        {boisen_exploration_1936}
\bibfield{author}{\bibinfo{person}{A.~T. Boisen}.} \bibinfo{year}{1936}\natexlab{}.
\newblock \bibinfo{booktitle}{\emph{The exploration of the inner world. {A} study of mental disorder and religious experience}}.
\newblock \bibinfo{publisher}{Willett, Clark}, \bibinfo{address}{Oxford, England}.
\newblock
\newblock
\shownote{Pages: xi, 322}.


\bibitem[Bonelli et~al\mbox{.}(2012)]%
        {bonelli_religious_2012}
\bibfield{author}{\bibinfo{person}{Raphael Bonelli}, \bibinfo{person}{Rachel~E. Dew}, \bibinfo{person}{Harold~G. Koenig}, \bibinfo{person}{David~H. Rosmarin}, {and} \bibinfo{person}{Sasan Vasegh}.} \bibinfo{year}{2012}\natexlab{}.
\newblock \showarticletitle{Religious and {Spiritual} {Factors} in {Depression}: {Review} and {Integration} of the {Research}}.
\newblock \bibinfo{journal}{\emph{Depression Research and Treatment}}  \bibinfo{volume}{2012} (\bibinfo{year}{2012}), \bibinfo{pages}{1--8}.
\newblock
\showISSN{2090-1321, 2090-133X}
\urldef\tempurl%
\url{https://doi.org/10.1155/2012/962860}
\showDOI{\tempurl}


\bibitem[Borneman et~al\mbox{.}(2010)]%
        {borneman_evaluation_2010}
\bibfield{author}{\bibinfo{person}{Tami Borneman}, \bibinfo{person}{Betty Ferrell}, {and} \bibinfo{person}{Christina~M. Puchalski}.} \bibinfo{year}{2010}\natexlab{}.
\newblock \showarticletitle{Evaluation of the {FICA} {Tool} for {Spiritual} {Assessment}}.
\newblock \bibinfo{journal}{\emph{Journal of Pain and Symptom Management}} \bibinfo{volume}{40}, \bibinfo{number}{2} (\bibinfo{date}{Aug.} \bibinfo{year}{2010}), \bibinfo{pages}{163--173}.
\newblock
\showISSN{0885-3924}
\urldef\tempurl%
\url{https://doi.org/10.1016/j.jpainsymman.2009.12.019}
\showDOI{\tempurl}


\bibitem[Braam and Koenig(2019)]%
        {braam_religion_2019}
\bibfield{author}{\bibinfo{person}{Arjan~W. Braam} {and} \bibinfo{person}{Harold~G. Koenig}.} \bibinfo{year}{2019}\natexlab{}.
\newblock \showarticletitle{Religion, spirituality and depression in prospective studies: {A} systematic review}.
\newblock \bibinfo{journal}{\emph{Journal of Affective Disorders}}  \bibinfo{volume}{257} (\bibinfo{date}{Oct.} \bibinfo{year}{2019}), \bibinfo{pages}{428--438}.
\newblock
\showISSN{01650327}
\urldef\tempurl%
\url{https://doi.org/10.1016/j.jad.2019.06.063}
\showDOI{\tempurl}


\bibitem[Brown(2023)]%
        {brown_will_2023}
\bibfield{author}{\bibinfo{person}{Elizabeth Brown}.} \bibinfo{year}{2023}\natexlab{}.
\newblock \showarticletitle{Will {AI} ever become spiritual? {A} {Hospital} {Chaplaincy} perspective}.
\newblock \bibinfo{journal}{\emph{Practical Theology}} \bibinfo{volume}{16}, \bibinfo{number}{6} (\bibinfo{date}{Nov.} \bibinfo{year}{2023}), \bibinfo{pages}{801--813}.
\newblock
\showISSN{1756-073X}
\urldef\tempurl%
\url{https://doi.org/10.1080/1756073X.2023.2242940}
\showDOI{\tempurl}
\newblock
\shownote{Publisher: Routledge \_eprint: https://doi.org/10.1080/1756073X.2023.2242940}.


\bibitem[Buklijaš(2008)]%
        {buklijas_medicine_2008}
\bibfield{author}{\bibinfo{person}{Tatjana Buklijaš}.} \bibinfo{year}{2008}\natexlab{}.
\newblock \showarticletitle{Medicine and {Society} in the {Medieval} {Hospital}}.
\newblock \bibinfo{journal}{\emph{Croatian medical journal}} \bibinfo{volume}{49}, \bibinfo{number}{2} (\bibinfo{date}{April} \bibinfo{year}{2008}), \bibinfo{pages}{151--154}.
\newblock
\showISSN{0353-9504}
\urldef\tempurl%
\url{https://doi.org/10.3325/cmj.2008.2.151}
\showDOI{\tempurl}


\bibitem[Buntinx-Krieg et~al\mbox{.}(2017)]%
        {buntinx-krieg_dermatology_2017}
\bibfield{author}{\bibinfo{person}{Talayesa Buntinx-Krieg}, \bibinfo{person}{Joseph Caravaglio}, \bibinfo{person}{Renee Domozych}, {and} \bibinfo{person}{Robert~P. Dellavalle}.} \bibinfo{year}{2017}\natexlab{}.
\newblock \showarticletitle{Dermatology on {Reddit}: elucidating trends in dermatologic communications on the world wide web}.
\newblock \bibinfo{journal}{\emph{Dermatology Online Journal}} \bibinfo{volume}{23}, \bibinfo{number}{7} (\bibinfo{year}{2017}).
\newblock
\urldef\tempurl%
\url{https://doi.org/10.5070/D3237035730}
\showDOI{\tempurl}


\bibitem[Bush et~al\mbox{.}(2018)]%
        {bush_systematic_2018}
\bibfield{author}{\bibinfo{person}{Ruth~A Bush}, \bibinfo{person}{Alexa Pérez}, \bibinfo{person}{Tanja Baum}, \bibinfo{person}{Caroline Etland}, {and} \bibinfo{person}{Cynthia~D Connelly}.} \bibinfo{year}{2018}\natexlab{}.
\newblock \showarticletitle{A systematic review of the use of the electronic health record for patient identification, communication, and clinical support in palliative care}.
\newblock \bibinfo{journal}{\emph{JAMIA Open}} \bibinfo{volume}{1}, \bibinfo{number}{2} (\bibinfo{date}{Oct.} \bibinfo{year}{2018}), \bibinfo{pages}{294--303}.
\newblock
\showISSN{2574-2531}
\urldef\tempurl%
\url{https://doi.org/10.1093/jamiaopen/ooy028}
\showDOI{\tempurl}


\bibitem[Calder et~al\mbox{.}(2023)]%
        {calder_improving_2023}
\bibfield{author}{\bibinfo{person}{Steve Calder}, \bibinfo{person}{Matthew Andreotta}, \bibinfo{person}{Thomas Morris}, {and} \bibinfo{person}{Mustafa Atee}.} \bibinfo{year}{2023}\natexlab{}.
\newblock \showarticletitle{Improving quality in pastoral care using the {Pastoral} {Care} {Activity} {Tracker} ({PCAT}): {A} feasibility study of a digital tool within an {Australian} healthcare organization}.
\newblock \bibinfo{journal}{\emph{Journal of Health Care Chaplaincy}} \bibinfo{volume}{29}, \bibinfo{number}{4} (\bibinfo{date}{Oct.} \bibinfo{year}{2023}), \bibinfo{pages}{353--367}.
\newblock
\showISSN{0885-4726}
\urldef\tempurl%
\url{https://doi.org/10.1080/08854726.2022.2091837}
\showDOI{\tempurl}
\newblock
\shownote{Publisher: Routledge \_eprint: https://doi.org/10.1080/08854726.2022.2091837}.


\bibitem[Campbell et~al\mbox{.}(2023)]%
        {campbell_standardized_2023}
\bibfield{author}{\bibinfo{person}{Duane Campbell}, \bibinfo{person}{Jeanene~(Gigi) Robison}, {and} \bibinfo{person}{Judi~Allyn Godsey}.} \bibinfo{year}{2023}\natexlab{}.
\newblock \showarticletitle{Standardized {Spiritual} {Screening} {Increases} {Chaplain} {Referrals} {Through} the {EMR}: {A} {Nurse}-{Chaplain} {Collaboration} for {Holistic} {Acute} {Healthcare}}.
\newblock \bibinfo{journal}{\emph{Journal of Holistic Nursing}} \bibinfo{volume}{41}, \bibinfo{number}{1} (\bibinfo{date}{March} \bibinfo{year}{2023}), \bibinfo{pages}{30--39}.
\newblock
\showISSN{0898-0101}
\urldef\tempurl%
\url{https://doi.org/10.1177/08980101221079463}
\showDOI{\tempurl}
\newblock
\shownote{Publisher: SAGE Publications Inc}.


\bibitem[Care(2014)]%
        {advocate_health_care_chaplaincy_2014}
\bibfield{author}{\bibinfo{person}{Advocate~Health Care}.} \bibinfo{year}{2014}\natexlab{}.
\newblock \bibinfo{title}{Chaplaincy {Taxonomy} {User}’s {Guide} (ver. 1)}.
\newblock
\newblock


\bibitem[Chancellor et~al\mbox{.}(2021)]%
        {chancellor_suicide_2021}
\bibfield{author}{\bibinfo{person}{Stevie Chancellor}, \bibinfo{person}{Steven~A. Sumner}, \bibinfo{person}{Corinne David-Ferdon}, \bibinfo{person}{Tahirah Ahmad}, {and} \bibinfo{person}{Munmun~De Choudhury}.} \bibinfo{year}{2021}\natexlab{}.
\newblock \showarticletitle{Suicide {Risk} and {Protective} {Factors} in {Online} {Support} {Forum} {Posts}: {Annotation} {Scheme} {Development} and {Validation} {Study}}.
\newblock \bibinfo{journal}{\emph{JMIR Mental Health}} \bibinfo{volume}{8}, \bibinfo{number}{11} (\bibinfo{date}{Nov.} \bibinfo{year}{2021}), \bibinfo{pages}{e24471}.
\newblock
\urldef\tempurl%
\url{https://doi.org/10.2196/24471}
\showDOI{\tempurl}
\newblock
\shownote{Company: JMIR Mental Health Distributor: JMIR Mental Health Institution: JMIR Mental Health Label: JMIR Mental Health Publisher: JMIR Publications Inc., Toronto, Canada}.


\bibitem[Chandrasekharan et~al\mbox{.}(2018)]%
        {chandrasekharan_internets_2018}
\bibfield{author}{\bibinfo{person}{Eshwar Chandrasekharan}, \bibinfo{person}{Mattia Samory}, \bibinfo{person}{Shagun Jhaver}, \bibinfo{person}{Hunter Charvat}, \bibinfo{person}{Amy Bruckman}, \bibinfo{person}{Cliff Lampe}, \bibinfo{person}{Jacob Eisenstein}, {and} \bibinfo{person}{Eric Gilbert}.} \bibinfo{year}{2018}\natexlab{}.
\newblock \showarticletitle{The {Internet}'s {Hidden} {Rules}: {An} {Empirical} {Study} of {Reddit} {Norm} {Violations} at {Micro}, {Meso}, and {Macro} {Scales}}.
\newblock \bibinfo{journal}{\emph{Proceedings of the ACM on Human-Computer Interaction}} \bibinfo{volume}{2}, \bibinfo{number}{CSCW} (\bibinfo{date}{Nov.} \bibinfo{year}{2018}), \bibinfo{pages}{1--25}.
\newblock
\showISSN{2573-0142}
\urldef\tempurl%
\url{https://doi.org/10.1145/3274301}
\showDOI{\tempurl}


\bibitem[Chaplain~Corps(2015)]%
        {chaplain_corps_us_2015}
\bibfield{author}{\bibinfo{person}{U.S.~Army Chaplain~Corps}.} \bibinfo{year}{2015}\natexlab{}.
\newblock \bibinfo{title}{U.{S}. {Army} {Chaplain} {Corps}}.
\newblock
\newblock
\urldef\tempurl%
\url{https://armyhistory.org/u-s-army-chaplain-corps/}
\showURL{%
\tempurl}


\bibitem[Choudhury and De(2014)]%
        {choudhury_mental_2014}
\bibfield{author}{\bibinfo{person}{Munmun~De Choudhury} {and} \bibinfo{person}{Sushovan De}.} \bibinfo{year}{2014}\natexlab{}.
\newblock \showarticletitle{Mental {Health} {Discourse} on reddit: {Self}-{Disclosure}, {Social} {Support}, and {Anonymity}}. In \bibinfo{booktitle}{\emph{Eighth {International} {AAAI} {Conference} on {Weblogs} and {Social} {Media}}}.
\newblock
\urldef\tempurl%
\url{https://www.aaai.org/ocs/index.php/ICWSM/ICWSM14/paper/view/8075}
\showURL{%
\tempurl}


\bibitem[Claisse and Durrant(2023)]%
        {claisse_keeping_2023}
\bibfield{author}{\bibinfo{person}{Caroline Claisse} {and} \bibinfo{person}{Abigail~C Durrant}.} \bibinfo{year}{2023}\natexlab{}.
\newblock \showarticletitle{‘{Keeping} our {Faith} {Alive}’: {Investigating} {Buddhism} {Practice} during {COVID}-19 to {Inform} {Design} for the {Online} {Community} {Practice} of {Faith}}. In \bibinfo{booktitle}{\emph{Proceedings of the 2023 {CHI} {Conference} on {Human} {Factors} in {Computing} {Systems}}}. \bibinfo{publisher}{ACM}, \bibinfo{address}{Hamburg Germany}, \bibinfo{pages}{1--19}.
\newblock
\showISBNx{978-1-4503-9421-5}
\urldef\tempurl%
\url{https://doi.org/10.1145/3544548.3581177}
\showDOI{\tempurl}


\bibitem[Colineau and Paris(2010)]%
        {colineau_talking_2010}
\bibfield{author}{\bibinfo{person}{Nathalie Colineau} {and} \bibinfo{person}{Cécile Paris}.} \bibinfo{year}{2010}\natexlab{}.
\newblock \showarticletitle{Talking about your health to strangers: understanding the use of online social networks by patients}.
\newblock \bibinfo{journal}{\emph{New Review of Hypermedia and Multimedia}} \bibinfo{volume}{16}, \bibinfo{number}{1-2} (\bibinfo{date}{April} \bibinfo{year}{2010}), \bibinfo{pages}{141--160}.
\newblock
\showISSN{1361-4568}
\urldef\tempurl%
\url{https://doi.org/10.1080/13614568.2010.496131}
\showDOI{\tempurl}
\newblock
\shownote{Publisher: Taylor \& Francis \_eprint: https://doi.org/10.1080/13614568.2010.496131}.


\bibitem[Coombs et~al\mbox{.}(2021)]%
        {coombs_barriers_2021}
\bibfield{author}{\bibinfo{person}{Nicholas~C. Coombs}, \bibinfo{person}{Wyatt~E. Meriwether}, \bibinfo{person}{James Caringi}, {and} \bibinfo{person}{Sophia~R. Newcomer}.} \bibinfo{year}{2021}\natexlab{}.
\newblock \showarticletitle{Barriers to healthcare access among {U}.{S}. adults with mental health challenges: {A} population-based study}.
\newblock \bibinfo{journal}{\emph{SSM - Population Health}}  \bibinfo{volume}{15} (\bibinfo{date}{Sept.} \bibinfo{year}{2021}), \bibinfo{pages}{100847}.
\newblock
\showISSN{2352-8273}
\urldef\tempurl%
\url{https://doi.org/10.1016/j.ssmph.2021.100847}
\showDOI{\tempurl}


\bibitem[Cooper et~al\mbox{.}(2023)]%
        {cooper_military_2023}
\bibfield{author}{\bibinfo{person}{Denise~C. Cooper}, \bibinfo{person}{Christian~A. Evans}, \bibinfo{person}{Sharmila~A. Chari}, \bibinfo{person}{Marjorie~S. Campbell}, {and} \bibinfo{person}{Tim Hoyt}.} \bibinfo{year}{2023}\natexlab{}.
\newblock \showarticletitle{Military chaplains and mental health clinicians: {Overlapping} roles and collaborative opportunities}.
\newblock \bibinfo{journal}{\emph{Psychological Services}} \bibinfo{volume}{20}, \bibinfo{number}{1} (\bibinfo{year}{2023}), \bibinfo{pages}{56--65}.
\newblock
\showISSN{1939-148X}
\urldef\tempurl%
\url{https://doi.org/10.1037/ser0000661}
\showDOI{\tempurl}
\newblock
\shownote{Place: US Publisher: Educational Publishing Foundation}.


\bibitem[Costantini et~al\mbox{.}(2021)]%
        {costantini_screening_2021}
\bibfield{author}{\bibinfo{person}{Luigi Costantini}, \bibinfo{person}{Cesira Pasquarella}, \bibinfo{person}{Anna Odone}, \bibinfo{person}{Maria~Eugenia Colucci}, \bibinfo{person}{Alessandra Costanza}, \bibinfo{person}{Gianluca Serafini}, \bibinfo{person}{Andrea Aguglia}, \bibinfo{person}{Martino Belvederi~Murri}, \bibinfo{person}{Vlasios Brakoulias}, \bibinfo{person}{Mario Amore}, \bibinfo{person}{S.~Nassir Ghaemi}, {and} \bibinfo{person}{Andrea Amerio}.} \bibinfo{year}{2021}\natexlab{}.
\newblock \showarticletitle{Screening for depression in primary care with {Patient} {Health} {Questionnaire}-9 ({PHQ}-9): {A} systematic review}.
\newblock \bibinfo{journal}{\emph{Journal of Affective Disorders}}  \bibinfo{volume}{279} (\bibinfo{date}{Jan.} \bibinfo{year}{2021}), \bibinfo{pages}{473--483}.
\newblock
\showISSN{0165-0327}
\urldef\tempurl%
\url{https://doi.org/10.1016/j.jad.2020.09.131}
\showDOI{\tempurl}


\bibitem[Coventry et~al\mbox{.}(2020)]%
        {coventry_psychological_2020}
\bibfield{author}{\bibinfo{person}{Peter~A. Coventry}, \bibinfo{person}{Nick Meader}, \bibinfo{person}{Hollie Melton}, \bibinfo{person}{Melanie Temple}, \bibinfo{person}{Holly Dale}, \bibinfo{person}{Kath Wright}, \bibinfo{person}{Marylène Cloitre}, \bibinfo{person}{Thanos Karatzias}, \bibinfo{person}{Jonathan Bisson}, \bibinfo{person}{Neil~P. Roberts}, \bibinfo{person}{Jennifer V.~E. Brown}, \bibinfo{person}{Corrado Barbui}, \bibinfo{person}{Rachel Churchill}, \bibinfo{person}{Karina Lovell}, \bibinfo{person}{Dean McMillan}, {and} \bibinfo{person}{Simon Gilbody}.} \bibinfo{year}{2020}\natexlab{}.
\newblock \showarticletitle{Psychological and pharmacological interventions for posttraumatic stress disorder and comorbid mental health problems following complex traumatic events: {Systematic} review and component network meta-analysis}.
\newblock \bibinfo{journal}{\emph{PLOS Medicine}} \bibinfo{volume}{17}, \bibinfo{number}{8} (\bibinfo{date}{Aug.} \bibinfo{year}{2020}), \bibinfo{pages}{e1003262}.
\newblock
\showISSN{1549-1676}
\urldef\tempurl%
\url{https://doi.org/10.1371/journal.pmed.1003262}
\showDOI{\tempurl}
\newblock
\shownote{Publisher: Public Library of Science}.


\bibitem[Cutrona and Suhr(1992)]%
        {cutrona_controllability_1992}
\bibfield{author}{\bibinfo{person}{Carolyn~E. Cutrona} {and} \bibinfo{person}{Julie~A. Suhr}.} \bibinfo{year}{1992}\natexlab{}.
\newblock \showarticletitle{Controllability of {Stressful} {Events} and {Satisfaction} {With} {Spouse} {Support} {Behaviors}}.
\newblock \bibinfo{journal}{\emph{Communication Research}} \bibinfo{volume}{19}, \bibinfo{number}{2} (\bibinfo{date}{April} \bibinfo{year}{1992}), \bibinfo{pages}{154--174}.
\newblock
\showISSN{0093-6502}
\urldef\tempurl%
\url{https://doi.org/10.1177/009365092019002002}
\showDOI{\tempurl}


\bibitem[Cyr et~al\mbox{.}(2019)]%
        {cyr_access_2019}
\bibfield{author}{\bibinfo{person}{Melissa~E. Cyr}, \bibinfo{person}{Anna~G. Etchin}, \bibinfo{person}{Barbara~J. Guthrie}, {and} \bibinfo{person}{James~C. Benneyan}.} \bibinfo{year}{2019}\natexlab{}.
\newblock \showarticletitle{Access to specialty healthcare in urban versus rural {US} populations: a systematic literature review}.
\newblock \bibinfo{journal}{\emph{BMC Health Services Research}} \bibinfo{volume}{19}, \bibinfo{number}{1} (\bibinfo{date}{Dec.} \bibinfo{year}{2019}), \bibinfo{pages}{974}.
\newblock
\showISSN{1472-6963}
\urldef\tempurl%
\url{https://doi.org/10.1186/s12913-019-4815-5}
\showDOI{\tempurl}


\bibitem[de~Diego-Cordero et~al\mbox{.}(2023)]%
        {de_diego-cordero_more_2023}
\bibfield{author}{\bibinfo{person}{Rocío de Diego-Cordero}, \bibinfo{person}{Ángeles~C. López-Tarrida}, \bibinfo{person}{Carmen Linero-Narváez}, {and} \bibinfo{person}{José~María Galán González-Serna}.} \bibinfo{year}{2023}\natexlab{}.
\newblock \showarticletitle{“{More} {Spiritual} {Health} {Professionals} {Provide} {Different} {Care}”: {A} {Qualitative} {Study} in the {Field} of {Mental} {Health}}.
\newblock \bibinfo{journal}{\emph{Healthcare}} \bibinfo{volume}{11}, \bibinfo{number}{3} (\bibinfo{date}{Jan.} \bibinfo{year}{2023}), \bibinfo{pages}{303}.
\newblock
\showISSN{2227-9032}
\urldef\tempurl%
\url{https://doi.org/10.3390/healthcare11030303}
\showDOI{\tempurl}


\bibitem[Denney(2018)]%
        {denney_prison_2018}
\bibfield{author}{\bibinfo{person}{Andrew~S. Denney}.} \bibinfo{year}{2018}\natexlab{}.
\newblock \showarticletitle{Prison {Chaplains}: {Perceptions} of {Criminality}, {Effective} {Prison} {Programming} {Characteristics}, and the {Role} of {Religion} in the {Desistance} from {Crime}}.
\newblock \bibinfo{journal}{\emph{American Journal of Criminal Justice}} \bibinfo{volume}{43}, \bibinfo{number}{3} (\bibinfo{date}{Sept.} \bibinfo{year}{2018}), \bibinfo{pages}{694--723}.
\newblock
\showISSN{1936-1351}
\urldef\tempurl%
\url{https://doi.org/10.1007/s12103-017-9425-3}
\showDOI{\tempurl}


\bibitem[Doyle et~al\mbox{.}(2024)]%
        {doyle_i_2024}
\bibfield{author}{\bibinfo{person}{Dylan~Thomas Doyle}, \bibinfo{person}{Charlie Blue~R. Brahm}, {and} \bibinfo{person}{Jed~R. Brubaker}.} \bibinfo{year}{2024}\natexlab{}.
\newblock \showarticletitle{"{I} hate you. {I} love you. {I}'m sorry. {I} miss you." {Understanding} {Online} {Grief} {Expression} {Through} {Suicide} {Bereavement} {Letter}-{Writing} {Practices}}.
\newblock \bibinfo{journal}{\emph{Proc. ACM Hum.-Comput. Interact.}} \bibinfo{volume}{8}, \bibinfo{number}{CSCW1} (\bibinfo{date}{April} \bibinfo{year}{2024}), \bibinfo{pages}{69:1--69:27}.
\newblock
\urldef\tempurl%
\url{https://doi.org/10.1145/3637346}
\showDOI{\tempurl}


\bibitem[Duimel et~al\mbox{.}(2023)]%
        {duimel_profiling_2023}
\bibfield{author}{\bibinfo{person}{Song L.~L. Duimel}, \bibinfo{person}{Annemiek~J. Linn}, \bibinfo{person}{Ellen M.~A. Smets}, \bibinfo{person}{Eline~S. Smit}, {and} \bibinfo{person}{Julia C.~M. van Weert}.} \bibinfo{year}{2023}\natexlab{}.
\newblock \showarticletitle{Profiling {Cancer} {Patients} {Based} on {Their} {Motives} for {Seeking} {Informational} and {Emotional} {Support} {Online}}.
\newblock \bibinfo{journal}{\emph{Health Communication}} \bibinfo{volume}{38}, \bibinfo{number}{14} (\bibinfo{date}{Dec.} \bibinfo{year}{2023}), \bibinfo{pages}{3223--3237}.
\newblock
\showISSN{1041-0236}
\urldef\tempurl%
\url{https://doi.org/10.1080/10410236.2022.2144287}
\showDOI{\tempurl}
\newblock
\shownote{Publisher: Routledge \_eprint: https://doi.org/10.1080/10410236.2022.2144287}.


\bibitem[Edenfield et~al\mbox{.}(2019)]%
        {edenfield_always_2019}
\bibfield{author}{\bibinfo{person}{Avery~C. Edenfield}, \bibinfo{person}{Jared~S. Colton}, {and} \bibinfo{person}{Steve Holmes}.} \bibinfo{year}{2019}\natexlab{}.
\newblock \showarticletitle{Always {Already} {Geopolitical}: {Trans} {Health} {Care} and {Global} {Tactical} {Technical} {Communication}}.
\newblock \bibinfo{journal}{\emph{Journal of Technical Writing and Communication}} \bibinfo{volume}{49}, \bibinfo{number}{4} (\bibinfo{date}{Oct.} \bibinfo{year}{2019}), \bibinfo{pages}{433--457}.
\newblock
\showISSN{0047-2816}
\urldef\tempurl%
\url{https://doi.org/10.1177/0047281619871211}
\showDOI{\tempurl}
\newblock
\shownote{Publisher: SAGE Publications Inc}.


\bibitem[Feldman(2011)]%
        {feldman_spirit_2011}
\bibfield{author}{\bibinfo{person}{Kathryn~Levy Feldman}.} \bibinfo{year}{2011}\natexlab{}.
\newblock \showarticletitle{The {Spirit} of {Caring}}.
\newblock \bibinfo{journal}{\emph{THE PENNSYLVANIA GAZETTE}} (\bibinfo{year}{2011}).
\newblock


\bibitem[Ferrell et~al\mbox{.}(2020)]%
        {ferrell_urgency_2020}
\bibfield{author}{\bibinfo{person}{Betty~R. Ferrell}, \bibinfo{person}{George Handzo}, \bibinfo{person}{Tina Picchi}, \bibinfo{person}{Christina Puchalski}, {and} \bibinfo{person}{William~E. Rosa}.} \bibinfo{year}{2020}\natexlab{}.
\newblock \showarticletitle{The {Urgency} of {Spiritual} {Care}: {COVID}-19 and the {Critical} {Need} for {Whole}-{Person} {Palliation}}.
\newblock \bibinfo{journal}{\emph{Journal of Pain and Symptom Management}} \bibinfo{volume}{60}, \bibinfo{number}{3} (\bibinfo{date}{Sept.} \bibinfo{year}{2020}), \bibinfo{pages}{e7--e11}.
\newblock
\showISSN{0885-3924}
\urldef\tempurl%
\url{https://doi.org/10.1016/j.jpainsymman.2020.06.034}
\showDOI{\tempurl}


\bibitem[Fiesler et~al\mbox{.}(2018)]%
        {fiesler_reddit_2018}
\bibfield{author}{\bibinfo{person}{Casey Fiesler}, \bibinfo{person}{Jialun Jiang}, \bibinfo{person}{Joshua McCann}, \bibinfo{person}{Kyle Frye}, {and} \bibinfo{person}{Jed Brubaker}.} \bibinfo{year}{2018}\natexlab{}.
\newblock \showarticletitle{Reddit {Rules}! {Characterizing} an {Ecosystem} of {Governance}}.
\newblock \bibinfo{journal}{\emph{Proceedings of the International AAAI Conference on Web and Social Media}} \bibinfo{volume}{12}, \bibinfo{number}{1} (\bibinfo{date}{June} \bibinfo{year}{2018}).
\newblock
\showISSN{2334-0770}
\urldef\tempurl%
\url{https://ojs.aaai.org/index.php/ICWSM/article/view/15033}
\showURL{%
\tempurl}
\newblock
\shownote{Number: 1}.


\bibitem[Fitchett et~al\mbox{.}(2019)]%
        {fitchett_development_2019}
\bibfield{author}{\bibinfo{person}{George Fitchett}, \bibinfo{person}{Anna Pierson}, \bibinfo{person}{Christine Hoffmeyer}, \bibinfo{person}{Dirk Labuschagne}, \bibinfo{person}{Aoife Lee}, \bibinfo{person}{Stacie Levine}, \bibinfo{person}{Sean Omahony}, \bibinfo{person}{Karen Pugliese}, {and} \bibinfo{person}{Nancy Waite}.} \bibinfo{year}{2019}\natexlab{}.
\newblock \showarticletitle{Development of the {PC}-7, a {Quantifiable} {Assessment} of {Spiritual} {Concerns} of {Patients} {Receiving} {Palliative} {Care} {Near} the {End} of {Life}}.
\newblock \bibinfo{journal}{\emph{Journal of Palliative Medicine}}  \bibinfo{volume}{23} (\bibinfo{date}{Sept.} \bibinfo{year}{2019}).
\newblock
\urldef\tempurl%
\url{https://doi.org/10.1089/jpm.2019.0188}
\showDOI{\tempurl}


\bibitem[Fleischman(2016)]%
        {fleischman_telechaplaincy_2016}
\bibfield{author}{\bibinfo{person}{Judy Fleischman}.} \bibinfo{year}{2016}\natexlab{}.
\newblock \bibinfo{title}{Telechaplaincy: {Best} {Practices} for {Telehealth} {Chaplaincy} {Care}}.
\newblock
\newblock
\urldef\tempurl%
\url{https://static1.squarespace.com/static/53acafeee4b00b01999974b2/t/5ae902edf950b7f108f379e7/1525220078430/Fleischman_HCCN_2016Conf_Telechaplaincy+FINAL.pdf}
\showURL{%
\tempurl}


\bibitem[Fleischman(2017)]%
        {fleischman_telechaplaincy_2017}
\bibfield{author}{\bibinfo{person}{Judy Fleischman}.} \bibinfo{year}{2017}\natexlab{}.
\newblock \bibinfo{title}{{TeleChaplaincy}: {Expanding} the {Reach} of {Patient}-{Centered} {Care}}.
\newblock
\newblock
\urldef\tempurl%
\url{https://static1.squarespace.com/static/53acafeee4b00b01999974b2/t/5ae903712b6a28d38239d3c9/1525220210758/FleischmanJ+HCCN+2017+WorkshopB3+.pdf}
\showURL{%
\tempurl}


\bibitem[Ford and Tartaglia(2006)]%
        {ford_development_2006}
\bibfield{author}{\bibinfo{person}{Tim Ford} {and} \bibinfo{person}{Alexander Tartaglia}.} \bibinfo{year}{2006}\natexlab{}.
\newblock \showarticletitle{The {Development}, {Status}, and {Future} of {Healthcare} {Chaplaincy}:}.
\newblock \bibinfo{journal}{\emph{Southern Medical Journal}} \bibinfo{volume}{99}, \bibinfo{number}{6} (\bibinfo{date}{June} \bibinfo{year}{2006}), \bibinfo{pages}{675--679}.
\newblock
\showISSN{0038-4348}
\urldef\tempurl%
\url{https://doi.org/10.1097/01.smj.0000220893.37354.1e}
\showDOI{\tempurl}


\bibitem[Gemert-Pijnen et~al\mbox{.}(2011)]%
        {gemert-pijnen_holistic_2011}
\bibfield{author}{\bibinfo{person}{Julia EWC~van Gemert-Pijnen}, \bibinfo{person}{Nicol Nijland}, \bibinfo{person}{Maarten~van Limburg}, \bibinfo{person}{Hans~C. Ossebaard}, \bibinfo{person}{Saskia~M. Kelders}, \bibinfo{person}{Gunther Eysenbach}, {and} \bibinfo{person}{Erwin~R. Seydel}.} \bibinfo{year}{2011}\natexlab{}.
\newblock \showarticletitle{A {Holistic} {Framework} to {Improve} the {Uptake} and {Impact} of {eHealth} {Technologies}}.
\newblock \bibinfo{journal}{\emph{Journal of Medical Internet Research}} \bibinfo{volume}{13}, \bibinfo{number}{4} (\bibinfo{date}{Dec.} \bibinfo{year}{2011}), \bibinfo{pages}{e1672}.
\newblock
\urldef\tempurl%
\url{https://doi.org/10.2196/jmir.1672}
\showDOI{\tempurl}
\newblock
\shownote{Company: Journal of Medical Internet Research Distributor: Journal of Medical Internet Research Institution: Journal of Medical Internet Research Label: Journal of Medical Internet Research Publisher: JMIR Publications Inc., Toronto, Canada}.


\bibitem[Gijsberts et~al\mbox{.}(2019)]%
        {gijsberts_spiritual_2019}
\bibfield{author}{\bibinfo{person}{Marie-José H.~E. Gijsberts}, \bibinfo{person}{Anke~I. Liefbroer}, \bibinfo{person}{René Otten}, {and} \bibinfo{person}{Erik Olsman}.} \bibinfo{year}{2019}\natexlab{}.
\newblock \showarticletitle{Spiritual {Care} in {Palliative} {Care}: {A} {Systematic} {Review} of the {Recent} {European} {Literature}}.
\newblock \bibinfo{journal}{\emph{Medical Sciences}} \bibinfo{volume}{7}, \bibinfo{number}{2} (\bibinfo{date}{Feb.} \bibinfo{year}{2019}), \bibinfo{pages}{25}.
\newblock
\showISSN{2076-3271}
\urldef\tempurl%
\url{https://doi.org/10.3390/medsci7020025}
\showDOI{\tempurl}
\newblock
\shownote{Number: 2 Publisher: Multidisciplinary Digital Publishing Institute}.


\bibitem[Giles and Newbold(2011)]%
        {giles_self-_2011}
\bibfield{author}{\bibinfo{person}{David~C. Giles} {and} \bibinfo{person}{Julie Newbold}.} \bibinfo{year}{2011}\natexlab{}.
\newblock \showarticletitle{Self- and {Other}-{Diagnosis} in {User}-{Led} {Mental} {Health} {Online} {Communities}}.
\newblock \bibinfo{journal}{\emph{Qualitative Health Research}} \bibinfo{volume}{21}, \bibinfo{number}{3} (\bibinfo{date}{March} \bibinfo{year}{2011}), \bibinfo{pages}{419--428}.
\newblock
\showISSN{1049-7323}
\urldef\tempurl%
\url{https://doi.org/10.1177/1049732310381388}
\showDOI{\tempurl}
\newblock
\shownote{Publisher: SAGE Publications Inc}.


\bibitem[Gordon(2023)]%
        {gordon_military_2023}
\bibfield{author}{\bibinfo{person}{Bri Gordon}.} \bibinfo{year}{2023}\natexlab{}.
\newblock \bibinfo{title}{{MILITARY} {CHAPLAINS}: {THEIR} {VITAL} {ROLE} {IN} {OUR} {MILITARY}}.
\newblock
\newblock
\urldef\tempurl%
\url{https://militaryreach.auburn.edu/research-in-action-detail?resourceid=dcc75f4e-1c36-46b8-8904-93eb8691608d}
\showURL{%
\tempurl}


\bibitem[Hall and Powell(2021)]%
        {hall_valuing_2021}
\bibfield{author}{\bibinfo{person}{Eric~J. Hall} {and} \bibinfo{person}{Richard~A. Powell}.} \bibinfo{year}{2021}\natexlab{}.
\newblock \showarticletitle{Valuing the {Spiritual}}.
\newblock \bibinfo{journal}{\emph{Journal of Religion and Health}} \bibinfo{volume}{60}, \bibinfo{number}{3} (\bibinfo{year}{2021}), \bibinfo{pages}{1430--1435}.
\newblock
\showISSN{0022-4197}
\urldef\tempurl%
\url{https://doi.org/10.1007/s10943-021-01206-9}
\showDOI{\tempurl}


\bibitem[Handzo et~al\mbox{.}(2014)]%
        {handzo_outcomes_2014}
\bibfield{author}{\bibinfo{person}{George~F. Handzo}, \bibinfo{person}{Mark Cobb}, \bibinfo{person}{Cheryl Holmes}, \bibinfo{person}{Ewan Kelly}, {and} \bibinfo{person}{Shane Sinclair}.} \bibinfo{year}{2014}\natexlab{}.
\newblock \showarticletitle{Outcomes for {Professional} {Health} {Care} {Chaplaincy}: {An} {International} {Call} to {Action}}.
\newblock \bibinfo{journal}{\emph{Journal of Health Care Chaplaincy}} \bibinfo{volume}{20}, \bibinfo{number}{2} (\bibinfo{date}{April} \bibinfo{year}{2014}), \bibinfo{pages}{43--53}.
\newblock
\showISSN{0885-4726}
\urldef\tempurl%
\url{https://doi.org/10.1080/08854726.2014.902713}
\showDOI{\tempurl}
\newblock
\shownote{Publisher: Routledge \_eprint: https://doi.org/10.1080/08854726.2014.902713}.


\bibitem[Haque et~al\mbox{.}(2014)]%
        {haque_findings_2014}
\bibfield{author}{\bibinfo{person}{Munirul~M. Haque}, \bibinfo{person}{Ferdaus Kawsar}, \bibinfo{person}{Md. Adibuzzaman}, \bibinfo{person}{Sheikh~I. Ahamed}, \bibinfo{person}{Richard Love}, \bibinfo{person}{Rumana Dowla}, \bibinfo{person}{David Roe}, \bibinfo{person}{Tahmina Ferdousy}, {and} \bibinfo{person}{Reza Selim}.} \bibinfo{year}{2014}\natexlab{}.
\newblock \showarticletitle{Findings of mobile based palliative care system: towards formulating a generic framework for measuring {QoL}}. In \bibinfo{booktitle}{\emph{Proceedings of the 8th {International} {Conference} on {Pervasive} {Computing} {Technologies} for {Healthcare}}} \emph{(\bibinfo{series}{{PervasiveHealth} '14})}. \bibinfo{publisher}{ICST (Institute for Computer Sciences, Social-Informatics and Telecommunications Engineering)}, \bibinfo{address}{Brussels, BEL}, \bibinfo{pages}{1--8}.
\newblock
\showISBNx{978-1-63190-011-2}
\urldef\tempurl%
\url{https://doi.org/10.4108/icst.pervasivehealth.2014.254960}
\showDOI{\tempurl}


\bibitem[Ho et~al\mbox{.}(2018)]%
        {ho_spiritual_2018}
\bibfield{author}{\bibinfo{person}{Jim~Q. Ho}, \bibinfo{person}{Christopher~D. Nguyen}, \bibinfo{person}{Richard Lopes}, \bibinfo{person}{Stephen~C. Ezeji-Okoye}, {and} \bibinfo{person}{Ware~G. Kuschner}.} \bibinfo{year}{2018}\natexlab{}.
\newblock \showarticletitle{Spiritual {Care} in the {Intensive} {Care} {Unit}: {A} {Narrative} {Review}}.
\newblock \bibinfo{journal}{\emph{Journal of Intensive Care Medicine}} \bibinfo{volume}{33}, \bibinfo{number}{5} (\bibinfo{date}{May} \bibinfo{year}{2018}), \bibinfo{pages}{279--287}.
\newblock
\showISSN{0885-0666}
\urldef\tempurl%
\url{https://doi.org/10.1177/0885066617712677}
\showDOI{\tempurl}
\newblock
\shownote{Publisher: SAGE Publications Inc STM}.


\bibitem[Hodge(2006)]%
        {hodge_template_2006}
\bibfield{author}{\bibinfo{person}{David~R. Hodge}.} \bibinfo{year}{2006}\natexlab{}.
\newblock \showarticletitle{A template for spiritual assessment: a review of the {JCAHO} requirements and guidelines for implementation}.
\newblock \bibinfo{journal}{\emph{Social Work}} \bibinfo{volume}{51}, \bibinfo{number}{4} (\bibinfo{date}{Oct.} \bibinfo{year}{2006}), \bibinfo{pages}{317--326}.
\newblock
\showISSN{0037-8046}
\urldef\tempurl%
\url{https://doi.org/10.1093/sw/51.4.317}
\showDOI{\tempurl}


\bibitem[Howard and Cox(2008)]%
        {howard_collaborative_2008}
\bibfield{author}{\bibinfo{person}{Michael~D. Howard} {and} \bibinfo{person}{Ruth~P. Cox}.} \bibinfo{year}{2008}\natexlab{}.
\newblock \showarticletitle{Collaborative {Intervention}: {A} {Model} for {Coordinated} {Treatment} of {Mental} {Health} {Issues} within a {Ground} {Combat} {Unit}}.
\newblock \bibinfo{journal}{\emph{Military Medicine}} \bibinfo{volume}{173}, \bibinfo{number}{4} (\bibinfo{date}{April} \bibinfo{year}{2008}), \bibinfo{pages}{339--348}.
\newblock
\showISSN{0026-4075}
\urldef\tempurl%
\url{https://doi.org/10.7205/MILMED.173.4.339}
\showDOI{\tempurl}


\bibitem[Jacobs(2008)]%
        {jacobs_what_2008}
\bibfield{author}{\bibinfo{person}{Martha~R. Jacobs}.} \bibinfo{year}{2008}\natexlab{}.
\newblock \showarticletitle{What {Are} {We} {Doing} {Here}?: {Chaplains} in {Contemporary} {Health} {Care}}.
\newblock \bibinfo{journal}{\emph{Hastings Center Report}} \bibinfo{volume}{38}, \bibinfo{number}{6} (\bibinfo{year}{2008}), \bibinfo{pages}{15--18}.
\newblock
\showISSN{1552-146X}
\urldef\tempurl%
\url{https://doi.org/10.1353/hcr.0.0079}
\showDOI{\tempurl}


\bibitem[Jeuland et~al\mbox{.}(2017)]%
        {jeuland_chaplains_2017}
\bibfield{author}{\bibinfo{person}{Jane Jeuland}, \bibinfo{person}{George Fitchett}, \bibinfo{person}{Dena Schulman-Green}, {and} \bibinfo{person}{Jennifer Kapo}.} \bibinfo{year}{2017}\natexlab{}.
\newblock \showarticletitle{Chaplains {Working} in {Palliative} {Care}: {Who} {They} {Are} and {What} {They} {Do}}.
\newblock \bibinfo{journal}{\emph{Journal of Palliative Medicine}} \bibinfo{volume}{20}, \bibinfo{number}{5} (\bibinfo{date}{May} \bibinfo{year}{2017}), \bibinfo{pages}{502--508}.
\newblock
\showISSN{1557-7740}
\urldef\tempurl%
\url{https://doi.org/10.1089/jpm.2016.0308}
\showDOI{\tempurl}


\bibitem[Jhaver et~al\mbox{.}(2019)]%
        {jhaver_human-machine_2019}
\bibfield{author}{\bibinfo{person}{Shagun Jhaver}, \bibinfo{person}{Iris Birman}, \bibinfo{person}{Eric Gilbert}, {and} \bibinfo{person}{Amy Bruckman}.} \bibinfo{year}{2019}\natexlab{}.
\newblock \showarticletitle{Human-{Machine} {Collaboration} for {Content} {Regulation}: {The} {Case} of {Reddit} {Automoderator}}.
\newblock \bibinfo{journal}{\emph{ACM Trans. Comput.-Hum. Interact.}} \bibinfo{volume}{26}, \bibinfo{number}{5} (\bibinfo{date}{July} \bibinfo{year}{2019}), \bibinfo{pages}{31:1--31:35}.
\newblock
\showISSN{1073-0516}
\urldef\tempurl%
\url{https://doi.org/10.1145/3338243}
\showDOI{\tempurl}


\bibitem[Jiang et~al\mbox{.}(2022)]%
        {jiang_many_2022}
\bibfield{author}{\bibinfo{person}{Yueyi Jiang}, \bibinfo{person}{Yunfan Jiang}, \bibinfo{person}{Liu Leqi}, {and} \bibinfo{person}{Piotr Winkielman}.} \bibinfo{year}{2022}\natexlab{}.
\newblock \showarticletitle{Many {Ways} to {Be} {Lonely}: {Fine}-{Grained} {Characterization} of {Loneliness} and {Its} {Potential} {Changes} in {COVID}-19}.
\newblock \bibinfo{journal}{\emph{Proceedings of the International AAAI Conference on Web and Social Media}}  \bibinfo{volume}{16} (\bibinfo{date}{May} \bibinfo{year}{2022}), \bibinfo{pages}{405--416}.
\newblock
\showISSN{2334-0770}
\urldef\tempurl%
\url{https://doi.org/10.1609/icwsm.v16i1.19302}
\showDOI{\tempurl}


\bibitem[Johnson et~al\mbox{.}(2019)]%
        {johnson_psychometric_2019}
\bibfield{author}{\bibinfo{person}{Sverre~Urnes Johnson}, \bibinfo{person}{Pål~Gunnar Ulvenes}, \bibinfo{person}{Tuva Øktedalen}, {and} \bibinfo{person}{Asle Hoffart}.} \bibinfo{year}{2019}\natexlab{}.
\newblock \showarticletitle{Psychometric {Properties} of the {General} {Anxiety} {Disorder} 7-{Item} ({GAD}-7) {Scale} in a {Heterogeneous} {Psychiatric} {Sample}}.
\newblock \bibinfo{journal}{\emph{Frontiers in Psychology}}  \bibinfo{volume}{10} (\bibinfo{date}{Aug.} \bibinfo{year}{2019}).
\newblock
\showISSN{1664-1078}
\urldef\tempurl%
\url{https://doi.org/10.3389/fpsyg.2019.01713}
\showDOI{\tempurl}
\newblock
\shownote{Publisher: Frontiers}.


\bibitem[Kairam et~al\mbox{.}(2022)]%
        {kairam_social-ecological_2022}
\bibfield{author}{\bibinfo{person}{Sanjay~R. Kairam}, \bibinfo{person}{Melissa~C. Mercado}, {and} \bibinfo{person}{Steven~A. Sumner}.} \bibinfo{year}{2022}\natexlab{}.
\newblock \showarticletitle{A {Social}-{Ecological} {Approach} to {Modeling} {Sense} of {Virtual} {Community} ({SOVC}) in {Livestreaming} {Communities}}.
\newblock \bibinfo{journal}{\emph{Proceedings of the ACM on Human-Computer Interaction}} \bibinfo{volume}{6}, \bibinfo{number}{CSCW2} (\bibinfo{date}{Nov.} \bibinfo{year}{2022}), \bibinfo{pages}{356:1--356:35}.
\newblock
\urldef\tempurl%
\url{https://doi.org/10.1145/3555081}
\showDOI{\tempurl}


\bibitem[Kaur et~al\mbox{.}(2021)]%
        {kaur_sway_2021}
\bibfield{author}{\bibinfo{person}{Avleen Kaur}, \bibinfo{person}{C.~Estelle Smith}, {and} \bibinfo{person}{Loren Terveen}.} \bibinfo{year}{2021}\natexlab{}.
\newblock \showarticletitle{Sway {Together}, {Stay} {Together}: {Visualizing} {Spiritual} {Support} {Networks} {Through} the {SoulGarden} {Prototype}}. In \bibinfo{booktitle}{\emph{Companion {Publication} of the 2021 {Conference} on {Computer} {Supported} {Cooperative} {Work} and {Social} {Computing}}}.
\newblock
\urldef\tempurl%
\url{https://dl.acm.org/doi/fullHtml/10.1145/3462204.3481774}
\showURL{%
\tempurl}


\bibitem[Kavuluru et~al\mbox{.}(2016)]%
        {kavuluru_classification_2016}
\bibfield{author}{\bibinfo{person}{Ramakanth Kavuluru}, \bibinfo{person}{María Ramos-Morales}, \bibinfo{person}{Tara Holaday}, \bibinfo{person}{Amanda~G. Williams}, \bibinfo{person}{Laura Haye}, {and} \bibinfo{person}{Julie Cerel}.} \bibinfo{year}{2016}\natexlab{}.
\newblock \showarticletitle{Classification of {Helpful} {Comments} on {Online} {Suicide} {Watch} {Forums}}. In \bibinfo{booktitle}{\emph{Proceedings of the 7th {ACM} {International} {Conference} on {Bioinformatics}, {Computational} {Biology}, and {Health} {Informatics}}} \emph{(\bibinfo{series}{{BCB} '16})}. \bibinfo{publisher}{Association for Computing Machinery}, \bibinfo{address}{New York, NY, USA}, \bibinfo{pages}{32--40}.
\newblock
\showISBNx{978-1-4503-4225-4}
\urldef\tempurl%
\url{https://doi.org/10.1145/2975167.2975170}
\showDOI{\tempurl}


\bibitem[Kawsar et~al\mbox{.}(2012)]%
        {kawsar_e-esas_2012}
\bibfield{author}{\bibinfo{person}{Ferdaus~A. Kawsar}, \bibinfo{person}{Md~Munirul Haque}, \bibinfo{person}{Mohammad Adibuzzaman}, \bibinfo{person}{Sheikh~Iqbal Ahamed}, \bibinfo{person}{Md Uddin}, \bibinfo{person}{Richard Love}, \bibinfo{person}{David Roe}, \bibinfo{person}{Rumana Dowla}, \bibinfo{person}{T. Ferdousy}, \bibinfo{person}{Reza Selim}, {and} \bibinfo{person}{Syed Hossain}.} \bibinfo{year}{2012}\natexlab{}.
\newblock \showarticletitle{e-{ESAS}: improving quality of life for breast cancer patients in developing countries}. In \bibinfo{booktitle}{\emph{Proceedings of the 2nd {ACM} international workshop on {Pervasive} {Wireless} {Healthcare}}} \emph{(\bibinfo{series}{{MobileHealth} '12})}. \bibinfo{publisher}{Association for Computing Machinery}, \bibinfo{address}{New York, NY, USA}, \bibinfo{pages}{9--14}.
\newblock
\showISBNx{978-1-4503-1292-9}
\urldef\tempurl%
\url{https://doi.org/10.1145/2248341.2248346}
\showDOI{\tempurl}


\bibitem[Kelly et~al\mbox{.}(2024)]%
        {kelly_single-arm_2024}
\bibfield{author}{\bibinfo{person}{Elizabeth~Palmer Kelly}, \bibinfo{person}{Maryanna Klatt}, \bibinfo{person}{Jacqueline Caputo}, {and} \bibinfo{person}{Timothy~M. Pawlik}.} \bibinfo{year}{2024}\natexlab{}.
\newblock \showarticletitle{A single-arm pilot of {MyInspiration}: a novel digital resource to support spiritual needs of patients undergoing cancer-directed surgery}.
\newblock \bibinfo{journal}{\emph{Supportive Care in Cancer: Official Journal of the Multinational Association of Supportive Care in Cancer}} \bibinfo{volume}{32}, \bibinfo{number}{5} (\bibinfo{date}{April} \bibinfo{year}{2024}), \bibinfo{pages}{289}.
\newblock
\showISSN{1433-7339}
\urldef\tempurl%
\url{https://doi.org/10.1007/s00520-024-08496-1}
\showDOI{\tempurl}


\bibitem[Kestenbaum et~al\mbox{.}(2022)]%
        {kestenbaum_spiritual_2022}
\bibfield{author}{\bibinfo{person}{Allison Kestenbaum}, \bibinfo{person}{Kelly~A. McEniry}, \bibinfo{person}{Sarah Friedman}, \bibinfo{person}{Jennifer Kent}, \bibinfo{person}{Joseph~D. Ma}, {and} \bibinfo{person}{Eric~J. Roeland}.} \bibinfo{year}{2022}\natexlab{}.
\newblock \showarticletitle{Spiritual {AIM}: assessment and documentation of spiritual needs in patients with cancer}.
\newblock \bibinfo{journal}{\emph{Journal of Health Care Chaplaincy}} \bibinfo{volume}{28}, \bibinfo{number}{4} (\bibinfo{year}{2022}), \bibinfo{pages}{566--577}.
\newblock
\showISSN{1528-6916}
\urldef\tempurl%
\url{https://doi.org/10.1080/08854726.2021.2008170}
\showDOI{\tempurl}


\bibitem[Kitson et~al\mbox{.}(2024)]%
        {kitson_supporting_2024}
\bibfield{author}{\bibinfo{person}{Alexandra Kitson}, \bibinfo{person}{Petr Slovak}, {and} \bibinfo{person}{Alissa~N. Antle}.} \bibinfo{year}{2024}\natexlab{}.
\newblock \showarticletitle{Supporting {Cognitive} {Reappraisal} {With} {Digital} {Technology}: {A} {Content} {Analysis} and {Scoping} {Review} of {Challenges}, {Interventions}, and {Future} {Directions}}. In \bibinfo{booktitle}{\emph{Proceedings of the {CHI} {Conference} on {Human} {Factors} in {Computing} {Systems}}}. \bibinfo{publisher}{ACM}, \bibinfo{address}{Honolulu HI USA}, \bibinfo{pages}{1--17}.
\newblock
\showISBNx{9798400703300}
\urldef\tempurl%
\url{https://doi.org/10.1145/3613904.3642488}
\showDOI{\tempurl}


\bibitem[Klitzman et~al\mbox{.}(2023)]%
        {klitzman_muslim_2023}
\bibfield{author}{\bibinfo{person}{Robert Klitzman}, \bibinfo{person}{Gabrielle Di~Sapia~Natarelli}, \bibinfo{person}{Elizaveta Garbuzova}, \bibinfo{person}{Stephanie Sinnappan}, {and} \bibinfo{person}{Jay Al-Hashimi}.} \bibinfo{year}{2023}\natexlab{}.
\newblock \showarticletitle{Muslim patients in the {U}.{S}. confronting challenges regarding end-of-life and palliative care: the experiences and roles of hospital chaplains}.
\newblock \bibinfo{journal}{\emph{BMC Palliative Care}} \bibinfo{volume}{22}, \bibinfo{number}{1} (\bibinfo{date}{March} \bibinfo{year}{2023}), \bibinfo{pages}{28}.
\newblock
\showISSN{1472-684X}
\urldef\tempurl%
\url{https://doi.org/10.1186/s12904-023-01144-1}
\showDOI{\tempurl}


\bibitem[Kluender et~al\mbox{.}(2021)]%
        {kluender_medical_2021}
\bibfield{author}{\bibinfo{person}{Raymond Kluender}, \bibinfo{person}{Neale Mahoney}, \bibinfo{person}{Francis Wong}, {and} \bibinfo{person}{Wesley Yin}.} \bibinfo{year}{2021}\natexlab{}.
\newblock \showarticletitle{Medical {Debt} in the {US}, 2009-2020}.
\newblock \bibinfo{journal}{\emph{JAMA}} \bibinfo{volume}{326}, \bibinfo{number}{3} (\bibinfo{date}{July} \bibinfo{year}{2021}), \bibinfo{pages}{250--256}.
\newblock
\showISSN{0098-7484}
\urldef\tempurl%
\url{https://doi.org/10.1001/jama.2021.8694}
\showDOI{\tempurl}


\bibitem[Kruzan et~al\mbox{.}(2022)]%
        {kruzan_social_2022}
\bibfield{author}{\bibinfo{person}{Kaylee~Payne Kruzan}, \bibinfo{person}{Kofoworola D.~A. Williams}, \bibinfo{person}{Jonah Meyerhoff}, \bibinfo{person}{Dong~Whi Yoo}, \bibinfo{person}{Linda~C. O'Dwyer}, \bibinfo{person}{Munmun De~Choudhury}, {and} \bibinfo{person}{David~C. Mohr}.} \bibinfo{year}{2022}\natexlab{}.
\newblock \showarticletitle{Social media-based interventions for adolescent and young adult mental health: {A} scoping review}.
\newblock \bibinfo{journal}{\emph{Internet Interventions}}  \bibinfo{volume}{30} (\bibinfo{date}{Dec.} \bibinfo{year}{2022}), \bibinfo{pages}{100578}.
\newblock
\showISSN{2214-7829}
\urldef\tempurl%
\url{https://doi.org/10.1016/j.invent.2022.100578}
\showDOI{\tempurl}


\bibitem[Kuzheleva-Sagan and Suchkova(2016)]%
        {kuzheleva-sagan_designing_2016}
\bibfield{author}{\bibinfo{person}{Irina~P. Kuzheleva-Sagan} {and} \bibinfo{person}{Natalya~A. Suchkova}.} \bibinfo{year}{2016}\natexlab{}.
\newblock \showarticletitle{Designing trust in the {Internet} services}.
\newblock \bibinfo{journal}{\emph{AI \& SOCIETY}} \bibinfo{volume}{31}, \bibinfo{number}{3} (\bibinfo{date}{Aug.} \bibinfo{year}{2016}), \bibinfo{pages}{381--392}.
\newblock
\showISSN{1435-5655}
\urldef\tempurl%
\url{https://doi.org/10.1007/s00146-015-0622-7}
\showDOI{\tempurl}


\bibitem[Kwon et~al\mbox{.}(2024)]%
        {kwon_spiritual_2024}
\bibfield{author}{\bibinfo{person}{Soonho Kwon}, \bibinfo{person}{Dong~Whi Yoo}, {and} \bibinfo{person}{Younah Kang}.} \bibinfo{year}{2024}\natexlab{}.
\newblock \showarticletitle{Spiritual {AI}: {Exploring} the {Possibilities} of a {Human}-{AI} {Interaction} {Beyond} {Productive} {Goals}}. In \bibinfo{booktitle}{\emph{Extended {Abstracts} of the {CHI} {Conference} on {Human} {Factors} in {Computing} {Systems}}}. \bibinfo{publisher}{ACM}, \bibinfo{address}{Honolulu HI USA}, \bibinfo{pages}{1--8}.
\newblock
\showISBNx{9798400703317}
\urldef\tempurl%
\url{https://doi.org/10.1145/3613905.3650743}
\showDOI{\tempurl}


\bibitem[Labuschagne et~al\mbox{.}(2024)]%
        {labuschagne_development_2024}
\bibfield{author}{\bibinfo{person}{Dirk Labuschagne}, \bibinfo{person}{Patricia Palmer}, \bibinfo{person}{Allison DeLaney}, \bibinfo{person}{Kristen Schenk}, {and} \bibinfo{person}{George Fitchett}.} \bibinfo{year}{2024}\natexlab{}.
\newblock \showarticletitle{Development of the {ONC}-5: {A} quantifiable assessment of spiritual concerns for adult oncology patients}.
\newblock \bibinfo{journal}{\emph{Journal of Health Care Chaplaincy}} \bibinfo{volume}{0}, \bibinfo{number}{0} (\bibinfo{year}{2024}), \bibinfo{pages}{1--16}.
\newblock
\showISSN{0885-4726}
\urldef\tempurl%
\url{https://doi.org/10.1080/08854726.2024.2399992}
\showDOI{\tempurl}
\newblock
\shownote{Publisher: Routledge \_eprint: https://doi.org/10.1080/08854726.2024.2399992}.


\bibitem[Langarizadeh et~al\mbox{.}(2017)]%
        {langarizadeh_telemental_2017}
\bibfield{author}{\bibinfo{person}{Mostafa Langarizadeh}, \bibinfo{person}{Mohsen~S. Tabatabaei}, \bibinfo{person}{Kamran Tavakol}, \bibinfo{person}{Majid Naghipour}, \bibinfo{person}{Alireza Rostami}, {and} \bibinfo{person}{Fatemeh Moghbeli}.} \bibinfo{year}{2017}\natexlab{}.
\newblock \showarticletitle{Telemental {Health} {Care}, an {Effective} {Alternative} to {Conventional} {Mental} {Care}: a {Systematic} {Review}}.
\newblock \bibinfo{journal}{\emph{Acta Informatica Medica}} \bibinfo{volume}{25}, \bibinfo{number}{4} (\bibinfo{date}{Dec.} \bibinfo{year}{2017}), \bibinfo{pages}{240--246}.
\newblock
\showISSN{0353-8109}
\urldef\tempurl%
\url{https://doi.org/10.5455/aim.2017.25.240-246}
\showDOI{\tempurl}


\bibitem[Laranjeira et~al\mbox{.}(2023)]%
        {laranjeira_perceived_2023}
\bibfield{author}{\bibinfo{person}{Carlos Laranjeira}, \bibinfo{person}{Maria~Anjos Dixe}, {and} \bibinfo{person}{Ana Querido}.} \bibinfo{year}{2023}\natexlab{}.
\newblock \showarticletitle{Perceived {Barriers} to {Providing} {Spiritual} {Care} in {Palliative} {Care} among {Professionals}: {A} {Portuguese} {Cross}-{Sectional} {Study}}.
\newblock \bibinfo{journal}{\emph{International Journal of Environmental Research and Public Health}} \bibinfo{volume}{20}, \bibinfo{number}{12} (\bibinfo{date}{June} \bibinfo{year}{2023}), \bibinfo{pages}{6121}.
\newblock
\showISSN{1661-7827}
\urldef\tempurl%
\url{https://doi.org/10.3390/ijerph20126121}
\showDOI{\tempurl}


\bibitem[Lee and Hawkins(2010)]%
        {lee_why_2010}
\bibfield{author}{\bibinfo{person}{Sun~Young Lee} {and} \bibinfo{person}{Robert Hawkins}.} \bibinfo{year}{2010}\natexlab{}.
\newblock \showarticletitle{Why {Do} {Patients} {Seek} an {Alternative} {Channel}? {The} {Effects} of {Unmet} {Needs} on {Patients}' {Health}-{Related} {Internet} {Use}}.
\newblock \bibinfo{journal}{\emph{Journal of Health Communication}} \bibinfo{volume}{15}, \bibinfo{number}{2} (\bibinfo{date}{March} \bibinfo{year}{2010}), \bibinfo{pages}{152--166}.
\newblock
\showISSN{1081-0730}
\urldef\tempurl%
\url{https://doi.org/10.1080/10810730903528033}
\showDOI{\tempurl}
\newblock
\shownote{Publisher: Taylor \& Francis \_eprint: https://doi.org/10.1080/10810730903528033}.


\bibitem[Li et~al\mbox{.}(2022)]%
        {li_measuring_2022}
\bibfield{author}{\bibinfo{person}{Hanlin Li}, \bibinfo{person}{Brent Hecht}, {and} \bibinfo{person}{Stevie Chancellor}.} \bibinfo{year}{2022}\natexlab{}.
\newblock \showarticletitle{Measuring the {Monetary} {Value} of {Online} {Volunteer} {Work}}.
\newblock \bibinfo{journal}{\emph{Proceedings of the International AAAI Conference on Web and Social Media}}  \bibinfo{volume}{16} (\bibinfo{date}{May} \bibinfo{year}{2022}), \bibinfo{pages}{596--606}.
\newblock
\showISSN{2334-0770}
\urldef\tempurl%
\url{https://doi.org/10.1609/icwsm.v16i1.19318}
\showDOI{\tempurl}


\bibitem[Liu et~al\mbox{.}(2022)]%
        {liu_time--event_2022}
\bibfield{author}{\bibinfo{person}{Xueying Liu}, \bibinfo{person}{Shiaofen Fang}, \bibinfo{person}{George Mohler}, \bibinfo{person}{Joan Carlson}, {and} \bibinfo{person}{Yunyu Xiao}.} \bibinfo{year}{2022}\natexlab{}.
\newblock \showarticletitle{Time-to-event modeling of subreddits transitions to r/{SuicideWatch}}. In \bibinfo{booktitle}{\emph{2022 21st {IEEE} {International} {Conference} on {Machine} {Learning} and {Applications} ({ICMLA})}}. \bibinfo{pages}{974--979}.
\newblock
\urldef\tempurl%
\url{https://doi.org/10.1109/ICMLA55696.2022.00163}
\showDOI{\tempurl}


\bibitem[Long et~al\mbox{.}(2024)]%
        {long_spirituality_2024}
\bibfield{author}{\bibinfo{person}{Katelyn N.~G. Long}, \bibinfo{person}{Xavier Symons}, \bibinfo{person}{Tyler~J. VanderWeele}, \bibinfo{person}{Tracy~A. Balboni}, \bibinfo{person}{David~H. Rosmarin}, \bibinfo{person}{Christina Puchalski}, \bibinfo{person}{Teresa Cutts}, \bibinfo{person}{Gary~R. Gunderson}, \bibinfo{person}{Ellen Idler}, \bibinfo{person}{Doug Oman}, \bibinfo{person}{Michael~J. Balboni}, \bibinfo{person}{Laura~S. Tuach}, {and} \bibinfo{person}{Howard~K. Koh}.} \bibinfo{year}{2024}\natexlab{}.
\newblock \showarticletitle{Spirituality {As} {A} {Determinant} {Of} {Health}: {Emerging} {Policies}, {Practices}, {And} {Systems}}.
\newblock \bibinfo{journal}{\emph{Health Affairs}} \bibinfo{volume}{43}, \bibinfo{number}{6} (\bibinfo{date}{June} \bibinfo{year}{2024}), \bibinfo{pages}{783--790}.
\newblock
\showISSN{0278-2715}
\urldef\tempurl%
\url{https://doi.org/10.1377/hlthaff.2023.01643}
\showDOI{\tempurl}
\newblock
\shownote{Publisher: Health Affairs}.


\bibitem[Low et~al\mbox{.}(2020)]%
        {low_natural_2020}
\bibfield{author}{\bibinfo{person}{Daniel~M. Low}, \bibinfo{person}{Laurie Rumker}, \bibinfo{person}{Tanya Talkar}, \bibinfo{person}{John Torous}, \bibinfo{person}{Guillermo Cecchi}, {and} \bibinfo{person}{Satrajit~S. Ghosh}.} \bibinfo{year}{2020}\natexlab{}.
\newblock \showarticletitle{Natural {Language} {Processing} {Reveals} {Vulnerable} {Mental} {Health} {Support} {Groups} and {Heightened} {Health} {Anxiety} on {Reddit} {During} {COVID}-19: {Observational} {Study}}.
\newblock \bibinfo{journal}{\emph{Journal of Medical Internet Research}} \bibinfo{volume}{22}, \bibinfo{number}{10} (\bibinfo{date}{Oct.} \bibinfo{year}{2020}), \bibinfo{pages}{e22635}.
\newblock
\urldef\tempurl%
\url{https://doi.org/10.2196/22635}
\showDOI{\tempurl}
\newblock
\shownote{Company: Journal of Medical Internet Research Distributor: Journal of Medical Internet Research Institution: Journal of Medical Internet Research Label: Journal of Medical Internet Research Publisher: JMIR Publications Inc., Toronto, Canada}.


\bibitem[Lowry et~al\mbox{.}(2014)]%
        {lowry_picture_2014}
\bibfield{author}{\bibinfo{person}{Paul~Benjamin Lowry}, \bibinfo{person}{David~W. Wilson}, {and} \bibinfo{person}{William~L. Haig}.} \bibinfo{year}{2014}\natexlab{}.
\newblock \showarticletitle{A {Picture} is {Worth} a {Thousand} {Words}: {Source} {Credibility} {Theory} {Applied} to {Logo} and {Website} {Design} for {Heightened} {Credibility} and {Consumer} {Trust}}.
\newblock \bibinfo{journal}{\emph{International Journal of Human–Computer Interaction}} \bibinfo{volume}{30}, \bibinfo{number}{1} (\bibinfo{date}{Jan.} \bibinfo{year}{2014}), \bibinfo{pages}{63--93}.
\newblock
\showISSN{1044-7318}
\urldef\tempurl%
\url{https://doi.org/10.1080/10447318.2013.839899}
\showDOI{\tempurl}
\newblock
\shownote{Publisher: Taylor \& Francis \_eprint: https://doi.org/10.1080/10447318.2013.839899}.


\bibitem[Ma et~al\mbox{.}(2017)]%
        {ma_write_2017}
\bibfield{author}{\bibinfo{person}{Haiwei Ma}, \bibinfo{person}{C.~Estelle Smith}, \bibinfo{person}{Lu He}, \bibinfo{person}{Saumik Narayanan}, \bibinfo{person}{Robert~A. Giaquinto}, \bibinfo{person}{Roni Evans}, \bibinfo{person}{Linda Hanson}, {and} \bibinfo{person}{Svetlana Yarosh}.} \bibinfo{year}{2017}\natexlab{}.
\newblock \showarticletitle{Write for {Life}: {Persisting} in {Online} {Health} {Communities} {Through} {Expressive} {Writing} and {Social} {Support}}.
\newblock \bibinfo{journal}{\emph{Proc. ACM Hum.-Comput. Interact.}} \bibinfo{volume}{1}, \bibinfo{number}{CSCW} (\bibinfo{date}{Dec.} \bibinfo{year}{2017}), \bibinfo{pages}{73:1--73:24}.
\newblock
\showISSN{2573-0142}
\urldef\tempurl%
\url{https://doi.org/10.1145/3134708}
\showDOI{\tempurl}


\bibitem[MacCaull et~al\mbox{.}(2010)]%
        {maccaull_using_2010}
\bibfield{author}{\bibinfo{person}{Wendy MacCaull}, \bibinfo{person}{Heather Jewers}, {and} \bibinfo{person}{Markus Latzel}.} \bibinfo{year}{2010}\natexlab{}.
\newblock \showarticletitle{Using an interdisciplinary approach to develop a knowledge-driven careflow management system for collaborative patient-centred palliative care}. In \bibinfo{booktitle}{\emph{Proceedings of the 1st {ACM} {International} {Health} {Informatics} {Symposium}}} \emph{(\bibinfo{series}{{IHI} '10})}. \bibinfo{publisher}{Association for Computing Machinery}, \bibinfo{address}{New York, NY, USA}, \bibinfo{pages}{507--511}.
\newblock
\showISBNx{978-1-4503-0030-8}
\urldef\tempurl%
\url{https://doi.org/10.1145/1882992.1883073}
\showDOI{\tempurl}


\bibitem[Markum et~al\mbox{.}(2024)]%
        {markum_mediating_2024}
\bibfield{author}{\bibinfo{person}{Robert~B. Markum}, \bibinfo{person}{Sara Wolf}, \bibinfo{person}{Caroline Claisse}, {and} \bibinfo{person}{Michael Hoefer}.} \bibinfo{year}{2024}\natexlab{}.
\newblock \showarticletitle{Mediating the {Sacred}: {Configuring} a {Design} {Space} for {Religious} and {Spiritual} {Tangible} {Interactive} {Artifacts}}. In \bibinfo{booktitle}{\emph{Proceedings of the {Eighteenth} {International} {Conference} on {Tangible}, {Embedded}, and {Embodied} {Interaction}}} \emph{(\bibinfo{series}{{TEI} '24})}. \bibinfo{publisher}{Association for Computing Machinery}, \bibinfo{address}{New York, NY, USA}, \bibinfo{pages}{1--22}.
\newblock
\showISBNx{9798400704024}
\urldef\tempurl%
\url{https://doi.org/10.1145/3623509.3633353}
\showDOI{\tempurl}


\bibitem[Massey et~al\mbox{.}(2015)]%
        {massey_what_2015}
\bibfield{author}{\bibinfo{person}{Kevin Massey}, \bibinfo{person}{Marilyn~JD Barnes}, \bibinfo{person}{Dana Villines}, \bibinfo{person}{Julie~D Goldstein}, \bibinfo{person}{Anna Lee~Hisey Pierson}, \bibinfo{person}{Cheryl Scherer}, \bibinfo{person}{Betty~Vander Laan}, {and} \bibinfo{person}{Wm~Thomas Summerfelt}.} \bibinfo{year}{2015}\natexlab{}.
\newblock \showarticletitle{What do {I} do? {Developing} a taxonomy of chaplaincy activities and interventions for spiritual care in intensive care unit palliative care}.
\newblock \bibinfo{journal}{\emph{BMC Palliative Care}}  \bibinfo{volume}{14} (\bibinfo{date}{April} \bibinfo{year}{2015}), \bibinfo{pages}{10}.
\newblock
\showISSN{1472-684X}
\urldef\tempurl%
\url{https://doi.org/10.1186/s12904-015-0008-0}
\showDOI{\tempurl}


\bibitem[Massey and Summerfelt(2012)]%
        {massey_growing_2012}
\bibfield{author}{\bibinfo{person}{Kevin Massey} {and} \bibinfo{person}{Wm~Thomas Summerfelt}.} \bibinfo{year}{2012}\natexlab{}.
\newblock \bibinfo{title}{Growing the {Field} of {Chaplaincy} in {Palliative} {Care}: “{What} {Do} {I} {Do}? {Developing} a {Taxonomy} of {Chaplaincy} {Activities} and {Interventions} for {Spiritual} {Care} in {ICU} {Palliative} {Care}.”}.
\newblock
\newblock


\bibitem[McKee(1988)]%
        {mckee_holistic_1988}
\bibfield{author}{\bibinfo{person}{Janet McKee}.} \bibinfo{year}{1988}\natexlab{}.
\newblock \showarticletitle{Holistic health and the critique of western medicine}.
\newblock \bibinfo{journal}{\emph{Social Science \& Medicine}} \bibinfo{volume}{26}, \bibinfo{number}{8} (\bibinfo{date}{Jan.} \bibinfo{year}{1988}), \bibinfo{pages}{775--784}.
\newblock
\showISSN{0277-9536}
\urldef\tempurl%
\url{https://doi.org/10.1016/0277-9536(88)90171-2}
\showDOI{\tempurl}


\bibitem[McLellan et~al\mbox{.}(2022)]%
        {mclellan_talking_2022}
\bibfield{author}{\bibinfo{person}{Ailsa McLellan}, \bibinfo{person}{Keana Schmidt-Waselenchuk}, \bibinfo{person}{Kari Duerksen}, {and} \bibinfo{person}{Erica Woodin}.} \bibinfo{year}{2022}\natexlab{}.
\newblock \showarticletitle{Talking back to mental health stigma: {An} exploration of {YouTube} comments on anti-stigma videos}.
\newblock \bibinfo{journal}{\emph{Computers in Human Behavior}}  \bibinfo{volume}{131} (\bibinfo{date}{June} \bibinfo{year}{2022}), \bibinfo{pages}{107214}.
\newblock
\showISSN{0747-5632}
\urldef\tempurl%
\url{https://doi.org/10.1016/j.chb.2022.107214}
\showDOI{\tempurl}


\bibitem[Morgan et~al\mbox{.}(2016)]%
        {morgan_help-seeking_2016}
\bibfield{author}{\bibinfo{person}{Jessica~Kelley Morgan}, \bibinfo{person}{Laurel Hourani}, \bibinfo{person}{Marian~E. Lane}, {and} \bibinfo{person}{Stephen Tueller}.} \bibinfo{year}{2016}\natexlab{}.
\newblock \showarticletitle{Help-{Seeking} {Behaviors} {Among} {Active}-{Duty} {Military} {Personnel}: {Utilization} of {Chaplains} and {Other} {Mental} {Health} {Service} {Providers}}.
\newblock \bibinfo{journal}{\emph{Journal of Health Care Chaplaincy}} \bibinfo{volume}{22}, \bibinfo{number}{3} (\bibinfo{date}{July} \bibinfo{year}{2016}), \bibinfo{pages}{102--117}.
\newblock
\showISSN{0885-4726}
\urldef\tempurl%
\url{https://doi.org/10.1080/08854726.2016.1171598}
\showDOI{\tempurl}
\newblock
\shownote{Publisher: Routledge \_eprint: https://www.tandfonline.com/doi/pdf/10.1080/08854726.2016.1171598}.


\bibitem[Morris and Picard(2014)]%
        {morris_crowd-powered_2014}
\bibfield{author}{\bibinfo{person}{Robert~R. Morris} {and} \bibinfo{person}{Rosalind Picard}.} \bibinfo{year}{2014}\natexlab{}.
\newblock \showarticletitle{Crowd-powered positive psychological interventions}.
\newblock \bibinfo{journal}{\emph{The Journal of Positive Psychology}} \bibinfo{volume}{9}, \bibinfo{number}{6} (\bibinfo{date}{Nov.} \bibinfo{year}{2014}), \bibinfo{pages}{509--516}.
\newblock
\showISSN{1743-9760, 1743-9779}
\urldef\tempurl%
\url{https://doi.org/10.1080/17439760.2014.913671}
\showDOI{\tempurl}


\bibitem[Morris et~al\mbox{.}(2015)]%
        {morris_efficacy_2015}
\bibfield{author}{\bibinfo{person}{Robert~R. Morris}, \bibinfo{person}{Stephen~M. Schueller}, {and} \bibinfo{person}{Rosalind~W. Picard}.} \bibinfo{year}{2015}\natexlab{}.
\newblock \showarticletitle{Efficacy of a {Web}-{Based}, {Crowdsourced} {Peer}-{To}-{Peer} {Cognitive} {Reappraisal} {Platform} for {Depression}: {Randomized} {Controlled} {Trial}}.
\newblock \bibinfo{journal}{\emph{Journal of Medical Internet Research}} \bibinfo{volume}{17}, \bibinfo{number}{3} (\bibinfo{year}{2015}), \bibinfo{pages}{e72}.
\newblock
\urldef\tempurl%
\url{https://doi.org/10.2196/jmir.4167}
\showDOI{\tempurl}
\newblock
\shownote{Company: Journal of Medical Internet Research Distributor: Journal of Medical Internet Research Institution: Journal of Medical Internet Research Label: Journal of Medical Internet Research Publisher: JMIR Publications Inc., Toronto, Canada}.


\bibitem[Mount(1993)]%
        {mount_whole_1993}
\bibfield{author}{\bibinfo{person}{Balfour Mount}.} \bibinfo{year}{1993}\natexlab{}.
\newblock \showarticletitle{Whole person care: {Beyond} psychosocial and physical needs}.
\newblock \bibinfo{journal}{\emph{American Journal of Hospice and Palliative Medicine®}} \bibinfo{volume}{10}, \bibinfo{number}{1} (\bibinfo{date}{Jan.} \bibinfo{year}{1993}), \bibinfo{pages}{28--37}.
\newblock
\showISSN{1049-9091}
\urldef\tempurl%
\url{https://doi.org/10.1177/104990919301000109}
\showDOI{\tempurl}
\newblock
\shownote{Publisher: SAGE Publications Inc}.


\bibitem[Muller(2014)]%
        {muller_curiosity_2014}
\bibfield{author}{\bibinfo{person}{Michael Muller}.} \bibinfo{year}{2014}\natexlab{}.
\newblock \showarticletitle{Curiosity, {Creativity}, and {Surprise} as {Analytic} {Tools}: {Grounded} {Theory} {Method}}.
\newblock In \bibinfo{booktitle}{\emph{Ways of {Knowing} in {HCI}}}, \bibfield{editor}{\bibinfo{person}{Judith~S. Olson} {and} \bibinfo{person}{Wendy~A. Kellogg}} (Eds.). \bibinfo{publisher}{Springer}, \bibinfo{address}{New York, NY}, \bibinfo{pages}{25--48}.
\newblock
\showISBNx{978-1-4939-0378-8}
\urldef\tempurl%
\url{https://doi.org/10.1007/978-1-4939-0378-8_2}
\showDOI{\tempurl}


\bibitem[Nambisan(2011)]%
        {nambisan_information_2011}
\bibfield{author}{\bibinfo{person}{Priya Nambisan}.} \bibinfo{year}{2011}\natexlab{}.
\newblock \showarticletitle{Information seeking and social support in online health communities: impact on patients' perceived empathy}.
\newblock \bibinfo{journal}{\emph{Journal of the American Medical Informatics Association}} \bibinfo{volume}{18}, \bibinfo{number}{3} (\bibinfo{date}{May} \bibinfo{year}{2011}), \bibinfo{pages}{298--304}.
\newblock
\showISSN{1067-5027}
\urldef\tempurl%
\url{https://doi.org/10.1136/amiajnl-2010-000058}
\showDOI{\tempurl}


\bibitem[Nash et~al\mbox{.}(2019)]%
        {nash_adapting_2019}
\bibfield{author}{\bibinfo{person}{Paul Nash}, \bibinfo{person}{Emma Roberts}, \bibinfo{person}{Sally Nash}, \bibinfo{person}{Kathryn Darby}, {and} \bibinfo{person}{Aftab~Ahmed Parwaz}.} \bibinfo{year}{2019}\natexlab{}.
\newblock \showarticletitle{Adapting the {Advocate} {Health} {Care} {Taxonomy} of {Chaplaincy} for a {Pediatric} {Hospital} {Context}: {A} {Pilot} {Study}}.
\newblock \bibinfo{journal}{\emph{Journal of Health Care Chaplaincy}} \bibinfo{volume}{25}, \bibinfo{number}{2} (\bibinfo{year}{2019}), \bibinfo{pages}{61--75}.
\newblock
\showISSN{1528-6916}
\urldef\tempurl%
\url{https://doi.org/10.1080/08854726.2018.1473911}
\showDOI{\tempurl}


\bibitem[Naslund et~al\mbox{.}(2020)]%
        {naslund_social_2020}
\bibfield{author}{\bibinfo{person}{John~A. Naslund}, \bibinfo{person}{Ameya Bondre}, \bibinfo{person}{John Torous}, {and} \bibinfo{person}{Kelly~A. Aschbrenner}.} \bibinfo{year}{2020}\natexlab{}.
\newblock \showarticletitle{Social {Media} and {Mental} {Health}: {Benefits}, {Risks}, and {Opportunities} for {Research} and {Practice}}.
\newblock \bibinfo{journal}{\emph{Journal of Technology in Behavioral Science}} \bibinfo{volume}{5}, \bibinfo{number}{3} (\bibinfo{date}{Sept.} \bibinfo{year}{2020}), \bibinfo{pages}{245--257}.
\newblock
\showISSN{2366-5963}
\urldef\tempurl%
\url{https://doi.org/10.1007/s41347-020-00134-x}
\showDOI{\tempurl}


\bibitem[Nieuwsma et~al\mbox{.}(2014)]%
        {nieuwsma_collaborating_2014}
\bibfield{author}{\bibinfo{person}{Jason~A. Nieuwsma}, \bibinfo{person}{George~L. Jackson}, \bibinfo{person}{Mark~B. DeKraai}, \bibinfo{person}{Denise~J. Bulling}, \bibinfo{person}{William~C. Cantrell}, \bibinfo{person}{Jeffrey~E. Rhodes}, \bibinfo{person}{Mark~J. Bates}, \bibinfo{person}{Keith Ethridge}, \bibinfo{person}{Marian~E. Lane}, \bibinfo{person}{Wendy~N. Tenhula}, \bibinfo{person}{Sonja~V. Batten}, {and} \bibinfo{person}{Keith~G. Meador}.} \bibinfo{year}{2014}\natexlab{}.
\newblock \showarticletitle{Collaborating {Across} the {Departments} of {Veterans} {Affairs} and {Defense} to {Integrate} {Mental} {Health} and {Chaplaincy} {Services}}.
\newblock \bibinfo{journal}{\emph{Journal of General Internal Medicine}} \bibinfo{volume}{29}, \bibinfo{number}{4} (\bibinfo{date}{Dec.} \bibinfo{year}{2014}), \bibinfo{pages}{885--894}.
\newblock
\showISSN{1525-1497}
\urldef\tempurl%
\url{https://doi.org/10.1007/s11606-014-3032-5}
\showDOI{\tempurl}


\bibitem[Nolan(2011)]%
        {nolan_hope_2011}
\bibfield{author}{\bibinfo{person}{Steve Nolan}.} \bibinfo{year}{2011}\natexlab{}.
\newblock \showarticletitle{Hope beyond (redundant) hope: how chaplains work with dying patients}.
\newblock \bibinfo{journal}{\emph{Palliative Medicine}} \bibinfo{volume}{25}, \bibinfo{number}{1} (\bibinfo{date}{Jan.} \bibinfo{year}{2011}), \bibinfo{pages}{21--25}.
\newblock
\showISSN{0269-2163, 1477-030X}
\urldef\tempurl%
\url{https://doi.org/10.1177/0269216310380297}
\showDOI{\tempurl}


\bibitem[Norman(2002)]%
        {norman_personal_2002}
\bibfield{author}{\bibinfo{person}{James~E. Norman}.} \bibinfo{year}{2002}\natexlab{}.
\newblock \showarticletitle{Personal difficulties and educational underachievement: {The} contribution of pastoral care and school {Chaplaincy}}.
\newblock \bibinfo{journal}{\emph{Irish Educational Studies}} (\bibinfo{date}{March} \bibinfo{year}{2002}).
\newblock
\urldef\tempurl%
\url{https://doi.org/10.1080/0332331020210106}
\showDOI{\tempurl}
\newblock
\shownote{Publisher: Taylor \& Francis Group}.


\bibitem[Nourriz et~al\mbox{.}(2024)]%
        {nourriz_design_2024}
\bibfield{author}{\bibinfo{person}{Shadi Nourriz}, \bibinfo{person}{Alemitu Bezabih}, {and} \bibinfo{person}{C.~Estelle Smith}.} \bibinfo{year}{2024}\natexlab{}.
\newblock \showarticletitle{On the {Design} {Risks} of {Empathy} {Fatigue}}. In \bibinfo{booktitle}{\emph{Proceedings of the 3rd {Empathy}-{Centric} {Design} {Workshop}: {Scrutinizing} {Empathy} {Beyond} the {Individual}}} \emph{(\bibinfo{series}{{EmpathiCH} '24})}. \bibinfo{publisher}{Association for Computing Machinery}, \bibinfo{address}{Honolulu HI USA}, \bibinfo{pages}{34--39}.
\newblock
\showISBNx{9798400717888}
\urldef\tempurl%
\url{https://doi.org/10.1145/3661790.3661798}
\showDOI{\tempurl}


\bibitem[Nurmi et~al\mbox{.}(2020)]%
        {nurmi_engaging_2020}
\bibfield{author}{\bibinfo{person}{Johanna Nurmi}, \bibinfo{person}{Keegan Knittle}, \bibinfo{person}{Todor Ginchev}, \bibinfo{person}{Fida Khattak}, \bibinfo{person}{Christopher Helf}, \bibinfo{person}{Patrick Zwickl}, \bibinfo{person}{Carmina Castellano-Tejedor}, \bibinfo{person}{Pilar Lusilla-Palacios}, \bibinfo{person}{Jose Costa-Requena}, \bibinfo{person}{Niklas Ravaja}, {and} \bibinfo{person}{Ari Haukkala}.} \bibinfo{year}{2020}\natexlab{}.
\newblock \showarticletitle{Engaging {Users} in the {Behavior} {Change} {Process} {With} {Digitalized} {Motivational} {Interviewing} and {Gamification}: {Development} and {Feasibility} {Testing} of the {Precious} {App}}.
\newblock \bibinfo{journal}{\emph{JMIR mHealth and uHealth}} \bibinfo{volume}{8}, \bibinfo{number}{1} (\bibinfo{date}{Jan.} \bibinfo{year}{2020}), \bibinfo{pages}{e12884}.
\newblock
\urldef\tempurl%
\url{https://doi.org/10.2196/12884}
\showDOI{\tempurl}
\newblock
\shownote{Company: JMIR mHealth and uHealth Distributor: JMIR mHealth and uHealth Institution: JMIR mHealth and uHealth Label: JMIR mHealth and uHealth Publisher: JMIR Publications Inc., Toronto, Canada}.


\bibitem[O'Leary et~al\mbox{.}(2024)]%
        {oleary_something_2024}
\bibfield{author}{\bibinfo{person}{Teresa~K. O'Leary}, \bibinfo{person}{Michael~K Paasche-Orlow}, {and} \bibinfo{person}{Timothy Bickmore}.} \bibinfo{year}{2024}\natexlab{}.
\newblock \showarticletitle{`{Something} {I} {Can} {Lean} {On}': {A} {Qualitative} {Evaluation} of a {Virtual} {Palliative} {Care} {Counselor} for {Patients} with {Life}-{Limiting} {Illnesses}}. In \bibinfo{booktitle}{\emph{Extended {Abstracts} of the 2024 {CHI} {Conference} on {Human} {Factors} in {Computing} {Systems}}} \emph{(\bibinfo{series}{{CHI} {EA} '24})}. \bibinfo{publisher}{Association for Computing Machinery}, \bibinfo{address}{New York, NY, USA}, \bibinfo{pages}{1--7}.
\newblock
\showISBNx{9798400703317}
\urldef\tempurl%
\url{https://doi.org/10.1145/3613905.3651106}
\showDOI{\tempurl}


\bibitem[Olsman(2022)]%
        {olsman_witnesses_2022}
\bibfield{author}{\bibinfo{person}{Erik Olsman}.} \bibinfo{year}{2022}\natexlab{}.
\newblock \showarticletitle{Witnesses of hope in times of despair: chaplains in palliative care. {A} qualitative study}.
\newblock \bibinfo{journal}{\emph{Journal of Health Care Chaplaincy}} \bibinfo{volume}{28}, \bibinfo{number}{1} (\bibinfo{date}{Jan.} \bibinfo{year}{2022}), \bibinfo{pages}{29--40}.
\newblock
\showISSN{0885-4726, 1528-6916}
\urldef\tempurl%
\url{https://doi.org/10.1080/08854726.2020.1727602}
\showDOI{\tempurl}


\bibitem[Otis(2009)]%
        {otis_overview_2009}
\bibfield{author}{\bibinfo{person}{Pauletta Otis}.} \bibinfo{year}{2009}\natexlab{}.
\newblock \showarticletitle{An {Overview} of the {U}.s. {Military} {Chaplaincy}: {A} {Ministry} of {Presence} and {Practice}}.
\newblock \bibinfo{journal}{\emph{The Review of Faith \& International Affairs}} \bibinfo{volume}{7}, \bibinfo{number}{4} (\bibinfo{date}{Dec.} \bibinfo{year}{2009}), \bibinfo{pages}{3--15}.
\newblock
\showISSN{1557-0274}
\urldef\tempurl%
\url{https://doi.org/10.1080/15570274.2009.9523410}
\showDOI{\tempurl}
\newblock
\shownote{Publisher: Routledge \_eprint: https://doi.org/10.1080/15570274.2009.9523410}.


\bibitem[Paal et~al\mbox{.}(2015)]%
        {paal_spiritual_2015}
\bibfield{author}{\bibinfo{person}{Piret Paal}, \bibinfo{person}{Yousef Helo}, {and} \bibinfo{person}{Eckhard Frick}.} \bibinfo{year}{2015}\natexlab{}.
\newblock \showarticletitle{Spiritual {Care} {Training} {Provided} to {Healthcare} {Professionals}: {A} {Systematic} {Review}}.
\newblock \bibinfo{journal}{\emph{Journal of Pastoral Care \& Counseling}} \bibinfo{volume}{69}, \bibinfo{number}{1} (\bibinfo{date}{March} \bibinfo{year}{2015}), \bibinfo{pages}{19--30}.
\newblock
\showISSN{1542-3050}
\urldef\tempurl%
\url{https://doi.org/10.1177/1542305015572955}
\showDOI{\tempurl}
\newblock
\shownote{Publisher: SAGE Publications Ltd STM}.


\bibitem[Pesut et~al\mbox{.}(2016)]%
        {pesut_health_2016}
\bibfield{author}{\bibinfo{person}{Barbara Pesut}, \bibinfo{person}{Shane Sinclair}, \bibinfo{person}{George Fitchett}, \bibinfo{person}{Madeleine Greig}, {and} \bibinfo{person}{Sarah~E. Koss}.} \bibinfo{year}{2016}\natexlab{}.
\newblock \showarticletitle{Health {Care} {Chaplaincy}: {A} {Scoping} {Review} of the {Evidence} 2009–2014}.
\newblock \bibinfo{journal}{\emph{Journal of Health Care Chaplaincy}} \bibinfo{volume}{22}, \bibinfo{number}{2} (\bibinfo{date}{April} \bibinfo{year}{2016}), \bibinfo{pages}{67--84}.
\newblock
\showISSN{0885-4726}
\urldef\tempurl%
\url{https://doi.org/10.1080/08854726.2015.1133185}
\showDOI{\tempurl}
\newblock
\shownote{Publisher: Routledge \_eprint: https://www.tandfonline.com/doi/pdf/10.1080/08854726.2015.1133185}.


\bibitem[Pinho et~al\mbox{.}(2024)]%
        {pinho_editorial_2024}
\bibfield{author}{\bibinfo{person}{Lara Guedes~de Pinho}, \bibinfo{person}{César Fonseca}, \bibinfo{person}{Łukasz Gawęda}, \bibinfo{person}{Manuel Lopes}, {and} \bibinfo{person}{Brooke~C. Schneider}.} \bibinfo{year}{2024}\natexlab{}.
\newblock \showarticletitle{Editorial: {Non}-pharmacological interventions for mental disorders}.
\newblock \bibinfo{journal}{\emph{Frontiers in Psychology}}  \bibinfo{volume}{15} (\bibinfo{date}{March} \bibinfo{year}{2024}).
\newblock
\showISSN{1664-1078}
\urldef\tempurl%
\url{https://doi.org/10.3389/fpsyg.2024.1363348}
\showDOI{\tempurl}
\newblock
\shownote{Publisher: Frontiers}.


\bibitem[Plummer(1996)]%
        {plummer_chaplaincy_1996}
\bibfield{author}{\bibinfo{person}{David~B. Plummer}.} \bibinfo{year}{1996}\natexlab{}.
\newblock \showarticletitle{Chaplaincy: {The} {Greatest} {Story} {Never} {Told}}.
\newblock \bibinfo{journal}{\emph{Journal of Pastoral Care}} \bibinfo{volume}{50}, \bibinfo{number}{1} (\bibinfo{date}{March} \bibinfo{year}{1996}), \bibinfo{pages}{1--11}.
\newblock
\showISSN{0022-3409}
\urldef\tempurl%
\url{https://doi.org/10.1177/002234099605000101}
\showDOI{\tempurl}
\newblock
\shownote{Publisher: SAGE Publications}.


\bibitem[Poncin et~al\mbox{.}(2020)]%
        {poncin_what_2020}
\bibfield{author}{\bibinfo{person}{Emmanuelle Poncin}, \bibinfo{person}{Brigitte Niquille}, \bibinfo{person}{Guy Jobin}, \bibinfo{person}{Charles Benaim}, {and} \bibinfo{person}{Etienne Rochat}.} \bibinfo{year}{2020}\natexlab{}.
\newblock \showarticletitle{What {Motivates} {Healthcare} {Professionals}' {Referrals} to {Chaplains}, and {How} to {Help} {Them} {Formulate} {Referrals} that {Accurately} {Reflect} {Patients}' {Spiritual} {Needs}?}
\newblock \bibinfo{journal}{\emph{Journal of Health Care Chaplaincy}} \bibinfo{volume}{26}, \bibinfo{number}{1} (\bibinfo{year}{2020}), \bibinfo{pages}{1--15}.
\newblock
\showISSN{1528-6916}
\urldef\tempurl%
\url{https://doi.org/10.1080/08854726.2019.1582211}
\showDOI{\tempurl}


\bibitem[Prinster et~al\mbox{.}(2024)]%
        {prinster_community_2024}
\bibfield{author}{\bibinfo{person}{Gale~H. Prinster}, \bibinfo{person}{C.~Estelle Smith}, \bibinfo{person}{Chenhao Tan}, {and} \bibinfo{person}{Brian~C. Keegan}.} \bibinfo{year}{2024}\natexlab{}.
\newblock \showarticletitle{Community {Archetypes}: {An} {Empirical} {Framework} for {Guiding} {Research} {Methodologies} to {Reflect} {User} {Experiences} of {Sense} of {Virtual} {Community} on {Reddit}}.
\newblock \bibinfo{journal}{\emph{Proceedings of the ACM on Human-Computer Interaction}} \bibinfo{volume}{8}, \bibinfo{number}{CSCW1} (\bibinfo{date}{April} \bibinfo{year}{2024}), \bibinfo{pages}{33:1--33:33}.
\newblock
\urldef\tempurl%
\url{https://doi.org/10.1145/3637310}
\showDOI{\tempurl}


\bibitem[Proferes et~al\mbox{.}(2021)]%
        {proferes_studying_2021}
\bibfield{author}{\bibinfo{person}{Nicholas Proferes}, \bibinfo{person}{Naiyan Jones}, \bibinfo{person}{Sarah Gilbert}, \bibinfo{person}{Casey Fiesler}, {and} \bibinfo{person}{Michael Zimmer}.} \bibinfo{year}{2021}\natexlab{}.
\newblock \showarticletitle{Studying {Reddit}: {A} {Systematic} {Overview} of {Disciplines}, {Approaches}, {Methods}, and {Ethics}}.
\newblock \bibinfo{journal}{\emph{Social Media + Society}} \bibinfo{volume}{7}, \bibinfo{number}{2} (\bibinfo{date}{April} \bibinfo{year}{2021}), \bibinfo{pages}{20563051211019004}.
\newblock
\showISSN{2056-3051}
\urldef\tempurl%
\url{https://doi.org/10.1177/20563051211019004}
\showDOI{\tempurl}
\newblock
\shownote{Publisher: SAGE Publications Ltd}.


\bibitem[Puchalski et~al\mbox{.}(2009)]%
        {puchalski_improving_2009}
\bibfield{author}{\bibinfo{person}{Christina Puchalski}, \bibinfo{person}{Betty Ferrell}, \bibinfo{person}{Rose Virani}, \bibinfo{person}{Shirley Otis-Green}, \bibinfo{person}{Pamela Baird}, \bibinfo{person}{Janet Bull}, \bibinfo{person}{Harvey Chochinov}, \bibinfo{person}{George Handzo}, \bibinfo{person}{Holly Nelson-Becker}, \bibinfo{person}{Maryjo Prince-Paul}, \bibinfo{person}{Karen Pugliese}, {and} \bibinfo{person}{Daniel Sulmasy}.} \bibinfo{year}{2009}\natexlab{}.
\newblock \showarticletitle{Improving the {Quality} of {Spiritual} {Care} as a {Dimension} of {Palliative} {Care}: {The} {Report} of the {Consensus} {Conference}}.
\newblock \bibinfo{journal}{\emph{Journal of Palliative Medicine}} \bibinfo{volume}{12}, \bibinfo{number}{10} (\bibinfo{date}{Oct.} \bibinfo{year}{2009}), \bibinfo{pages}{885--904}.
\newblock
\showISSN{1557-7740}
\urldef\tempurl%
\url{https://doi.org/10.1089/jpm.2009.0142}
\showDOI{\tempurl}


\bibitem[{Raymond De Vries} et~al\mbox{.}(2008)]%
        {raymond_de_vries_lost_2008}
\bibfield{author}{\bibinfo{person}{{Raymond De Vries}}, \bibinfo{person}{{Nancy Berlinger}}, {and} \bibinfo{person}{{Wendy Cadge}}.} \bibinfo{year}{2008}\natexlab{}.
\newblock \showarticletitle{Lost in {Translation}: {The} {Chaplain}'s {Role} in {Health} {Care}}.
\newblock \bibinfo{journal}{\emph{Hastings Center Report}} \bibinfo{volume}{38}, \bibinfo{number}{6} (\bibinfo{year}{2008}), \bibinfo{pages}{23--27}.
\newblock
\showISSN{1552-146X}
\urldef\tempurl%
\url{https://doi.org/10.1353/hcr.0.0081}
\showDOI{\tempurl}


\bibitem[Record et~al\mbox{.}(2018)]%
        {record_i_2018}
\bibfield{author}{\bibinfo{person}{Rachael~A. Record}, \bibinfo{person}{Will~R. Silberman}, \bibinfo{person}{Joshua~E. Santiago}, {and} \bibinfo{person}{Taewook Ham}.} \bibinfo{year}{2018}\natexlab{}.
\newblock \showarticletitle{I {Sought} {It}, {I} {Reddit}: {Examining} {Health} {Information} {Engagement} {Behaviors} among {Reddit} {Users}}.
\newblock \bibinfo{journal}{\emph{Journal of Health Communication}} \bibinfo{volume}{23}, \bibinfo{number}{5} (\bibinfo{date}{May} \bibinfo{year}{2018}), \bibinfo{pages}{470--476}.
\newblock
\showISSN{1081-0730}
\urldef\tempurl%
\url{https://doi.org/10.1080/10810730.2018.1465493}
\showDOI{\tempurl}
\newblock
\shownote{Publisher: Taylor \& Francis \_eprint: https://doi.org/10.1080/10810730.2018.1465493}.


\bibitem[Rice-Jalloh(2022)]%
        {rice-jalloh_zooming_2022}
\bibfield{author}{\bibinfo{person}{K.~Monet Rice-Jalloh}.} \bibinfo{year}{2022}\natexlab{}.
\newblock \showarticletitle{Zooming to the {Scene}: {Higher} {Education} {Chaplaincy} and {Hybrid} {Digital} {Care}}.
\newblock \bibinfo{journal}{\emph{Journal of Pastoral Theology}} \bibinfo{volume}{32}, \bibinfo{number}{1} (\bibinfo{date}{Jan.} \bibinfo{year}{2022}), \bibinfo{pages}{36--40}.
\newblock
\showISSN{1064-9867}
\urldef\tempurl%
\url{https://doi.org/10.1080/10649867.2022.2059228}
\showDOI{\tempurl}
\newblock
\shownote{Publisher: Routledge \_eprint: https://doi.org/10.1080/10649867.2022.2059228}.


\bibitem[Rosmarin et~al\mbox{.}(2013)]%
        {rosmarin_religious_2013}
\bibfield{author}{\bibinfo{person}{David~H. Rosmarin}, \bibinfo{person}{Joseph~S. Bigda-Peyton}, \bibinfo{person}{Dost Öngur}, \bibinfo{person}{Kenneth~I. Pargament}, {and} \bibinfo{person}{Thröstur Björgvinsson}.} \bibinfo{year}{2013}\natexlab{}.
\newblock \showarticletitle{Religious coping among psychotic patients: {Relevance} to suicidality and treatment outcomes}.
\newblock \bibinfo{journal}{\emph{Psychiatry Research}} \bibinfo{volume}{210}, \bibinfo{number}{1} (\bibinfo{date}{Nov.} \bibinfo{year}{2013}), \bibinfo{pages}{182--187}.
\newblock
\showISSN{01651781}
\urldef\tempurl%
\url{https://doi.org/10.1016/j.psychres.2013.03.023}
\showDOI{\tempurl}


\bibitem[Rushton(2014)]%
        {rushton_what_2014}
\bibfield{author}{\bibinfo{person}{Lucy Rushton}.} \bibinfo{year}{2014}\natexlab{}.
\newblock \showarticletitle{What are the barriers to spiritual care in a hospital setting?}
\newblock \bibinfo{journal}{\emph{British Journal of Nursing}} \bibinfo{volume}{23}, \bibinfo{number}{7} (\bibinfo{date}{April} \bibinfo{year}{2014}), \bibinfo{pages}{370--374}.
\newblock
\showISSN{0966-0461}
\urldef\tempurl%
\url{https://doi.org/10.12968/bjon.2014.23.7.370}
\showDOI{\tempurl}
\newblock
\shownote{Publisher: Mark Allen Group}.


\bibitem[Rykkje et~al\mbox{.}(2022)]%
        {rykkje_educational_2022}
\bibfield{author}{\bibinfo{person}{Linda Rykkje}, \bibinfo{person}{Margrethe~Bakstad Søvik}, \bibinfo{person}{Linda Ross}, \bibinfo{person}{Wilfred McSherry}, \bibinfo{person}{Pamela Cone}, {and} \bibinfo{person}{Tove Giske}.} \bibinfo{year}{2022}\natexlab{}.
\newblock \showarticletitle{Educational interventions and strategies for spiritual care in nursing and healthcare students and staff: {A} scoping review}.
\newblock \bibinfo{journal}{\emph{Journal of Clinical Nursing}} \bibinfo{volume}{31}, \bibinfo{number}{11-12} (\bibinfo{year}{2022}), \bibinfo{pages}{1440--1464}.
\newblock
\showISSN{1365-2702}
\urldef\tempurl%
\url{https://doi.org/10.1111/jocn.16067}
\showDOI{\tempurl}
\newblock
\shownote{\_eprint: https://onlinelibrary.wiley.com/doi/pdf/10.1111/jocn.16067}.


\bibitem[Scardigno and Mininni(2020)]%
        {scardigno_authentic_2020}
\bibfield{author}{\bibinfo{person}{Rosa Scardigno} {and} \bibinfo{person}{Giuseppe Mininni}.} \bibinfo{year}{2020}\natexlab{}.
\newblock \showarticletitle{An authentic feeling? {Religious} experience through {Q}\&{A} websites}.
\newblock \bibinfo{journal}{\emph{Archive for the Psychology of Religion}} \bibinfo{volume}{42}, \bibinfo{number}{2} (\bibinfo{date}{July} \bibinfo{year}{2020}), \bibinfo{pages}{211--231}.
\newblock
\showISSN{0084-6724}
\urldef\tempurl%
\url{https://doi.org/10.1177/0084672420917451}
\showDOI{\tempurl}
\newblock
\shownote{Publisher: SAGE Publications Ltd}.


\bibitem[Schneider(2022)]%
        {schneider_admins_2022}
\bibfield{author}{\bibinfo{person}{Nathan Schneider}.} \bibinfo{year}{2022}\natexlab{}.
\newblock \showarticletitle{Admins, mods, and benevolent dictators for life: {The} implicit feudalism of online communities}.
\newblock \bibinfo{journal}{\emph{New Media \& Society}} \bibinfo{volume}{24}, \bibinfo{number}{9} (\bibinfo{date}{Sept.} \bibinfo{year}{2022}), \bibinfo{pages}{1965--1985}.
\newblock
\showISSN{1461-4448}
\urldef\tempurl%
\url{https://doi.org/10.1177/1461444820986553}
\showDOI{\tempurl}
\newblock
\shownote{Publisher: SAGE Publications}.


\bibitem[Schroeder et~al\mbox{.}(2018)]%
        {schroeder_pocket_2018}
\bibfield{author}{\bibinfo{person}{Jessica Schroeder}, \bibinfo{person}{Chelsey Wilkes}, \bibinfo{person}{Kael Rowan}, \bibinfo{person}{Arturo Toledo}, \bibinfo{person}{Ann Paradiso}, \bibinfo{person}{Mary Czerwinski}, \bibinfo{person}{Gloria Mark}, {and} \bibinfo{person}{Marsha~M. Linehan}.} \bibinfo{year}{2018}\natexlab{}.
\newblock \showarticletitle{Pocket {Skills}: {A} {Conversational} {Mobile} {Web} {App} {To} {Support} {Dialectical} {Behavioral} {Therapy}}. In \bibinfo{booktitle}{\emph{Proceedings of the 2018 {CHI} {Conference} on {Human} {Factors} in {Computing} {Systems}}}. \bibinfo{publisher}{ACM}, \bibinfo{address}{Montreal QC Canada}, \bibinfo{pages}{1--15}.
\newblock
\showISBNx{978-1-4503-5620-6}
\urldef\tempurl%
\url{https://doi.org/10.1145/3173574.3173972}
\showDOI{\tempurl}


\bibitem[Schulte et~al\mbox{.}(2020)]%
        {schulte_hospital_2020}
\bibfield{author}{\bibinfo{person}{Britta Schulte}, \bibinfo{person}{Philipp Graf}, \bibinfo{person}{Lena Franzkowiak}, {and} \bibinfo{person}{Eva Hornecker}.} \bibinfo{year}{2020}\natexlab{}.
\newblock \showarticletitle{Hospital {Beds}, {Robot} {Priests} and {Huggables}: {A} ({Fictional}) {Review} of {Commercially} {Available} {Care} {Robots}}. In \bibinfo{booktitle}{\emph{Proceedings of the 11th {Nordic} {Conference} on {Human}-{Computer} {Interaction}: {Shaping} {Experiences}, {Shaping} {Society}}} \emph{(\bibinfo{series}{{NordiCHI} '20})}. \bibinfo{publisher}{Association for Computing Machinery}, \bibinfo{address}{New York, NY, USA}, \bibinfo{pages}{1--11}.
\newblock
\showISBNx{978-1-4503-7579-5}
\urldef\tempurl%
\url{https://doi.org/10.1145/3419249.3420127}
\showDOI{\tempurl}


\bibitem[Seales(2012)]%
        {seales_corporate_2012}
\bibfield{author}{\bibinfo{person}{Chad~E. Seales}.} \bibinfo{year}{2012}\natexlab{}.
\newblock \showarticletitle{Corporate {Chaplaincy} and the {American} {Workplace}}.
\newblock \bibinfo{journal}{\emph{Religion Compass}} \bibinfo{volume}{6}, \bibinfo{number}{3} (\bibinfo{year}{2012}), \bibinfo{pages}{195--203}.
\newblock
\showISSN{1749-8171}
\urldef\tempurl%
\url{https://doi.org/10.1111/j.1749-8171.2012.00340.x}
\showDOI{\tempurl}
\newblock
\shownote{\_eprint: https://onlinelibrary.wiley.com/doi/pdf/10.1111/j.1749-8171.2012.00340.x}.


\bibitem[Seering et~al\mbox{.}(2022)]%
        {seering_metaphors_2022}
\bibfield{author}{\bibinfo{person}{Joseph Seering}, \bibinfo{person}{Geoff Kaufman}, {and} \bibinfo{person}{Stevie Chancellor}.} \bibinfo{year}{2022}\natexlab{}.
\newblock \showarticletitle{Metaphors in moderation}.
\newblock \bibinfo{journal}{\emph{New Media \& Society}} \bibinfo{volume}{24}, \bibinfo{number}{3} (\bibinfo{date}{March} \bibinfo{year}{2022}), \bibinfo{pages}{621--640}.
\newblock
\showISSN{1461-4448, 1461-7315}
\urldef\tempurl%
\url{https://doi.org/10.1177/1461444820964968}
\showDOI{\tempurl}


\bibitem[Shah et~al\mbox{.}(2022)]%
        {shah_modeling_2022}
\bibfield{author}{\bibinfo{person}{Raj~Sanjay Shah}, \bibinfo{person}{Faye Holt}, \bibinfo{person}{Shirley~Anugrah Hayati}, \bibinfo{person}{Aastha Agarwal}, \bibinfo{person}{Yi-Chia Wang}, \bibinfo{person}{Robert~E. Kraut}, {and} \bibinfo{person}{Diyi Yang}.} \bibinfo{year}{2022}\natexlab{}.
\newblock \showarticletitle{Modeling {Motivational} {Interviewing} {Strategies} on an {Online} {Peer}-to-{Peer} {Counseling} {Platform}}.
\newblock \bibinfo{journal}{\emph{Proceedings of the ACM on Human-Computer Interaction}} \bibinfo{volume}{6}, \bibinfo{number}{CSCW2} (\bibinfo{date}{Nov.} \bibinfo{year}{2022}), \bibinfo{pages}{1--24}.
\newblock
\showISSN{2573-0142}
\urldef\tempurl%
\url{https://doi.org/10.1145/3555640}
\showDOI{\tempurl}


\bibitem[Shields et~al\mbox{.}(2015)]%
        {shields_spiritual_2015}
\bibfield{author}{\bibinfo{person}{Michele Shields}, \bibinfo{person}{Allison Kestenbaum}, {and} \bibinfo{person}{Laura~B. Dunn}.} \bibinfo{year}{2015}\natexlab{}.
\newblock \showarticletitle{Spiritual {AIM} and the work of the chaplain: {A} model for assessing spiritual needs and outcomes in relationship}.
\newblock \bibinfo{journal}{\emph{Palliative and Supportive Care}} \bibinfo{volume}{13}, \bibinfo{number}{1} (\bibinfo{date}{Feb.} \bibinfo{year}{2015}), \bibinfo{pages}{75--89}.
\newblock
\showISSN{1478-9515, 1478-9523}
\urldef\tempurl%
\url{https://doi.org/10.1017/S1478951513001120}
\showDOI{\tempurl}


\bibitem[Sinclair et~al\mbox{.}(2012)]%
        {sinclair_spiritual_2012}
\bibfield{author}{\bibinfo{person}{Shane Sinclair}, \bibinfo{person}{Shelley~Raffin Bouchal}, \bibinfo{person}{Harvey Chochinov}, \bibinfo{person}{Neil Hagen}, {and} \bibinfo{person}{Susan McClement}.} \bibinfo{year}{2012}\natexlab{}.
\newblock \showarticletitle{Spiritual care: how to do it}.
\newblock \bibinfo{journal}{\emph{BMJ Supportive \& Palliative Care}} \bibinfo{volume}{2}, \bibinfo{number}{4} (\bibinfo{date}{Dec.} \bibinfo{year}{2012}), \bibinfo{pages}{319--327}.
\newblock
\showISSN{2045-435X, 2045-4368}
\urldef\tempurl%
\url{https://doi.org/10.1136/bmjspcare-2011-000191}
\showDOI{\tempurl}


\bibitem[Slovak and Munson(2024)]%
        {slovak_hci_2024}
\bibfield{author}{\bibinfo{person}{Petr Slovak} {and} \bibinfo{person}{Sean~A. Munson}.} \bibinfo{year}{2024}\natexlab{}.
\newblock \showarticletitle{{HCI} {Contributions} in {Mental} {Health}: {A} {Modular} {Framework} to {Guide} {Psychosocial} {Intervention} {Design}}. In \bibinfo{booktitle}{\emph{Proceedings of the {CHI} {Conference} on {Human} {Factors} in {Computing} {Systems}}}. \bibinfo{publisher}{ACM}, \bibinfo{address}{Honolulu HI USA}, \bibinfo{pages}{1--21}.
\newblock
\showISBNx{9798400703300}
\urldef\tempurl%
\url{https://doi.org/10.1145/3613904.3642624}
\showDOI{\tempurl}


\bibitem[Smith(2022)]%
        {smith_sacred_2022}
\bibfield{author}{\bibinfo{person}{C.~Estelle Smith}.} \bibinfo{year}{2022}\natexlab{}.
\newblock \showarticletitle{Sacred be thy tech: thoughts (and prayers) on integrating spirituality in technology for health and well-being}.
\newblock \bibinfo{journal}{\emph{Interactions}} \bibinfo{volume}{29}, \bibinfo{number}{4} (\bibinfo{date}{July} \bibinfo{year}{2022}), \bibinfo{pages}{68--72}.
\newblock
\showISSN{1072-5520, 1558-3449}
\urldef\tempurl%
\url{https://doi.org/10.1145/3543893}
\showDOI{\tempurl}


\bibitem[Smith et~al\mbox{.}(2022)]%
        {smith_impact_2022}
\bibfield{author}{\bibinfo{person}{C.~Estelle Smith}, \bibinfo{person}{Irfanul Alam}, \bibinfo{person}{Chenhao Tan}, \bibinfo{person}{Brian~C. Keegan}, {and} \bibinfo{person}{Anita~L. Blanchard}.} \bibinfo{year}{2022}\natexlab{}.
\newblock \showarticletitle{The {Impact} of {Governance} {Bots} on {Sense} of {Virtual} {Community}: {Development} and {Validation} of the {GOV}-{BOTs} {Scale}}.
\newblock \bibinfo{journal}{\emph{Proceedings of the ACM on Human-Computer Interaction}} \bibinfo{volume}{6}, \bibinfo{number}{CSCW2} (\bibinfo{date}{Nov.} \bibinfo{year}{2022}), \bibinfo{pages}{462:1--462:30}.
\newblock
\urldef\tempurl%
\url{https://doi.org/10.1145/3555563}
\showDOI{\tempurl}


\bibitem[Smith et~al\mbox{.}(2021a)]%
        {smith_what_2021}
\bibfield{author}{\bibinfo{person}{C.~Estelle Smith}, \bibinfo{person}{Avleen Kaur}, \bibinfo{person}{Katie~Z. Gach}, \bibinfo{person}{Loren Terveen}, \bibinfo{person}{Mary~Jo Kreitzer}, {and} \bibinfo{person}{Susan O'Conner-Von}.} \bibinfo{year}{2021}\natexlab{a}.
\newblock \showarticletitle{What is {Spiritual} {Support} and {How} {Might} {It} {Impact} the {Design} of {Online} {Communities}?}
\newblock \bibinfo{journal}{\emph{Proceedings of the ACM on Human-Computer Interaction}} \bibinfo{volume}{5}, \bibinfo{number}{CSCW1} (\bibinfo{date}{April} \bibinfo{year}{2021}), \bibinfo{pages}{1--42}.
\newblock
\showISSN{2573-0142}
\urldef\tempurl%
\url{https://doi.org/10.1145/3449117}
\showDOI{\tempurl}


\bibitem[Smith et~al\mbox{.}(2024)]%
        {smith_governance_2024}
\bibfield{author}{\bibinfo{person}{C.~Estelle Smith}, \bibinfo{person}{Shamika Klassen}, \bibinfo{person}{Gale~H. Prinster}, \bibinfo{person}{Chenhao Tan}, {and} \bibinfo{person}{Brian~C. Keegan}.} \bibinfo{year}{2024}\natexlab{}.
\newblock \showarticletitle{Governance of the {Black} {Experience} on {Reddit}: r/{BlackPeopleTwitter} as a {Case} {Study} in {Supporting} {Sense} of {Virtual} {Community} for {Black} {Users}}.
\newblock \bibinfo{journal}{\emph{Proceedings of the ACM on Human-Computer Interaction}} \bibinfo{number}{CSCW2} (\bibinfo{year}{2024}).
\newblock


\bibitem[Smith et~al\mbox{.}(2021b)]%
        {smith_effective_2021}
\bibfield{author}{\bibinfo{person}{C.~Estelle Smith}, \bibinfo{person}{William Lane}, \bibinfo{person}{Hannah Miller~Hillberg}, \bibinfo{person}{Daniel Kluver}, \bibinfo{person}{Loren Terveen}, {and} \bibinfo{person}{Svetlana Yarosh}.} \bibinfo{year}{2021}\natexlab{b}.
\newblock \showarticletitle{Effective {Strategies} for {Crowd}-{Powered} {Cognitive} {Reappraisal} {Systems}: {A} {Field} {Deployment} of the {Flip}*{Doubt} {Web} {Application} for {Mental} {Health}}.
\newblock \bibinfo{journal}{\emph{Proceedings of the ACM on Human-Computer Interaction}} \bibinfo{volume}{5}, \bibinfo{number}{CSCW2} (\bibinfo{date}{Oct.} \bibinfo{year}{2021}), \bibinfo{pages}{1--37}.
\newblock
\showISSN{2573-0142}
\urldef\tempurl%
\url{https://doi.org/10.1145/3479561}
\showDOI{\tempurl}


\bibitem[Smith et~al\mbox{.}(2020)]%
        {smith_i_2020}
\bibfield{author}{\bibinfo{person}{C.~Estelle Smith}, \bibinfo{person}{Zachary Levonian}, \bibinfo{person}{Haiwei Ma}, \bibinfo{person}{Robert Giaquinto}, \bibinfo{person}{Gemma Lein-Mcdonough}, \bibinfo{person}{Zixuan Li}, \bibinfo{person}{Susan O’conner-Von}, {and} \bibinfo{person}{Svetlana Yarosh}.} \bibinfo{year}{2020}\natexlab{}.
\newblock \showarticletitle{"{I} {Cannot} {Do} {All} of {This} {Alone}": {Exploring} {Instrumental} and {Prayer} {Support} in {Online} {Health} {Communities}}.
\newblock \bibinfo{journal}{\emph{ACM Transactions on Computer-Human Interaction}} \bibinfo{volume}{27}, \bibinfo{number}{5} (\bibinfo{date}{Aug.} \bibinfo{year}{2020}), \bibinfo{pages}{38:1--38:41}.
\newblock
\showISSN{1073-0516}
\urldef\tempurl%
\url{https://doi.org/10.1145/3402855}
\showDOI{\tempurl}


\bibitem[Smith et~al\mbox{.}(2023)]%
        {smith_thoughts_2023}
\bibfield{author}{\bibinfo{person}{C.~Estelle Smith}, \bibinfo{person}{Hannah Miller~Hillberg}, {and} \bibinfo{person}{Zachary Levonian}.} \bibinfo{year}{2023}\natexlab{}.
\newblock \showarticletitle{Thoughts and {Prayers}: {How} the {Release} of {New} {Reactions} on {CaringBridge} {Reshapes} {Supportive} {Communication} in {Health} {Crises}}.
\newblock \bibinfo{journal}{\emph{Proceedings of the ACM on Human-Computer Interaction}} \bibinfo{volume}{7}, \bibinfo{number}{CSCW2} (\bibinfo{date}{Sept.} \bibinfo{year}{2023}), \bibinfo{pages}{1--39}.
\newblock
\showISSN{2573-0142}
\urldef\tempurl%
\url{https://doi.org/10.1145/3610035}
\showDOI{\tempurl}


\bibitem[Sprik et~al\mbox{.}(2022)]%
        {sprik_chaplains_2022}
\bibfield{author}{\bibinfo{person}{Petra~J. Sprik}, \bibinfo{person}{Angela Janssen~Keenan}, \bibinfo{person}{Danielle Boselli}, {and} \bibinfo{person}{Daniel~H. Grossoehme}.} \bibinfo{year}{2022}\natexlab{}.
\newblock \showarticletitle{Chaplains and telechaplaincy: best practices, strengths, weaknesses—a national study}.
\newblock \bibinfo{journal}{\emph{Journal of Health Care Chaplaincy}} \bibinfo{volume}{0}, \bibinfo{number}{0} (\bibinfo{date}{Jan.} \bibinfo{year}{2022}), \bibinfo{pages}{1--23}.
\newblock
\showISSN{0885-4726}
\urldef\tempurl%
\url{https://doi.org/10.1080/08854726.2022.2026103}
\showDOI{\tempurl}
\newblock
\shownote{Publisher: Routledge \_eprint: https://doi.org/10.1080/08854726.2022.2026103}.


\bibitem[Thieme et~al\mbox{.}(2023)]%
        {thieme_designing_2023}
\bibfield{author}{\bibinfo{person}{Anja Thieme}, \bibinfo{person}{Maryann Hanratty}, \bibinfo{person}{Maria Lyons}, \bibinfo{person}{Jorge Palacios}, \bibinfo{person}{Rita~Faia Marques}, \bibinfo{person}{Cecily Morrison}, {and} \bibinfo{person}{Gavin Doherty}.} \bibinfo{year}{2023}\natexlab{}.
\newblock \showarticletitle{Designing {Human}-centered {AI} for {Mental} {Health}: {Developing} {Clinically} {Relevant} {Applications} for {Online} {CBT} {Treatment}}.
\newblock \bibinfo{journal}{\emph{ACM Transactions on Computer-Human Interaction}} \bibinfo{volume}{30}, \bibinfo{number}{2} (\bibinfo{date}{April} \bibinfo{year}{2023}), \bibinfo{pages}{1--50}.
\newblock
\showISSN{1073-0516, 1557-7325}
\urldef\tempurl%
\url{https://doi.org/10.1145/3564752}
\showDOI{\tempurl}


\bibitem[Timmins et~al\mbox{.}(2018)]%
        {timmins_role_2018}
\bibfield{author}{\bibinfo{person}{Fiona Timmins}, \bibinfo{person}{Sílvia Caldeira}, \bibinfo{person}{Maryanne Murphy}, \bibinfo{person}{Nicolas Pujol}, \bibinfo{person}{Greg Sheaf}, \bibinfo{person}{Elizabeth Weathers}, \bibinfo{person}{Jacqueline Whelan}, {and} \bibinfo{person}{Bernadette Flanagan}.} \bibinfo{year}{2018}\natexlab{}.
\newblock \showarticletitle{The {Role} of the {Healthcare} {Chaplain}: {A} {Literature} {Review}}.
\newblock \bibinfo{journal}{\emph{Journal of Health Care Chaplaincy}} \bibinfo{volume}{24}, \bibinfo{number}{3} (\bibinfo{date}{July} \bibinfo{year}{2018}), \bibinfo{pages}{87--106}.
\newblock
\showISSN{0885-4726, 1528-6916}
\urldef\tempurl%
\url{https://doi.org/10.1080/08854726.2017.1338048}
\showDOI{\tempurl}


\bibitem[Trothen(2022)]%
        {trothen_replika_2022}
\bibfield{author}{\bibinfo{person}{Tracy~J. Trothen}.} \bibinfo{year}{2022}\natexlab{}.
\newblock \showarticletitle{Replika: {Spiritual} {Enhancement} {Technology}?}
\newblock \bibinfo{journal}{\emph{Religions}} \bibinfo{volume}{13}, \bibinfo{number}{4} (\bibinfo{date}{April} \bibinfo{year}{2022}), \bibinfo{pages}{275}.
\newblock
\showISSN{2077-1444}
\urldef\tempurl%
\url{https://doi.org/10.3390/rel13040275}
\showDOI{\tempurl}
\newblock
\shownote{Number: 4 Publisher: Multidisciplinary Digital Publishing Institute}.


\bibitem[Urban et~al\mbox{.}(2009)]%
        {urban_online_2009}
\bibfield{author}{\bibinfo{person}{Glen~L. Urban}, \bibinfo{person}{Cinda Amyx}, {and} \bibinfo{person}{Antonio Lorenzon}.} \bibinfo{year}{2009}\natexlab{}.
\newblock \showarticletitle{Online {Trust}: {State} of the {Art}, {New} {Frontiers}, and {Research} {Potential}}.
\newblock \bibinfo{journal}{\emph{Journal of Interactive Marketing}} \bibinfo{volume}{23}, \bibinfo{number}{2} (\bibinfo{date}{May} \bibinfo{year}{2009}), \bibinfo{pages}{179--190}.
\newblock
\showISSN{1094-9968}
\urldef\tempurl%
\url{https://doi.org/10.1016/j.intmar.2009.03.001}
\showDOI{\tempurl}
\newblock
\shownote{Publisher: SAGE Publications}.


\bibitem[van Meerveld et~al\mbox{.}(2017)]%
        {van_meerveld_role_2017}
\bibfield{author}{\bibinfo{person}{Anne van Meerveld}, \bibinfo{person}{Donna Wielinga}, \bibinfo{person}{Els Versluis}, {and} \bibinfo{person}{Linda V.}} \bibinfo{year}{2017}\natexlab{}.
\newblock \bibinfo{title}{The role of anonymity in discussing social anxiety online: {A} comparative platform analysis of {Facebook} and {Reddit}}.
\newblock
\newblock
\urldef\tempurl%
\url{https://mastersofmedia.hum.uva.nl/blog/2017/10/24/the-role-of-anonymity-in-discussing-social-anxiety-online-a-comparative-platform-analysis-of-facebook-and-reddit/}
\showURL{%
\tempurl}


\bibitem[VandeCreek and Lucas(2014)]%
        {vandecreek_discipline_2014}
\bibfield{author}{\bibinfo{person}{Larry VandeCreek} {and} \bibinfo{person}{Arthur~M. Lucas}.} \bibinfo{year}{2014}\natexlab{}.
\newblock \bibinfo{booktitle}{\emph{The {Discipline} for {Pastoral} {Care} {Giving}: {Foundations} for {Outcome} {Oriented} {Chaplaincy}}}.
\newblock \bibinfo{publisher}{Routledge}.
\newblock
\showISBNx{978-1-317-78883-6}
\newblock
\shownote{Google-Books-ID: Vi\_GBQAAQBAJ}.


\bibitem[Waller et~al\mbox{.}(2023)]%
        {waller_collegiate_2023}
\bibfield{author}{\bibinfo{person}{Steven Waller}, \bibinfo{person}{Lars Dzikus}, {and} \bibinfo{person}{Robin Hardin}.} \bibinfo{year}{2023}\natexlab{}.
\newblock \showarticletitle{Collegiate {Sport} {Chaplaincy}: {Problems} and {Promise}}.
\newblock \bibinfo{journal}{\emph{Journal of Issues in Intercollegiate Athletics}} \bibinfo{volume}{1}, \bibinfo{number}{1} (\bibinfo{date}{Dec.} \bibinfo{year}{2023}).
\newblock
\urldef\tempurl%
\url{https://scholarcommons.sc.edu/jiia/vol1/iss1/18}
\showURL{%
\tempurl}


\bibitem[Wang et~al\mbox{.}(2021)]%
        {wang_online_2021}
\bibfield{author}{\bibinfo{person}{Xiaohui Wang}, \bibinfo{person}{Jingyuan Shi}, {and} \bibinfo{person}{Hanxiao Kong}.} \bibinfo{year}{2021}\natexlab{}.
\newblock \showarticletitle{Online {Health} {Information} {Seeking}: {A} {Review} and {Meta}-{Analysis}}.
\newblock \bibinfo{journal}{\emph{Health Communication}} \bibinfo{volume}{36}, \bibinfo{number}{10} (\bibinfo{date}{Aug.} \bibinfo{year}{2021}), \bibinfo{pages}{1163--1175}.
\newblock
\showISSN{1041-0236}
\urldef\tempurl%
\url{https://doi.org/10.1080/10410236.2020.1748829}
\showDOI{\tempurl}
\newblock
\shownote{Publisher: Routledge \_eprint: https://doi.org/10.1080/10410236.2020.1748829}.


\bibitem[Wang and Emurian(2005)]%
        {wang_overview_2005}
\bibfield{author}{\bibinfo{person}{Ye~Diana Wang} {and} \bibinfo{person}{Henry~H. Emurian}.} \bibinfo{year}{2005}\natexlab{}.
\newblock \showarticletitle{An overview of online trust: {Concepts}, elements, and implications}.
\newblock \bibinfo{journal}{\emph{Computers in Human Behavior}} \bibinfo{volume}{21}, \bibinfo{number}{1} (\bibinfo{date}{Jan.} \bibinfo{year}{2005}), \bibinfo{pages}{105--125}.
\newblock
\showISSN{0747-5632}
\urldef\tempurl%
\url{https://doi.org/10.1016/j.chb.2003.11.008}
\showDOI{\tempurl}


\bibitem[Wang-Qin and Hong-Lin(2018)]%
        {wang-qin_nurse_2018}
\bibfield{author}{\bibinfo{person}{Shen Wang-Qin} {and} \bibinfo{person}{Chen Hong-Lin}.} \bibinfo{year}{2018}\natexlab{}.
\newblock \showarticletitle{Nurse {Staffs}' {Knowledge}, {Attitudes}, and {Practices} toward {Palliative} {Care} in {China}: a {Cross}-{Sectional} {Survey}}. In \bibinfo{booktitle}{\emph{Proceedings of the 2nd {International} {Conference} on {Medical} and {Health} {Informatics}}} \emph{(\bibinfo{series}{{ICMHI} '18})}. \bibinfo{publisher}{Association for Computing Machinery}, \bibinfo{address}{New York, NY, USA}, \bibinfo{pages}{238--243}.
\newblock
\showISBNx{978-1-4503-6389-1}
\urldef\tempurl%
\url{https://doi.org/10.1145/3239438.3239455}
\showDOI{\tempurl}


\bibitem[Weinstein(1965)]%
        {weinstein_stigma_1965}
\bibfield{author}{\bibinfo{person}{Eugene Weinstein}.} \bibinfo{year}{1965}\natexlab{}.
\newblock \showarticletitle{Stigma: {Notes} on the {Management} of {Spoiled} {Identity}. {Erving} {Goffman}}.
\newblock \bibinfo{journal}{\emph{Amer. J. Sociology}} \bibinfo{volume}{70}, \bibinfo{number}{5} (\bibinfo{date}{March} \bibinfo{year}{1965}), \bibinfo{pages}{636--636}.
\newblock
\showISSN{0002-9602}
\urldef\tempurl%
\url{https://doi.org/10.1086/223949}
\showDOI{\tempurl}
\newblock
\shownote{Publisher: The University of Chicago Press}.


\bibitem[White et~al\mbox{.}(2023)]%
        {white_provision_2023}
\bibfield{author}{\bibinfo{person}{Kelsey~B. White}, \bibinfo{person}{Shoou-Yih~Daniel Lee}, \bibinfo{person}{J’Aime~C. Jennings}, \bibinfo{person}{Seyed Karimi}, \bibinfo{person}{Christopher~E. Johnson}, {and} \bibinfo{person}{George Fitchett}.} \bibinfo{year}{2023}\natexlab{}.
\newblock \showarticletitle{Provision of chaplaincy services in {U}.{S}. hospitals: {A} strategic conformity perspective}.
\newblock \bibinfo{journal}{\emph{Health Care Management Review}} \bibinfo{volume}{48}, \bibinfo{number}{4} (\bibinfo{date}{Dec.} \bibinfo{year}{2023}), \bibinfo{pages}{342}.
\newblock
\showISSN{0361-6274}
\urldef\tempurl%
\url{https://doi.org/10.1097/HMR.0000000000000382}
\showDOI{\tempurl}


\bibitem[Winiger(2022)]%
        {winiger_changing_2022}
\bibfield{author}{\bibinfo{person}{Fabian Winiger}.} \bibinfo{year}{2022}\natexlab{}.
\newblock \showarticletitle{The changing face of spiritual care: current developments in telechaplaincy.}
\newblock \bibinfo{journal}{\emph{Journal of Health Care Chaplaincy}} \bibinfo{volume}{0}, \bibinfo{number}{0} (\bibinfo{date}{March} \bibinfo{year}{2022}), \bibinfo{pages}{1--18}.
\newblock
\showISSN{0885-4726}
\urldef\tempurl%
\url{https://doi.org/10.1080/08854726.2022.2040895}
\showDOI{\tempurl}
\newblock
\shownote{Publisher: Routledge \_eprint: https://doi.org/10.1080/08854726.2022.2040895}.


\bibitem[Winiger and Sprik(2023)]%
        {winiger_navigating_2023}
\bibfield{author}{\bibinfo{person}{Fabian Winiger} {and} \bibinfo{person}{Petra Sprik}.} \bibinfo{year}{2023}\natexlab{}.
\newblock \showarticletitle{Navigating challenges in telechaplaincy: {A} thematic analysis of an international conference}.
\newblock \bibinfo{journal}{\emph{Journal of Health Care Chaplaincy}} (\bibinfo{date}{Dec.} \bibinfo{year}{2023}), \bibinfo{pages}{1--16}.
\newblock
\showISSN{0885-4726, 1528-6916}
\urldef\tempurl%
\url{https://doi.org/10.1080/08854726.2023.2294680}
\showDOI{\tempurl}


\bibitem[Woggon et~al\mbox{.}(2022)]%
        {woggon_development_2022}
\bibfield{author}{\bibinfo{person}{Frank Woggon}, \bibinfo{person}{Matthew Arlyck}, \bibinfo{person}{Stephenie Maddox~Hill}, {and} \bibinfo{person}{Leslie Small~Stokes}.} \bibinfo{year}{2022}\natexlab{}.
\newblock \showarticletitle{The development of an outcome oriented and research informed spiritual care assessment and documentation form for the electronic health record in an adult hospital setting}.
\newblock \bibinfo{journal}{\emph{Journal of Health Care Chaplaincy}} \bibinfo{volume}{28}, \bibinfo{number}{3} (\bibinfo{date}{July} \bibinfo{year}{2022}), \bibinfo{pages}{400--414}.
\newblock
\showISSN{0885-4726}
\urldef\tempurl%
\url{https://doi.org/10.1080/08854726.2021.1916333}
\showDOI{\tempurl}
\newblock
\shownote{Publisher: Routledge \_eprint: https://doi.org/10.1080/08854726.2021.1916333}.


\bibitem[Wolf et~al\mbox{.}(2022)]%
        {wolf_spirituality_2022}
\bibfield{author}{\bibinfo{person}{Sara Wolf}, \bibinfo{person}{Frauke Moerike}, \bibinfo{person}{Simon Luthe}, \bibinfo{person}{Ilona Nord}, {and} \bibinfo{person}{Jörn Hurtienne}.} \bibinfo{year}{2022}\natexlab{}.
\newblock \showarticletitle{Spirituality at the {Breakfast} {Table}: {Experiences} of {Christian} {Online} {Worship} {Services}}. In \bibinfo{booktitle}{\emph{Extended {Abstracts} of the 2022 {CHI} {Conference} on {Human} {Factors} in {Computing} {Systems}}} \emph{(\bibinfo{series}{{CHI} {EA} '22})}. \bibinfo{publisher}{Association for Computing Machinery}, \bibinfo{address}{New York, NY, USA}, \bibinfo{pages}{1--7}.
\newblock
\showISBNx{978-1-4503-9156-6}
\urldef\tempurl%
\url{https://doi.org/10.1145/3491101.3519856}
\showDOI{\tempurl}


\bibitem[Wolf et~al\mbox{.}(2023)]%
        {wolf_god-i-box_2023}
\bibfield{author}{\bibinfo{person}{Sara Wolf}, \bibinfo{person}{Benedikt Steinmüller}, \bibinfo{person}{Frauke Mörike}, \bibinfo{person}{Simon Luthe}, {and} \bibinfo{person}{Jörn Hurtienne}.} \bibinfo{year}{2023}\natexlab{}.
\newblock \showarticletitle{The {God}-{I}-{Box}: {Iteratively} {Provotyping} {Technology}-{Mediated} {Worship} {Services}}. In \bibinfo{booktitle}{\emph{Proceedings of the 2023 {ACM} {Designing} {Interactive} {Systems} {Conference}}} \emph{(\bibinfo{series}{{DIS} '23})}. \bibinfo{publisher}{Association for Computing Machinery}, \bibinfo{address}{New York, NY, USA}, \bibinfo{pages}{1710--1723}.
\newblock
\showISBNx{978-1-4503-9893-0}
\urldef\tempurl%
\url{https://doi.org/10.1145/3563657.3596029}
\showDOI{\tempurl}


\bibitem[Wolf and Feldbauer-Durstmüller(2018)]%
        {wolf_workplace_2018}
\bibfield{author}{\bibinfo{person}{Tanja Wolf} {and} \bibinfo{person}{Birgit Feldbauer-Durstmüller}.} \bibinfo{year}{2018}\natexlab{}.
\newblock \showarticletitle{Workplace chaplaincy: a literature review}.
\newblock \bibinfo{journal}{\emph{Journal of Management, Spirituality \& Religion}} \bibinfo{volume}{15}, \bibinfo{number}{1} (\bibinfo{date}{Jan.} \bibinfo{year}{2018}), \bibinfo{pages}{38--63}.
\newblock
\urldef\tempurl%
\url{https://doi.org/10.1080/14766086.2017.1385514}
\showDOI{\tempurl}


\bibitem[Yaden et~al\mbox{.}(2022)]%
        {yaden_meta-analysis_2022}
\bibfield{author}{\bibinfo{person}{David~B. Yaden}, \bibinfo{person}{Cassondra~L. Batz-Barbarich}, \bibinfo{person}{Vincent Ng}, \bibinfo{person}{Hoda Vaziri}, \bibinfo{person}{Jessica~N. Gladstone}, \bibinfo{person}{James~O. Pawelski}, {and} \bibinfo{person}{Louis Tay}.} \bibinfo{year}{2022}\natexlab{}.
\newblock \showarticletitle{A {Meta}-{Analysis} of {Religion}/{Spirituality} and {Life} {Satisfaction}}.
\newblock \bibinfo{journal}{\emph{Journal of Happiness Studies}} \bibinfo{volume}{23}, \bibinfo{number}{8} (\bibinfo{date}{Dec.} \bibinfo{year}{2022}), \bibinfo{pages}{4147--4163}.
\newblock
\showISSN{1389-4978, 1573-7780}
\urldef\tempurl%
\url{https://doi.org/10.1007/s10902-022-00558-7}
\showDOI{\tempurl}


\bibitem[Yang(2019)]%
        {yang_role_2019}
\bibfield{author}{\bibinfo{person}{Andrew~Chixiao Yang}.} \bibinfo{year}{2019}\natexlab{}.
\newblock \showarticletitle{The {Role} of {Color} in {Palliative} {Care} for {Children}}. In \bibinfo{booktitle}{\emph{Proceedings of the 2019 11th {International} {Conference} on {Bioinformatics} and {Biomedical} {Technology}}} \emph{(\bibinfo{series}{{ICBBT} '19})}. \bibinfo{publisher}{Association for Computing Machinery}, \bibinfo{address}{New York, NY, USA}, \bibinfo{pages}{27--31}.
\newblock
\showISBNx{978-1-4503-6231-3}
\urldef\tempurl%
\url{https://doi.org/10.1145/3340074.3340098}
\showDOI{\tempurl}


\bibitem[Young(2022)]%
        {young_virtual_2022}
\bibfield{author}{\bibinfo{person}{William Young}.} \bibinfo{year}{2022}\natexlab{}.
\newblock \showarticletitle{Virtual {Pastor}: {Virtualization}, {AI}, and {Pastoral} {Care}}.
\newblock \bibinfo{journal}{\emph{Theology and Science}} \bibinfo{volume}{20}, \bibinfo{number}{1} (\bibinfo{date}{Jan.} \bibinfo{year}{2022}), \bibinfo{pages}{6--22}.
\newblock
\showISSN{1474-6700}
\urldef\tempurl%
\url{https://doi.org/10.1080/14746700.2021.2012915}
\showDOI{\tempurl}
\newblock
\shownote{Publisher: Routledge \_eprint: https://doi.org/10.1080/14746700.2021.2012915}.


\bibitem[Zhang et~al\mbox{.}(2020)]%
        {zhang_policykit_2020}
\bibfield{author}{\bibinfo{person}{Amy~X. Zhang}, \bibinfo{person}{Grant Hugh}, {and} \bibinfo{person}{Michael~S. Bernstein}.} \bibinfo{year}{2020}\natexlab{}.
\newblock \showarticletitle{{PolicyKit}: {Building} {Governance} in {Online} {Communities}}. In \bibinfo{booktitle}{\emph{Proceedings of the 33rd {Annual} {ACM} {Symposium} on {User} {Interface} {Software} and {Technology}}} \emph{(\bibinfo{series}{{UIST} '20})}. \bibinfo{publisher}{Association for Computing Machinery}, \bibinfo{address}{New York, NY, USA}, \bibinfo{pages}{365--378}.
\newblock
\showISBNx{978-1-4503-7514-6}
\urldef\tempurl%
\url{https://doi.org/10.1145/3379337.3415858}
\showDOI{\tempurl}


\bibitem[Zhao et~al\mbox{.}(2022)]%
        {zhao_online_2022}
\bibfield{author}{\bibinfo{person}{Yuxiang~Chris Zhao}, \bibinfo{person}{Mengyuan Zhao}, {and} \bibinfo{person}{Shijie Song}.} \bibinfo{year}{2022}\natexlab{}.
\newblock \showarticletitle{Online {Health} {Information} {Seeking} {Among} {Patients} {With} {Chronic} {Conditions}: {Integrating} the {Health} {Belief} {Model} and {Social} {Support} {Theory}}.
\newblock \bibinfo{journal}{\emph{Journal of Medical Internet Research}} \bibinfo{volume}{24}, \bibinfo{number}{11} (\bibinfo{date}{Nov.} \bibinfo{year}{2022}), \bibinfo{pages}{e42447}.
\newblock
\urldef\tempurl%
\url{https://doi.org/10.2196/42447}
\showDOI{\tempurl}
\newblock
\shownote{Company: Journal of Medical Internet Research Distributor: Journal of Medical Internet Research Institution: Journal of Medical Internet Research Label: Journal of Medical Internet Research Publisher: JMIR Publications Inc., Toronto, Canada}.


\bibitem[Zhu et~al\mbox{.}(2024)]%
        {zhu_user_2024}
\bibfield{author}{\bibinfo{person}{Jianfeng Zhu}, \bibinfo{person}{Ruoming Jin}, \bibinfo{person}{Deric~R. Kenne}, \bibinfo{person}{NhatHai Phan}, {and} \bibinfo{person}{Wei-Shinn Ku}.} \bibinfo{year}{2024}\natexlab{}.
\newblock \showarticletitle{User {Dynamics} and {Thematic} {Exploration} in r/{Depression} {During} the {COVID}-19 {Pandemic}: {Insights} {From} {Overlapping} r/{SuicideWatch} {Users}}.
\newblock \bibinfo{journal}{\emph{Journal of Medical Internet Research}} \bibinfo{volume}{26}, \bibinfo{number}{1} (\bibinfo{date}{May} \bibinfo{year}{2024}), \bibinfo{pages}{e53968}.
\newblock
\urldef\tempurl%
\url{https://doi.org/10.2196/53968}
\showDOI{\tempurl}
\newblock
\shownote{Company: Journal of Medical Internet Research Distributor: Journal of Medical Internet Research Institution: Journal of Medical Internet Research Label: Journal of Medical Internet Research Publisher: JMIR Publications Inc., Toronto, Canada}.


\bibitem[Zou et~al\mbox{.}(2024)]%
        {zou_self-disclosure_2024}
\bibfield{author}{\bibinfo{person}{Wenxue Zou}, \bibinfo{person}{Lu Tang}, \bibinfo{person}{Mi Zhou}, {and} \bibinfo{person}{Xinyu Zhang}.} \bibinfo{year}{2024}\natexlab{}.
\newblock \showarticletitle{Self-disclosure and received social support among women experiencing infertility on reddit: {A} natural language processing approach}.
\newblock \bibinfo{journal}{\emph{Computers in Human Behavior}}  \bibinfo{volume}{154} (\bibinfo{date}{May} \bibinfo{year}{2024}), \bibinfo{pages}{108159}.
\newblock
\showISSN{0747-5632}
\urldef\tempurl%
\url{https://doi.org/10.1016/j.chb.2024.108159}
\showDOI{\tempurl}


\end{thebibliography}


\appendix
\section{Interview Questions}\label{sec:interviewquestions}

\subsection*{A.1\ Roles and Responsibilities.}
Can you please briefly describe your day-to-day responsibilities and tasks as a professional spiritual care provider?
\begin{itemize}
\item Are there particular illnesses or types of people/patient groups you tend to work with regularly?
    \item What types of technology have already been integrated (or not) within your practice?
    \item What challenges and opportunities has this technology (or lack thereof) introduced?
\end{itemize}

\subsection*{A.2 \ Social Media Usage and Impacts.}

What social media platforms do your clients use commonly?
\begin{itemize}
    \item What positive or beneficial behaviors or impacts have you observed?
    \item What negative or detrimental behaviors or impacts have you observed?
    \item Have you ever made any recommendations or referrals to use or to avoid social media platforms, based on these positive or negative impacts?
\end{itemize}

\subsection*{A.3 \ Perspectives on Spiritual Support Definition and Social Media Integration}

Researcher presents participant with a piece of paper with the definition of “spiritual support” from (Smith et al., 2021) printed on it, and explains how this definition was derived with stakeholders of the online health blogging platform, CaringBridge.org:
\begin{quote}
\textbf{Spiritual support} is an integral dimension that underlies and can be expressed through every category of social support, including informational, emotional, instrumental, network, esteem, and prayer support. This dimension creates a triadic relationship between a recipient, a provider, and the sacred or significant, with the purpose of helping recipients and providers experience a mutually positive presence with each other, and with the sacred or significant.
\end{quote}
\begin{itemize}
    \item Do you have any thoughts or observations about how users exchange spiritual support across social media platforms, according to the provided definition?
    \item What are your initial thoughts on the prospect of spiritual care providers turning to social media platforms in order to provide online spiritual support?

\end{itemize}

\section{User Testing Sessions}\label{sec:usertestingsessions}
 
\subsection*{B.1\ Researcher Summary and Live Demo of Reddit.}
Researcher navigates through various subreddit communities (e.g., lonely, PTSD, etc.), discussing their features and demonstrating the details live.

Any questions? 
\subsection*{B.2 \ Exploring Different Subreddits}
Researcher turns keyboard over to participant, guiding participants with the following prompts:
\begin{itemize}
\item Now, I’d like to ask you to explore a few support-related subreddits that you think have the most relevance to your work as a spiritual care provider.
    \item Let’s start by choosing one of the examples I showed you, or we can search for a subreddit on a topic you think could be valuable to your clients.
\begin{quote}
    [Participant searches for and clicks into a subreddit]
\end{quote}
    \item What are the first things you notice about this subreddit?
    \item Please reflect on the “community description” and the “rules” described in the sidebar. How do you feel that these UI/UX affordances are providing effective or ineffective guidance for online spiritual support?
    \item Please click through a few posts and comments. As you’re doing that, please describe and reflect upon all of your thoughts out loud.
        \begin{itemize}
        \item What aspects of these online interactions are \textbf{healthy } and \textbf{beneficial} v.s. \textbf{unhealthy} or \textbf{problematic} for users?
        \item How do you view the online environment as affecting these interactions?
        \item Would you recommend that your clients visit this subreddit? Why or why not? (What would need to change before you felt you could recommend that your clients visit this subreddit?)
        \end{itemize}
    \item Please choose one specific post written by a user who is expressing \textbf{distress or uncertainty}. In your professional capacity as a spiritual care provider, please skim the comments on this post (if any), and then describe a strategy for how to intervene effectively to provide meaningful spiritual support. 
    \item Now, compose a comment in response to the OP or to another commenter in thread, verbalizing your thoughts as you write.
        \begin{itemize}
        \item Participant should type a comment into the real commenting interface, however the comment should not actually be submitted.
        \item Researcher should screenshot the conversation and prospective comment.
        \end{itemize}
    (Repeat it once or twice with new subreddits, if times allows.)
   
\end{itemize}

\subsection*{B.3 \ Reflecting on Subreddits}
    \subsubsection*{B.3.1} \ In what ways might \textbf{professional spiritual care providers} intervene in these spaces to maximize benefits to users?
    \begin{itemize}
        \item How might we provide \textbf{training}, guidance, or technical features (e.g., algorithms, bots, specialized rules, threads, tags, or spaces, etc.) to support this?
        \item Are there any types of features or platform affordances \textbf{missing} that you need to participate in these spaces, in a professional capacity?
    \end{itemize}
    \subsubsection*{B.3.2} \ In what ways do \textbf{users} need to behave in order to improve the quality of spiritual support exchanged?
    \begin{itemize}
        \item How do you view the \textbf{anonymity} of user profiles in relation to the ability of Reddit to provide meaningful spiritual care?
        \item How might we provide \textbf{training}, guidance, or technical features (e.g., algorithms, bots, specialized rules, threads, tags, or spaces, etc.) to support this?
    \end{itemize}

\subsection*{B.4 \ Reflections on Reddit as an Online Spiritual Support Space}
Assuming that \textbf{Reddit} **is** a suitable site for online spiritual support...
    \subsubsection*{B.4.1} \ What are the features that make this a safe or unsafe space for users?
    \subsubsection*{B.4.2} \ Should researchers and/or practitioners invest more effort in supporting existing subreddits or creating new subreddits?
    \subsubsection*{B.4.3} \ How should subreddits be ideally designed to support your clients?
    \begin{itemize}
        \item Affiliated with \textbf{specific disease} states?
        \item \textbf{Geographic} areas? \textbf{Clinics}? Religious or spiritual organizations or communities?
        \item For patients? Caregivers and supporters? Or both in the same space?
        \item How large or small should they be? Private or public?
        \item Who should be \textbf{moderating}? How? With what rules, authority, and level of identification (e.g., user handles with real names and affiliations, anonymous, etc.)
    \end{itemize}

\subsection*{B.5 \ Suitable Platform for Spiritual Support}
Given what you’ve seen today...
 \subsubsection*{B.5.1} \ Do you feel that a platform like Reddit is a suitable for the provision of meaningful spiritual support?
 \subsubsection*{B.5.2} \ Would another platform be better? (E.g., Facebook, Instagram, Twitter, CaringBridge, etc.) Why or why not?

\subsection*{B.6 \ Designing Online Spaces for Effective Spiritual Support}
If you had the opportunity to tell designers of social media platforms anything else about how to create online spaces that offer better opportunities for the provision of safe, effective, and trustworthy spiritual support, what would that be?

\end{document}